\documentclass[conference]{IEEEtran}
\usepackage{blindtext, graphicx}
\usepackage{cite}
\usepackage{multicol}

\usepackage{amssymb}
\usepackage{graphicx}
\usepackage{epstopdf}
\usepackage{url}
\urldef{\mailsa}\path|jiwan.ninglekhu@gmail.com, ram.krishnan@utsa.edu|

\usepackage[]{graphicx}
\usepackage{tabu}
\usepackage{amsmath}
\usepackage{textcomp}
\usepackage{fixltx2e}
\usepackage{amssymb}
\usepackage{ltablex,booktabs}
\usepackage[T1]{fontenc}

\usepackage{tabulary}
\usepackage{xspace}
\usepackage{floatrow}
\usepackage{algorithm}
\usepackage{algorithmic}
\usepackage{hhline}
\makeatletter
\DeclareRobustCommand{\textsupsub}[2]{{%
  \m@th\ensuremath{%
    ^{\mbox{\fontsize\sf@size\z@#1}}%
    _{\mbox{\fontsize\sf@size\z@#2}}%
  }%
}}
\makeatother

\newcommand\tab[1][1.27cm]{\hspace*{#1}}
\newcommand\stab[1][0.635cm]{\hspace*{#1}}

\newcommand{\U}{USERS}
\newcommand{\R}{ROLES}
\newcommand{\T}{T}
\newcommand{\tH}{TH}
\newcommand{\AU}{AU}
\newcommand{\PA}{PA}
\newcommand{\TA}{TA}
\renewcommand{\P}{PERMS}
\newcommand{\AR}{AR}
\newcommand{\ARH}{ARH}
\newcommand{\AUA}{AUA}
\newcommand{\CR}{CR}
\newcommand{\RH}{RH}
\newcommand{\UATT}{UATT}
\newcommand{\AATT}{AATT}

\newcommand{\PATT}{PATT}

\newcommand{\OP}{AOP}
\newcommand{\UA}{UA}
\newcommand{\RN}{ROLES}
\newcommand{\RNH}{RH}

\newcommand{\exmm}{exp\_mob\_mem}
\newcommand{\immm}{imp\_mob\_mem}
\newcommand{\eximm}{exp\_immob\_mem}
\newcommand{\imimm}{imp\_immob\_mem}
\newcommand{\isauth}{is\_authorized}
\newcommand{\isord}{is\_ordered}
\newcommand{\OU}{ORGU}
\newcommand{\OUH}{OUH}
\newcommand{\UUA}{UUA}
\newcommand{\PPA}{PPA}
\renewcommand{\ss}{\textsubscript}

\renewcommand{\it}{\textit}

\newcommand{\UP}{UP}
\newcommand{\UPH}{UPH}
\newcommand{\UUPA}{UUPA}
\newcommand{\AUH}{AUH}
\newcommand{\upau}{userpool\_adminunit}
\newcommand{\aur}{adminunit\_role}
\newcommand{\au}{admin\_unit}
\newcommand{\tadu}{task\_adminu}

\newcommand{\ups}{userpools}
\newcommand{\uam}{user\_am}
\newcommand{\ram}{role\_am}
\newcommand{\oam}{object\_am}
\newcommand{\classp}{classp}
\renewcommand{\t}{\texttt}
\newcommand{\q}{\textquotesingle}
\newcommand{\OI}{O}
\newcommand{\scope}{Scope}
\newcommand{\adroles}{assigned\_roles}

\usepackage[margin=1in,left=1in, includefoot]{geometry}
\usepackage{fancyhdr}
\usepackage{setspace}

\usepackage{booktabs}

\usepackage{enumitem}

\usepackage{cite}

\begin{document}

\title{Attribute Based Administration of Role Based Access Control : 
 A Detailed Description}

\author{\IEEEauthorblockN{Jiwan L. Ninglekhu}
\IEEEauthorblockA{Department of Electrical and Computer Engineering \\
The University of Texas at San Antonio\\
San Antonio, Texas 78249\\
Email: jiwan.ninglekhu@gmail.com}
\and 
\IEEEauthorblockN{Ram Krishnan}
\IEEEauthorblockA{Department of Electrical and Computer Engineering\\ 
The University of Texas at San Antonio\\
San Antonio, Texas 78249\\
Email: ram.krishnan@utsa.edu}}

\maketitle

\begin{abstract} 
  
  
  Administrative Role Based Access Control (ARBAC) models deal with how to
  manage user-role assignments (URA), permission-role assignments (PRA),
  and role-role assignments (RRA).  A wide-variety of approaches have been
  proposed in the literature for URA, PRA and RRA. In this paper, we
  propose attribute-based administrative models that unify many prior approaches for URA
  and PRA. The motivating factor is that attributes of various RBAC entities
  such as admin users, regular users and permissions can be used to 
  administer URA and PRA in a highly-flexible manner. We develop an
  attribute-based URA model called AURA and an attribute-based PRA model
  called ARPA. We demonstrate that AURA and ARPA can express and unify many
  prior URA and PRA models.

%
\end{abstract}


\IEEEpeerreviewmaketitle

\section{Introduction}

Role-based access control (RBAC)~\cite{sandhu2000nist,sandhu-rbac96} is a
well-adopted access control model in enterprise
settings~\cite{Oconnor2010nist}, and a well-studied access control model in
the academic community~\cite{Fuchs2011748}.  However, administration of
user-role, permission-role and role-role assignments (often referred to as
Administrative RBAC or ARBAC) is both a critical and challenging
task~\cite{Sandhu97}. For example, ARBAC focusses on assigning/revoking
users to/from roles, permission to/from roles, etc.  Many approaches have
been proposed in the literature for ARBAC~\cite{Sandhu97, Sandhu99, Oh02,
uarbac, BiswasUni}. Most of these approaches are role-driven---for example,
in URA97~\cite{Sandhu97}, user-role assignment is determined based on
prerequisite roles of the target user. Similarly, in URA99~\cite{Sandhu99},
it is determined based on the target users' current membership in mobile
and/or immobile roles.

Attribute-Based Access Control (ABAC) has recently gained significant
attention because of its flexibility~\cite{kuhn2010adding, huNISTguide,
jin2012unified, biswas2017attribute, bhatt2017abac}. Moreover, it has
proven to have the ability to represent different access control
models~\cite{jin2012unified}, as well as application in different
technology domains such as cloud and Internet of Things
(IoT)~\cite{jin2014role, alshehri2016access}. However, using ABAC for
administrative purposes has not been thoroughly explored.  In this paper,
we investigate an attribute-based approach for administration of RBAC.  In
the context of ARBAC, attributes allow for more flexibility in specifying
the conditions under which users and permissions can be assigned to roles.
For instance, the notions of prerequisite roles in ARBAC97~\cite{Sandhu97},
mobility of roles in ARBAC99~\cite{Sandhu99}, and organization unit in
ARBAC02~\cite{Oh02} can be captured as user attributes. Similarly, the
notion of administrative roles in the above models and the notion of
administrative unit in Uni-ARBAC~\cite{BiswasUni} can be captured as
attributes of administrative users. 

This allows for the attribute-based models we develop to express any of
these ARBAC models and beyond. That is, it allows our attribute-based
models to express any combination of features from prior models, and new
features that are not \emph{intuitively} expressible in those prior models. Thus,
this work is motivated largely by two critical factors: (a) since
administrative RBAC has been fairly explored in the literature, it is
timely to explore unification of these works into a coherent model that can
be configured to express prior models and beyond, and (b) a unified model
can be analyzed \emph{once} for various desirable security properties, and a
\emph{single} codebase can be generated to express prior models and beyond.

The contributions of this paper are two-fold:

\begin{itemize}

  \item We develop an attribute-based administrative model for user-role
    assignment (\textbf{AURA}) and permission-role assignment
    (\textbf{ARPA}). AURA deals with assigning/revoking users to/from roles
    that is determined based on the attributes of the administrative user
    and those of the regular (target) user. ARPA deals with
    assigning/revoking permissions to/from roles that is determined based
    on the attributes of the administrative user and those of the target
    permission(s).
\item Demonstrate that AURA and ARPA are capable of expressing many prior
  approaches to URA and PRA such as ARBAC97,
  ARBAC99, ARBAC02, UARBAC, and Uni-ARBAC.

\end{itemize}

The remainder of this paper is organized as follows. In
Section~\ref{sec:relatedwork}, we discuss related work.  In
Section~\ref{sec:aarbac}, AURA and ARPA models are presented as units of 
attribute based administration of RBAC (AARBAC). Sections~\ref{sec:URAtranslations} and  
~\ref{sec:PRAtranslations}
present algorithms that translate prior URA and PRA instances into equivalent AURA
 and ARPA instances. We conclude in Section~\ref{sec:conclude}.

\section{Related work}\label{sec:relatedwork}

This paper focuses on attribute-based user-role assignment (AURA) and
attribute-based permission-role assignment (ARPA). In particular, we
scope-out role-role assignment as future work. Therefore, in the following
discussion, we limit to related works on URA and PRA.

ARBAC97~\cite{Sandhu97}, ARBAC99\cite{Sandhu99}, ARBAC02\cite{Oh02},
Uni-ARBAC\cite{BiswasUni} and UARBAC\cite{uarbac} are some of the prominent
prior works in administrative RBAC. All these models deal with user-role
and permission-role assignments. 

In all these models, the policy for assigning a user or a permission to a
role is specified based on an explicit and a fixed set of properties of the
relevant entities that are involved in the decision-making process, namely,
the administrative user, the target role, and the regular user (or the
permission) that is assigned to the role. For example, in URA97, the
properties that are used include the \emph{admin role} of the
administrative user and the \emph{current set of roles} of the regular
user. Similarly, in UARBAC, the properties include a relationship based on
\emph{access modes} between the admin user, the target role and the regular
user. However, because AURA and ARPA are based on non-explicit and a
varying set of properties (attributes) of the relevant entities that are
involved in the decion-making process, these models are
more flexible.


A closely related work is that of Al-Kahtani et. al~\cite{al2002model},
which presents a family of models for automated assignment of users to
roles based on user attributes.  The primary focus of their work is
user-role assignment based on user attributes.  Our models take a more
holistic approach to RBAC administration based on attributes of various
RBAC entities such as regular users, admin users and permissions.  The
major advantage of taking such an approach is that our models both subsume
prior approaches to RBAC administration, and allow for specification of
novel policies.

The benefits of integrating attributes into an RBAC operational model
has been investigated in 
the literature~\cite{kuhn2010adding,jin2012rabac,rajpoot2015attributes}.
However, our work focuses on advantages of using an attribute-based
approach for RBAC administration.
Also, attribute-based
access
control~\cite{yuan2005attributed,jin2012unified,huNISTguide,servos2014hgabac} has been
well-studied. Such prior ABAC works primarily focus on operational aspects
of access control---that is, making decisions when a user requests access
to an object.


\section{AARBAC: AURA and ARPA Models}\label{sec:aarbac}

\begin{figure}[t]
  \centering
  \includegraphics[width=\linewidth]{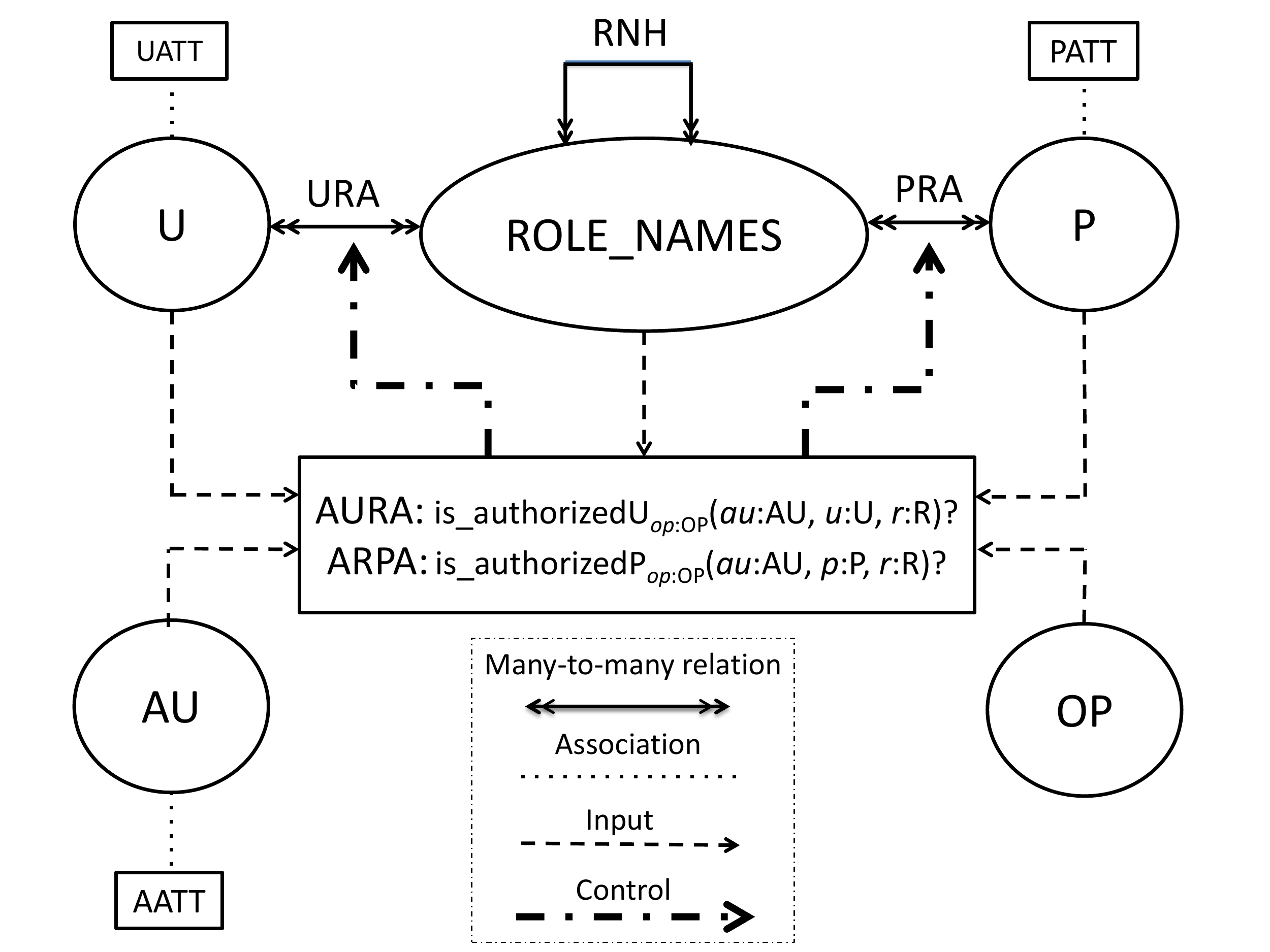}
  \caption{Attribute Based Administration of RBAC}
  \label{fig:fig1}
\end{figure}

In this section, we present our approach for attribute-based administrative
RBAC (AARBAC). We develop an attribute-based administrative model for
user-role assignment (called AURA) and an attribute-based administrative
model for permission-role assignment (called ARPA).  Figure~\ref{fig:fig1}
illustrates AURA and ARPA. The entities collectively involved in AURA and
ARPA include admin users and their associated attributes, regular users and
their attributes, permissions and their attributes, roles with a hierarchy,
and the operations.  In AURA, the admin users control the URA relation,
while in ARPA, they control the PRA relation.  Thus, in AURA, authorization
decisions for assigning a regular user to a role, which is an example of an
operation, is made based on attributes of the admin user and that of the
regular user. In ARPA, authorization decisions for assigning a permission
to a role is made based on attributes of the admin user and that of the
permission.  In the following subsections, the AURA and ARPA models are
presented in detail.

\begin{table*}[tbp]
\caption{AURA Model}
\centering
\label{tab:aura}
\begin{tabu} to 1\textwidth { | X[l] | }
\hline
-- \U\ is a finite set of regular users.\\
-- \AU\ is a finite set of administrative users.\\
-- \OP\ is a finite set of admin operations such as assign and revoke.\\ 
-- \R\ is a finite set of regular roles.\\
-- \RH\ $\subseteq$ \R\ $\times$ \R, a partial ordering on the set \R.\\

We assume a system maintained user attribute function called \it{roles}
that specifies the roles assigned to various regular users as follows:\\
-- \it{\adroles} : \U$\ $$\rightarrow$  2\textsuperscript{\R}\\

-- \UATT\ is a finite set of regular user attribute functions.\\
-- \AATT\ is a finite set of administrative user attribute functions.\\

-- For each \textit{att} in \UATT\ $\cup$ \AATT,
\scope(\it{att}) 
is a finite set of atomic values from which the range of the \\
\vspace*{-3ex}

\hspace{5pt} attribute function \it{att} is derived.
\vspace{1ex}

-- attType : \UATT$\ $$\cup$ \AATT$\ $$\rightarrow$ \{set, atomic\},
which specifies whether the range of a given attribute is atomic\\
\vspace{-2ex}

\hspace{5pt} or set valued.\\
-- Each attribute function maps elements in \U\ and \AU\ to atomic or set values.

\[
    \forall uatt \in \mbox{\UATT}.\ uatt : \mbox{\U} \rightarrow \left\{
                \begin{array}{ll}
                  \mbox{\scope}(uatt)\> \mbox{if} \> \mbox{attType}(uatt) = \mbox{atomic}\\
                  2\textsuperscript{\scope(\textit{uatt})}\> \mbox{if}\> \mbox{attType}(uatt) = \mbox{set}\\
                 
                \end{array}
              \right.
  \]
  \[
    \forall aatt \in \mbox{\AATT}.\ aatt : \mbox{\AU} \rightarrow \left\{
                \begin{array}{ll}
                  \mbox{\scope}(aatt)\> \mbox{if} \> \mbox{attType}(aatt) = \mbox{atomic}\\
                  2\textsuperscript{\scope(\textit{aatt})}\> \mbox{if}\> \mbox{attType}(aatt) = \mbox{set}\\
                 
                \end{array}
              \right.
  \]\\
-- \isord\ : \UATT\ $\cup$ \AATT\ $\rightarrow$ \{True, False\}, specifies
if the scope is ordered for each of the attributes.\\
-- For each \it{att} $\in$ \UATT\ $\cup$ \AATT,

\hspace{7pt}if \isord(\it{att}) = True, 
H\textsubscript{\it{att}} $\subseteq$ \scope(\it{att}) $\times$ \scope(\it{att}),
a partially ordered attribute hierarchy, and H\textsubscript{\it{att}} $\neq$
$\phi$,\\
\hspace{7pt}else, 
if \isord(\it{att}) = False, 
H\textsubscript{\it{att}} = $\phi$\\
\hspace{3pt}
(For some $att \in$ \UATT\ $\cup$ \AATT\
for which attType(\it{att}) = set and \isord(\it{att}) = True,
if
$\{a,b\}$, $\{c,d\}$ $\in$\\

\hspace{7pt}2\textsuperscript{\scope(\textit{att})}
(where $a, b, c, d \in$ \scope($att$)), 
we infer
$\{a,b\}$ $\geq$ $\{c,d\}$ if
$(a,c)$, $(a,d)$, $(b,c)$, $(b,d)$ $\in$ 
H\textsupsub{*}{\it{att}}.)
\vspace{3pt}\\

\hline
AURA model allows an administrator to perform an operation on a single user
or a set of users at a time. The authorization rule for performing an operation on a single user is as follows:\\

For each \it{op} in \OP, {\textbf{\isauth}}\textbf{U\textsubscript{\it{op}}}(\it{au}: \AU, \it{u} :
\U, \it{r} : \R) specifies if the admin user \it{au} is allowed to perform
the operation \it{op} (e.g. assign, revoke, etc.) between the
regular user \it{u} and the role \it{r}. 
This rule is written as a logical expression using attributes of the admin
user \it{au} and attributes of the 
regular user \it{u}.
\vspace{1em}\\
\hhline{|=|}
The authorization rule for performing an operation on a set of users is as follows:\\

For each \it{op} in \OP, {\textbf{\isauth}}\textbf{U\textsubscript{\it{op}}}(\it{au}: \AU, $\chi$ :
2\textsuperscript{\U}, \it{r} : \R) specifies if the admin user \it{au} is allowed to
perform the operation \it{op} (e.g. assign, revoke, etc.) between the users
in the set $\chi$ and the role \it{r}. 

Here $\chi$ is a set of users that can be specified using a set-builder
notation, whose rule is written using user attributes.
\vspace{1em}
\\
\hline
\end{tabu}
\end{table*}


Table~\ref{tab:aura} presents the formal AURA model. 
As illustrated in Figure~\ref{fig:fig1}, the entities involved in AURA
include regular users (\U), admin users (\AU), roles (\R) with a role
hierarchy (\RH), and admin operations (\OP). The goal of AURA is allow for
an admin user in \AU\ to perform an admin operation such as assign/revoke in
\OP\ between a regular user in \U\ and a role in \R, by using attributes of
various entities.
To meet this goal, we define a set of attribute functions for the regular users
(\UATT) and admin users (\AATT). 
One of the motivations for AURA is that we wanted AURA to have the ability
to capture the features of prior URA models such as URA97, URA99, and
URA02. As we will see, to this end, we only need to include attributes for
regular users and admin users. While one can envision attributes for other
entities in AURA such as attributes for \OP, we limit the scope of model design based on
the above-mentioned motivation.
In addition, we also assume a system maintained user
attribute function called \it{\adroles}, which maps each user to set of
roles currently assigned to them. Although the notion of roles can be
captured as a user attribute function in \UATT, we made this design choice in order to
reflect the fact that role is not an optional attribute in the context of
AURA.

The attribute functions (called simply attributes from now on) are defined
as a mapping from its domain (\U\ or \AU\ as the case may be) to its range.
The range of an attribute \it{att}, which can be atomic or set valued, is
derived from a specified set of scope of atomic values denoted
\scope(\it{att}). Whether an attribute is atomic or set valued is specified
by a function called attType. Also, the scope of an attribute can be either
ordered or unordered, which is specified by a function called \isord.
If an attribute \it{att} is ordered, we require that a corresponding
hierarchy, denoted H\textsubscript{\it{att}}, be specified on its scope
\scope(\it{att}). H\textsubscript{\it{att}} is a partial ordering on \scope(\it{att}).
Note that, even in the case of a set valued attribute \it{att}, the
hierarchy H\textsubscript{\it{att}} is specified on \scope(\it{att}) instead of
2\textsuperscript{\scope(\it{att})}. We infer the ordering between two set values given
an ordering on atomic values as explained in Table~\ref{tab:aura}. (Note
that 
H\textsupsub{*}{\it{att}}
denotes the reflexive transitive closure of
H\textsubscript{\it{att}}.)

AURA supports two ways to select a set of regular users for
assigning a role. The first
one allows an admin user to identify a single regular user, a role and perform an
operation such as assign. The second one allows an admin user to identify a
set of regular users, a role and perform an operation such as assign for
all those regular users. In this case, the selection criteria for the set
of regular users can be specified using a set-builder notation whose rule
is stated using the regular users' attributes. For example, 
{\isauth}U\textsubscript{\textbf{assign}}(\textbf{au},
\{\it{u} | \it{u} $\in$ \U\ $\wedge$ \textbf{aunit}
$\in$ \it{admin\_unit}($u$)\},
\textbf{r}) 
would specify a policy for an admin user \textbf{au} 
who identifies the set of all users who belong to
the admin unit \textbf{aunit} in order to assign a role \textbf{r} to
all those users.
Finally, the authorization rule is specified as a usual logical expression on
the attributes of admin users and those of regular users in question.
Examples can be seen in sections~\ref{sec:URAtranslations} and ~\ref{sec:PRAtranslations}.

\begin{table*}[tbp]
\caption{ARPA Model}
\centering
\label{tab:arpa}
\begin{tabu} to 1\textwidth { | X[l] | }
\hline
-- \AU\ is a finite set of administrative users.\\
-- \OP\ is a finite set of admin operations such as assign and revoke.\\ 
-- \R\ is a finite set of regular roles.\\
-- \RH\ $\subseteq$ \R\ $\times$ \R, a partial ordering on the set \R.\\
-- \P\ is a finite set of permissions.\\
-- \AATT\ is a finite set of administrative user attribute functions.\\
-- \PATT\ is a finite set of permission attribute functions.\\

-- For each \textit{att} in \AATT\ $\cup$ \PATT, 
\scope(\it{att}) 
is a finite set of atomic values from which the range of the \\
\vspace*{-3ex}

\hspace{5pt} attribute function \it{att} is derived.
\vspace{1ex}

-- attType : \AATT\ $\cup$ \PATT\ $\rightarrow$ \{set, atomic\}, 
which specifies whether the range of given attribute is atomic \\
\vspace{-2ex}

\hspace{5pt} or set valued. \\
-- Each attribute function maps elements in \AU\ and \P\ to atomic or set values.
   \[
    \forall aatt \in \mbox{\AATT}.\ aatt : \mbox{\AU} \rightarrow \left\{
                \begin{array}{ll}
                  \mbox{\scope}(aatt)\> \mbox{if} \> \mbox{attType}(aatt) = \mbox{atomic}\\
                  2\textsuperscript{\scope(\textit{aatt})}\> \mbox{if}\> \mbox{attType}(aatt) = \mbox{set}\\
                 
                \end{array}
              \right.
  \]\\
\[
    \forall patt \in \mbox{\PATT}.\ patt : \mbox{\P} \rightarrow \left\{
                \begin{array}{ll}
                  \mbox{\scope}(patt)\> \mbox{if} \> \mbox{attType}(patt) = \mbox{atomic}\\
                  2\textsuperscript{\scope(\textit{patt})}\> \mbox{if}\> \mbox{attType}(patt) = \mbox{set}\\
                 
                \end{array}
              \right.
  \]\\
  
-- \isord\ : \AATT\ $\cup$ \PATT\ $\rightarrow$ \{True, False\}, specifies 
if the scope is ordered for each of the attributes.\\
-- For each \it{att} $\in$ \AATT\ $\cup$ \PATT, \\
\hspace{7pt}if \isord(\it{att}) = True, 
H\textsubscript{\it{att}} $\subseteq$ \scope(\it{att}) $\times$ \scope(\it{att}), 
a partially ordered attribute hierarchy, and H\textsubscript{\it{att}} $\neq$ $\phi$\\
\hspace{7pt}else, 
if \isord(\it{att}) = False, 
H\textsubscript{\it{att}} = $\phi$ \\
\hspace{3pt}
(For some $att \in$ \PATT\ $\cup$ \AATT\
for which attType(\it{att}) = set and \isord(\it{att}) = True,
if
$\{a,b\}$, $\{c,d\}$ $\in$\\

\hspace{7pt}2\textsuperscript{\scope(\textit{att})}
(where $a, b, c, d \in$ \scope($att$)), 
we infer
$\{a,b\}$ $\geq$ $\{c,d\}$ if
$(a,c)$, $(a,d)$, $(b,c)$, $(b,d)$ $\in$ 
H\textsupsub{*}{\it{att}}.)
\vspace{3pt}\\

\hline
ARPA model allows an administrator to perform an operation on a single permission
or a set of permissions at a time. The authorization rule for performing an operation on a single permission is as follows:\\

For each \it{op} in \OP, {\textbf{\isauth}}\textbf{P\textsubscript{\it{op}}}(\it{au}: \AU, \it{p} : 
\P, \it{r} : \R) specifies if the admin user \it{au} is allowed to perform 
the operation \it{op} (e.g. assign, revoke, etc.) between the permission \it{p} and the role
 \it{r}. This rule is written as a logical expression using attributes of the admin user \it{au} and attributes of the permission \it{p}. 
\vspace{1em}\\
\hhline{|=|}
The authorization rule for performing an operation on a set of permissions is as follows.\\

For each \it{op} in \OP, {\textbf{\isauth}}\textbf{P\textsubscript{\it{op}}}(\it{au}: \AU, $\chi$ : 2\textsuperscript{\P}, \it{r} : \R) specifies if the admin user \it{au} is allowed to perform the operation \it{op} (e.g. assign, revoke, etc.) between the permissions in the set of $\chi$ and the role \it{r}. 

Here $\chi$ is a set of permissions that can be specified using a set-builder 
notation, whose rule is written using permission attributes.
\vspace{1em}
\\
\hline
\end{tabu}
\end{table*}

 The ARPA model presented in Table~\ref{tab:arpa} is very similar to the
 AURA model. The main difference is that since the focus of ARPA is about
 permission role assignment, it replaces regular users (\U) in AURA with 
 permissions (\P). Similarly, it replaces user attributes (\UATT) with permission
 attibutes (\PATT). Again, the motivation is that in order to capture the
 features of prior PRA models such
 as those in PRA02, UARBAC and Uni-ARBAC, we only need attributes for admin
 users and permissions. Similar to AURA, ARPA also supports two ways to select
 permissions to which an admin user can assign a role. A set of permissions
 can be selected using a set-builder notation whose rule is specified using
 permission attributes. Finally, the authorization rule is specified as a
 logical expression in the usual way over the attributes of the admin users
 and those of the permissions.

\section{Mapping Prior URA Models in AURA}\label{sec:URAtranslations}

In this section, we demonstrate that AURA can intuitively simulate
the features of prior URA models. In particular, we have developed
concrete algorithms that can convert any instance of URA97, URA99, URA02,
the URA model in UARBAC, and the URA model in Uni-ARBAC into an equivalent
instance of AURA. In the following sections also presents example instances 
of each of the prior URA models and their corresponding instances in 
AURA/ARPA model followed by a formal mapping alorithms.

\subsection{URA97 in AURA}
\subsubsection{URA97 Instance}\label{ura97instance}
In this section we present an example instance for URA97. \\
\underline{Sets and Functions}: 
\begin{itemize}
\item \U\ = \{\textbf{u\textsubscript1, u\textsubscript2, u\textsubscript3, u\textsubscript4}\} 
\item \R\ = \{\textbf{x\textsubscript{1}, x\textsubscript{2}, x\textsubscript{3}, x\textsubscript4, x\textsubscript5, x\textsubscript6}\}
\item \AR\ = \{\textbf{ar\textsubscript{1}, ar\textsubscript2}\}
\item \UA\ = \{(\textbf{u\textsubscript1, x\textsubscript1}), (\textbf{u\textsubscript1, x\textsubscript2}), (\textbf{u\textsubscript2, x\textsubscript3}), (\textbf{u\textsubscript2, x\textsubscript4})\}
\item \AUA\ = \{(\textbf{u\textsubscript3, ar\textsubscript1}), (\textbf{u\textsubscript4, ar\textsubscript2})\}
\item \RH\ = \{(\textbf{x\textsubscript1, x\textsubscript2}), (\textbf{x\textsubscript2, x\textsubscript3}), (\textbf{x\textsubscript3, x\textsubscript4}), (\textbf{x\textsubscript4, x\textsubscript5}), (\textbf{x\textsubscript5, x\textsubscript6})\}
\item \ARH\ = \{(\textbf{ar\textsubscript{1}, ar\textsubscript2})\}
\item \CR\ = \{\textbf{x\textsubscript{1} $\wedge$ x\textsubscript{2}}, \textbf{$\bar{\textrm{x}}$\textsubscript{1}} $\vee$ (\textbf{$\bar{\textrm{x}}$\textsubscript{2} $\wedge$ x\textsubscript{3}})\}
\end{itemize}

\noindent
Let \textit{cr\textsubscript{1}} = \textbf{x\textsubscript{1} $\wedge$ x\textsubscript{2}} and, \textit{cr\textsubscript{2}} = \textbf{$\bar{\textrm{x}}$\textsubscript{1} $\vee$ ($\bar{\textrm{x}}$\textsubscript{2} }$\wedge$ \textbf{x\textsubscript{3}}) be two prerequisite conditions. Prerequisite condition \it{cr\textsubscript{1}} is evaluated as follows:

\vspace{0.17cm}
\noindent
For any \textit{u} in \U\ undertaken for assignment, \\ 
($\exists$\textit{x} $\geq$ \textbf{x\textsubscript{1}}). (\textit{u}, \textit{x}) $\in$ \UA\ $\wedge$ ($\exists$\textit{x} $\geq$ \textbf{x\textsubscript{2}}). (\textit{u}, \textit{x}) $\in$ \UA

\vspace{0.17cm}
\noindent
\textit{cr\textsubscript{2}} is evaluated as follows:\\
For any \textit{u} in \U\ undertaken for assignment,\\
($\forall$\textit{x} $\geq$ \textbf{x\textsubscript{1}}). (\textit{u}, \textit{x}) $\notin$ \UA\ $\vee$ (($\forall$\textit{x} $\geq$ \textbf{x\textsubscript{2}}). (\textit{u}, \textit{x}) $\notin$ \UA\ $\wedge$ \\
($\exists$\textit{x} $\geq$ \textbf{x\textsubscript{3}}). (\textit{u}, \textit{x}) $\in$ \UA)

\vspace{0.17cm}
\noindent
Let \textit{can\_assign} and \textit{can\_revoke} be as follows:\\
\textit{can\_assign} = \{(\textbf{ar\textsubscript1}, \textit{cr\textsubscript1,} \{\textbf{x\textsubscript4, x\textsubscript5}\}), (\textbf{ar\textsubscript1}, \textit{cr\textsubscript2,} \{\textbf{x\textsubscript6}\})\}\\
\it{can\_revoke} = \{(\textbf{ar\textsubscript1}, \{\textbf{x\textsubscript4, x\textsubscript5, x\textsubscript6}\})\}\\
\subsubsection{Equivalent URA97 Instance in AURA}
In this segment AURA instance equivalent to aforementioned URA97 instance in presented based on the AURA model depicted in Table~\ref{tab:aura}.\\
\underline{Sets and functions}
\begin{itemize}
\item \U\ =\{\textbf{u\textsubscript1, u\textsubscript2, u\textsubscript3, u\textsubscript4}\}   
\item \AU\ = \{\textbf{u\textsubscript1, u\textsubscript2, u\textsubscript3, u\textsubscript4}\} 
\item \OP\ = \{\textbf{assign, revoke}\}
\item \R\ = \{\textbf{x\textsubscript{1}, x\textsubscript{2}, x\textsubscript3, x\textsubscript4, x\textsubscript5, x\textsubscript6}\}
\item \RH\ = \{(\textbf{x\textsubscript1, x\textsubscript2}), (\textbf{x\textsubscript2, x\textsubscript3}), (\textbf{x\textsubscript3, x\textsubscript4}), (\textbf{x\textsubscript4, x\textsubscript5}), (\textbf{x\textsubscript5, x\textsubscript6})\}
\item \it{\adroles}(\textbf{u\textsubscript1}) = \{\textbf{x\textsubscript1, x\textsubscript2}\}, 
\item[]\it{\adroles}(\textbf{u\textsubscript2}) = \{\textbf{x\textsubscript3, x\textsubscript4}\}
\item \UATT\ = \{\}
\item \AATT\ = \{\it{aroles}\}\\
\scope(\it{aroles}) = \{\textbf{ar\textsubscript1, ar\textsubscript2}\}, attType(\textit{aroles}) = set, \isord(\it{aroles}) = True, H\textsubscript{\it{aroles}} = \{(\textbf{ar\textsubscript1, ar\textsubscript2})\}
\item \it{aroles}(\textbf{u\textsubscript3}) = \{\textbf{ar\textsubscript1}\}, \it{aroles}(\textbf{u\textsubscript4}) = \{\textbf{ar\textsubscript2}\}
\end{itemize}
Authorization rules for user-role assignment and revocation for the given instance can be expressed respectively, as follows:\\
 -- {\isauth}U\textsubscript{\textbf{assign}}(\it{au} : \AU, \it{u} : \U, \it{r} : \R) \\
\hspace*{0.19cm}$\equiv$ (($\exists$\textit{ar} $\geq$ \textbf{ar\textsubscript1}). \textit{ar} $\in$ \textit{aroles}(\textit{au}) $\wedge$ 
\it{r} $\in$ \{\textbf{x\textsubscript4, x\textsubscript5}\} $\wedge$ \\
\hspace*{0.19cm}(($\exists$\textit{x} $\geq$ \textbf{x\textsubscript{1}}). \it{x} $\in$ \textit{roles}(\textit{u}) $\wedge$ ($\exists$\textit{x} $\geq$ \textbf{x\textsubscript{2}}). \it{x} $\in$ \textit{roles}(\textit{u})) $\vee$ \\
\hspace*{0.19cm}($\exists$\textit{ar} $\geq$ \textbf{ar\textsubscript1}). \textit{ar} $\in$ \textit{aroles}(\textit{au})  $\wedge$ \it{r} $\in$ \{\textbf{x\textsubscript6}\} $\wedge$ \\
\hspace*{0.19cm}(($\exists$\textit{x} $\geq$ \textbf{x\textsubscript{1}}). \it{x} $\notin$ \textit{roles}(\textit{u}) $\vee$ (($\exists$\textit{x} $\geq$ \textbf{x\textsubscript{2}}). \it{x} $\notin$ \textit{roles}(\textit{u}) $\wedge$ \\
\hspace*{0.19cm}($\exists$\textit{x} $\geq$ \textbf{x\textsubscript{3}}). \it{x} $\in$ \textit{roles}(\textit{u}))) 

\vspace*{0.2cm}
\noindent
 -- {\isauth}U\textsubscript{\textbf{revoke}}(\it{au} : \AU, \it{u} : \U, \it{r} : \R) \\
 \hspace*{0.19cm} $\equiv$   
 ($\exists$\textit{ar} $\geq$ \textbf{ar\textsubscript1}). \textit{ar} $\in$ \textit{aroles}(\textit{au}) $\wedge$ \it{r} $\in$ \{\textbf{x\textsubscript4, x\textsubscript5, x\textsubscript6}\}\\


\setcounter{algorithm}{0}
\begin{algorithm}
\caption{Map\textsubscript{URA97}}
\label{alg:alg1}
\begin{algorithmic} [1] 
\begin{spacing}{1.15}

\item[]\hspace{-17pt}\textbf{Input:} URA97 instance
\item[]\hspace{-17pt}\textbf{Output:} AURA instance 
\item[\textbf{Step 1:}]  \  /* Map basic sets and functions in AURA */
\item[] a. \U\textsuperscript{\t{A}} $\leftarrow$ \U\textsuperscript{\t97} ; \AU\ $\leftarrow$ \U\textsuperscript{\t{97}} \\
\item[] b. \OP\textsuperscript{\t{A}} $\leftarrow$ \{\textbf{assign, revoke}\}  
\item[] c. \R\textsuperscript{\t{A}} $\leftarrow$ \R\textsuperscript{\t{97}} ; \RH\textsuperscript{\t{A}} $\leftarrow$ \RH\textsuperscript{\t{97}} 
\item[] d. For each \it{u} $\in$ \U\textsuperscript{\t{A}}, \it{\adroles}\textsuperscript{\t{A}}(\it{u}) = $\phi$  
\item[] e. For each (\it{u, r}) $\in$ \UA\textsuperscript{\t{A}}, 
\item[] \stab \it{\adroles}\textsuperscript{\t{A}}(\it{u})\q\ = \it{\adroles}\textsuperscript{\t{A}}(\it{u}) $\cup$ \it{r}
\item[\textbf{Step 2:}]   \stab /* Map attribute functions in AURA */
\item[] a. \UATT\textsuperscript{\t{A}} $\leftarrow$ $\phi$ ; \AATT\textsuperscript{\t{A}} $\leftarrow$ \{\it{aroles}\} 
\item[] b. \scope\textsuperscript{\t{A}}(\it{aroles}) = \AR\textsuperscript{\t{97}} ; attType\textsuperscript{\t{A}}(\it{aroles}) = set 
\item[] c. \isord\textsuperscript{\t{A}}(\it{aroles}) = True ; H\textsupsub{\t{A}}{\it{aroles}} $\leftarrow$ \ARH\textsuperscript{\t{97}} 
\item[] d. For each \it{u} $\in$ \AU\textsuperscript{\t{A}}, \it{aroles}(\it{u}) = $\phi$  
\item[] e. For each (\it{u, ar}) in \AUA\textsuperscript{\t{97}}, 
\item[] \stab \it{aroles}(\it{u}) = \it{aroles}(\it{u}) $\cup$ \it{ar}
\item[\textbf{Step 3:}] \stab /* Construct assign rule in AURA */
\item[] a. assign\_formula = $\phi$ 
\item[] b. For each (\it{ar, cr, Z}) $\in$ \it{can\_assign}\textsuperscript{\t{97}}, 
\item[] \stab assign\_formula\textquotesingle\ = assign\_formula $\vee$ \\
\item[] \stab (($\exists$\textit{ar\textquotesingle\ $\geq$ ar}). \it{ar\textquotesingle} $\in$ \it{aroles}(\it{au}) $\wedge$ \it{r} $\in$ \it{Z} $\wedge$\\
\item[] \stab (\it{translate}(\it{cr})))
\item[] c. auth\_assign = {\isauth}U\textsubscript{\textbf{assign}}(\it{au} : \AU\textsuperscript{\t{A}}, \\
\item[] \stab \it{u} : \U\textsuperscript{\t{A}}, \it{r} : \R\textsuperscript{\t{A}}) $\equiv$ assign\_formula\textquotesingle
\item[\textbf{Step 4:}] \stab /* Construct revoke rule for AURA */
\item[] a. revoke\_formula = $\phi$
\item[] b. For each (\it{ar, Z}) $\in$ \it{can\_revoke}\textsuperscript{\t{97}}
\item[] \stab revoke\_formula\textquotesingle\ = revoke\_formula $\vee$ \\
\item[] \stab (($\exists$\textit{ar\textquotesingle\ $\geq$ ar}). \it{ar\textquotesingle} $\in$ \it{aroles}(\it{au}) $\wedge$ \it{r} $\in$ \it{Z})
\item[] c. auth\_revoke = {\isauth}U\textsubscript{\textbf{revoke}}(\it{au} : \AU\textsuperscript{\t{A}}, \\
\item[] \stab \it{u} : \U\textsuperscript{\t{A}}, \it{r} : \R\textsuperscript{\t{A}}) $\equiv$ revoke\_formula\textquotesingle
\vspace{-10pt}
\end{spacing}
\end{algorithmic}
\end{algorithm}
\floatname{algorithm}{Support routine for Algorithm}\label{support:ura97}
\setcounter{algorithm}{0}
\begin{algorithm}
\caption{\it{translate}\textsubscript{97}}
\begin{algorithmic} [1]
\item[]\hspace{-12pt}\textbf{Input:} A URA97 prerequisite condition, \it{cr}
\item[]\hspace{-12pt}\textbf{Output:} An equivalent sub-rule for AURA authorization assign rule.
\STATE \it{rule\_string} = $\phi$ 
\STATE For each \it{symbol} in \it{cr},
\STATE \stab \textbf{if} \it{symbol} is a role and in the form \it{x}
\item[] \tab \tab \stab (i.e., the user holds role \it{x}) 
\STATE \stab \stab \it{rule\_string\textquotesingle} = \it{rule\_string} +
\item[] \tab \tab   ($\exists$\it{x\textquotesingle}\ $\geq$ \it{x}). \it{x\textquotesingle}\ $\in$ \it{\adroles}\textsuperscript{\t{A}}(\it{u})
\STATE \stab \textbf{else if} \it{symbol} is a role and in the form $\bar{x}$
\item[] \tab \tab (i.e., the user doesn't hold role \it{x}) 
\STATE \stab \stab \it{rule\_string\textquotesingle} = \it{rule\_string} + 
\item[] \tab \tab ($\exists$\it{x\textquotesingle}\ $\geq$ \it{x}). \it{x\textquotesingle}\ $\notin$ \it{\adroles}\textsuperscript{\t{A}}(\it{u}) 
\STATE \stab \textbf{else}
\STATE \stab \stab \it{rule\_string\textquotesingle} = \it{rule\_string} + \it{symbol} \ 
\item[] \ \ \ /* where a \it{symbol} is a $\wedge$ or $\vee$ logical operator */
\STATE \stab \textbf{end if}
\end{algorithmic}
\end{algorithm}

\subsubsection{MAP\textsubscript{URA97}} \label{algosec:ura97}
Map\textsubscript{URA97} is an algorithm for mapping a URA97
instance into equivalent AURA instance. Sets and functions from URA97 and AURA are marked 
with superscripts \t{97} and \t{A}, respectively. Map\textsubscript{URA97} takes URA97 instance as its input. 
In particular, input for Map\textsubscript{URA97} fundamentally consists of \U\textsuperscript{\t{97}},
\R\textsuperscript{\t{97}}, \AR\textsuperscript{\t{97}}, \UA\textsuperscript{\t{97}} and \AUA\textsuperscript{\t{97}}, \RH\textsuperscript{\t{97}}, \ARH\textsuperscript{\t{97}}, \it{can\_assign}\textsuperscript{\t{97}} and
\it{can\_revoke}\textsuperscript{\t{97}}. 

Output from Map\textsubscript{URA97} algorithm is 
an equivalent AURA instance, with primarily consisting of \U\textsuperscript{\t{A}}, \AU\textsuperscript{\t{A}}, \OP\textsuperscript{\t{A}},
\R\textsuperscript{\t{A}}, \RH\textsuperscript{\t{A}}, For each \it{u} $\in$ \U\textsuperscript{\t{A}},
\it{\adroles}\textsuperscript{\t{A}}(\it{u}), \UATT\textsuperscript{\t{A}}, \AATT\textsuperscript{\t{A}}, For each
attribute \it{att} $\in$ \UATT\textsuperscript{\t{A}} $\cup$  \AATT\textsuperscript{\t{A}},
\scope\textsuperscript{\t{A}}(\it{att}), attType\textsuperscript{\t{A}}(\it{att}),
\isord\textsuperscript{\t{A}}(\it{att}) and H\textsuperscript{\t{A}}\textsubscript{\it{att}}, For each user
\it{u} $\in$ \U\textsuperscript{\t{A}}, and for each \it{att} $\in$ \UATT\textsuperscript{\t{A}}
$\cup$ \AATT\textsuperscript{\t{A}}, \it{att}(\it{u}), Authorization rule for assign
(auth\_assign) and Authorization rule for revoke (auth\_revoke). 

As indicated in Map\textsubscript{URA97}, there are four main steps for mapping. 
In Step 1, sets and functions from URA97 are mapped into AURA sets and functions. 
In Step 2, user attributes and administrative user attribute functions are
expressed. As we do not need user attributes in representing an equivalent AURA for URA97, \UATT\ is set to null. 
Admin user attribute \it{aroles} captures the notion of admin roles
in URA97. Step 3 involves constructing assign\_formula in AURA that is equivalent to
\it{can\_assign}\textsuperscript{\t{97}} in URA97. \it{can\_assign}\textsuperscript{\t{97}} is a set of triples. Each
triple bears information on whether an admin role can assign a candidate
user to a set of roles. 
%
Equivalent translation of \it{can\_assign}\textsuperscript{\t{97}} in AURA is given by {\isauth}U\textsubscript{\textbf{assign}}(\it{au} : \AU\textsuperscript{\t{A}}, \it{u} :
\U\textsuperscript{\t{A}}, \it{r} : \R\textsuperscript{\t{A}}). Similarly, In Step 4, revoke\_formula equivalent to \it{can\_revoke}\textsuperscript{\t{97}} is presented. A support routine \it{translate}\textsubscript{97} for Map\textsubscript{URA97} translates prerequisite condition. 


\setcounter{algorithm}{1}\label{algo:mapURA99}
\floatname{algorithm}{Algorithm}
\begin{algorithm}[tp]
\caption{Map\textsubscript{URA99}}
\begin{algorithmic} []
\begin{spacing}{1.2}
\item[] \hspace{-17pt}\textbf{Input:} URA99 instance 
\item[] \hspace{-17pt}\textbf{Output:} AURA instance
\item[\textbf{Step 1:}] \ \ /* Map basic sets and functions in AURA */
\item[] a. \U\textsuperscript{\t{A}} $\leftarrow$ \U\textsuperscript{\t{99}} ; \AU\textsuperscript{\t{A}} $\leftarrow$ \U\textsuperscript{\t{99}} 
\item[] b. \R\textsuperscript{\t{A}} $\leftarrow$ \R\textsuperscript{\t{99}} ; \RH\textsuperscript{\t{A}} $\leftarrow$ \RH\textsuperscript{\t{99}} 
\item[] c. \OP\textsuperscript{\t{A}} $\leftarrow$ \{\textbf{mob-assign, immob-assign, 
\item[] mob-revoke, immob-revoke}\} 
\item[\textbf{Step 2:}] \stab /* Map attribute functions in AURA */
\item[] a. \UATT\textsuperscript{\t{A}} = \{\it{\exmm, \immm,
\item[] \ \ \eximm, \imimm}\} 
\item[] b. \scope(\it{\exmm}) = \R\textsuperscript{\t{A}} 
\item[] c. attType(\it{\exmm}) = set 
\item[] d. is\_ordered(\it{\exmm}) = True
\item[] e. H\textsubscript{\it{\exmm}} = \RH\textsuperscript{\t{A}} ; For each \it{u} $\in$ \U\textsuperscript{\t{A}}, 
\item[]  \stab \it{\exmm}(\it{u}) = $\phi$
\item[] f. For each (\it{u,} M\it{r}) $\in$ UA\textsuperscript{\t{99}},
\item[] \ \  \it{\exmm}(\it{u}) = \it{\exmm}(\it{u}) $\cup$ \it{r} 
\item[] g. \scope(\it{\immm}) = \R\textsuperscript{\t{A}}\ 
\item[] h. attType(\it{\immm}) = set 
\item[] i. is\_ordered(\it{\immm}) = True 
\item[] j. H\textsubscript{\it{\immm}} = \RH\textsuperscript{\t{A}} ; For each \it{u} $\in$ \U\textsuperscript{\t{A}},
\item[] \stab \it{\immm}(\it{u}) = $\phi$ 
\item[] k. For each (\it{u,} M\it{r}) $\in$ UA\textsuperscript{\t{99}} and for each \it{r} > \it{r}\textquotesingle,
\item[] \stab \it{\immm}(\it{u}) = 
\item[] \tab \tab \it{\immm}(\it{u}) $\cup$ \it{r\textquotesingle}
\item[] l. \scope(\it{\eximm}) = \R\textsuperscript{\t{A}}
\item[] m. attType(\it{\eximm}) = set
\item[] n. is\_ordered(\it{\eximm}) = True
\item[] o. H\textsubscript{\it{\eximm}} = \RH\textsuperscript{\t{A}} 
\item[] p. For each \it{u} $\in$ \U\textsuperscript{\t{A}}, 
\item[] \tab \it{\eximm}(\it{u}) = $\phi$ 
\item[] q. For each (\it{u,} IM\it{r}) $\in$ UA\textsuperscript{\t{99}}, 
\item[] \stab \it{\eximm}(\it{u}) = 
\item[] \tab \tab \it{\eximm}(\it{u}) $\cup$ \it{r}
\item[] r. \scope(\it{\imimm}) = \R\textsuperscript{\t{A}} 
\item[] s. attType(\it{\imimm}) = set 
\item[] t. is\_ordered(\it{\imimm}) = True
\item[] u. H\textsubscript{\it{\imimm}} = \RH\textsuperscript{\t{A}} ; For each \it{u} $\in$ \U\textsuperscript{\t{A}}, 
\item[] \tab \it{\imimm}(\it{u}) = $\phi$
\item[] v. For each (\it{u,} IM\it{r}) $\in$ \UA\textsuperscript{\t{A}} and for each \it{r} > \it{r}\textquotesingle\ 
\item[] \stab \it{\imimm}(\it{u}) = 
\item[] \tab \tab \it{\imimm}(\it{u}) $\cup$ \it{r\textquotesingle}
\vspace{-0.5cm}
\end{spacing}
\end{algorithmic}
\end{algorithm}


\setcounter{algorithm}{1}
\floatname{algorithm}{Continuation of Algorithm}
\begin{algorithm}[]
\caption{Map\textsubscript{URA99}}   
\begin{algorithmic} []  
\begin{spacing}{1.2}

\item[] w. \AATT\textsuperscript{\t{A}} $\leftarrow$ \{\it{aroles}\} ; \scope(\it{aroles}) = \AR\textsuperscript{\t{99}}
\item[] x. attType(\it{aroles}) = set ; is\_ordered(\it{aroles}) = True 
\item[] y. H\textsubscript{\it{aroles}} = \RH\textsuperscript{\t{A}} ; For each \it{u} $\in$ \AU\textsuperscript{\t{A}}, \it{aroles}(\it{u}) = $\phi$ 
\item[] z. For each (\it{u, ar}) in \AUA\textsuperscript{\t{99}}, 
\item[] \tab \it{aroles}(\it{u}) = \it{aroles}(\it{u}) $\cup$ \it{ar}

\item[\textbf{Step 3:}] \tab /* Construct assign rule in AURA */
\item[] a. assign-mob-formula = $\phi$
\item[] b. For each (\it{ar, cr, Z}) $\in$ \it{can-assign-M}\textsuperscript{\t{99}}, 
\item[] \ \ \ \ assign-mob-formula\textquotesingle\ = \\
\item[] \stab assign-mob-formula $\vee$ (($\exists$\it{ar\textquotesingle\ $\geq$ ar}). \it{ar\textquotesingle} $\in$ \\
\item[] \stab \it{aroles}(\it{au}) $\wedge$ \it{r} $\in$ Z $\wedge$ (\it{translate}(\it{cr}, assign))) 
\item[] c. auth\_mob\_assign = 
\item[] \stab  {\isauth}U\textsubscript{\textbf{mob-assign}}(\it{au} : \AU\textsuperscript{\t{A}}, \it{u} : \U\textsuperscript{\t{A}}, \\
\item[] \stab \it{r} : \R\textsuperscript{\t{A}}) $\equiv$ assign-mob-formula\textquotesingle
 
\item[] d. assign-immob-formula = $\phi$
\item[] e. For each (\it{ar, cr, Z}) $\in$ \it{can-assign-IM}\textsuperscript{\t{99}}, 
\item[] \ \ \  assign-immob-formula\textquotesingle\ = \\
\item[] \stab assign-immob-formula  $\vee$ (($\exists$\it{ar\textquotesingle\ $\geq$ ar}). \it{ar\textquotesingle} 
\item[] \stab $\in$ \it{aroles}(\it{au}) $\wedge$ \it{r} $\in$ \it{Z} $\wedge$ (\it{translate}(\it{cr}, assign)))
\item[] f. auth\_immob\_assign = 
\item[] \stab {\isauth}U\textsubscript{\textbf{immob-assign}}(\it{au} : \AU\textsuperscript{\t{A}}, \it{u} : \U\textsuperscript{\t{A}}, 
\item[] \stab  \it{r} : \R\textsuperscript{\t{A}}) $\equiv$ assign-immob-formula\textquotesingle

\item[\textbf{Step 4:}] \stab /* Construct revoke rule in AURA */
\item[] a. revoke-mob-formula = $\phi$
\item[] b. For each (\it{ar, cr, Z}) $\in$ \it{can-revoke-M}\textsuperscript{\t{99}}, 
\item[] \stab revoke-mob-formula\textquotesingle\ = revoke-mob-formula $\vee$ \\
\item[] \stab(($\exists$\it{ar\textquotesingle\ $\geq$ ar}). \it{ar\textquotesingle} $\in$ \it{aroles}(\it{au}) $\wedge$ \it{r} $\in$ \it{Z} $\wedge$ 
\item[] \stab (\it{translate}(\it{cr}, revoke)))  
\item[] c. auth\_mob\_revoke = 
\item[] \stab {\isauth}U\textsubscript{\textbf{mob-revoke}}(\it{au} : \AU\textsuperscript{\t{A}}, \it{u} : \U\textsuperscript{\t{A}}, 
\item[] \stab \it{r} : \R\textsuperscript{\t{A}}) $\equiv$ revoke-mob-formula\textquotesingle
\item[] d. revoke-immob-formula = $\phi$
\item[] e. For each (\it{ar, cr, Z}) $\in$ \it{can-revoke-IM}\textsuperscript{\t{99}}, 
\item[] \stab revoke-immob-formula\textquotesingle\ = \\
\item[] \stab revoke-immob-formula $\vee$ (($\exists$\it{ar\textquotesingle\ $\geq$ ar}). \it{ar\textquotesingle} $\in$ 
\item[] \stab \it{aroles}(\it{au}) $\wedge$ \it{r} $\in$ \it{Z} $\wedge$ (\it{translate}(\it{cr}, revoke)))
\item[] f. auth\_immob\_revoke = 
\item[] \stab {\isauth}U\textsubscript{\textbf{immob-revoke}}(\it{au} : \AU\textsuperscript{\t{A}}, \it{u} : \U\textsuperscript{\t{A}}, 
\item[] \stab \it{r} : \R\textsuperscript{\t{A}}) $\equiv$ revoke-mob-formula\textquotesingle
\vspace{-0.5cm}
\end{spacing}
\end{algorithmic}
\end{algorithm}

\floatname{algorithm}{Support routine for algorithm}
\setcounter{algorithm}{1}
\begin{algorithm}
\label{algo:translate99}
\caption{\it{translate}\textsubscript{99}}
\begin{algorithmic} [1]
\begin{spacing}{}
\item[]\hspace{-17pt}\textbf{Input:} A URA99 prerequisite condition (\it{cr}), \\
\item[]\it{op} $\in$ \{\textbf{assign, revoke}\}
\item[]\hspace{-17pt}\textbf{Output:} An equivalent sub-rule for AURA authorization assign rule.
\STATE \it{rule\_string} = $\phi$ 
\STATE For each \it{symbol} in \it{cr}
\STATE \ \  \textbf{if} \it{op} = assign $\wedge$ \it{symbol} is a role \\
\item[] \stab \ \ \ and in the form \it{x} (i.e., the user holds role \it{x})
\STATE \ \ \ \ \it{rule\_string} = \it{rule\_string} + \it{x} $\in$ \it{\exmm}(\it{u})
\item[]\tab \stab \ $\vee$ (\it{x} $\in$ \it{\immm}(\it{u}) \\
\item[]\tab \stab \ $\wedge$ \it{x} $\notin$ \it{\eximm}(\it{u}))
\item[]\ \  \textbf{else if} \it{op} = revoke $\wedge$ \it{symbol} is a role 
\item[] \stab \ \ \ and in the form \it{x} (i.e., the user holds role \it{x})
\STATE \ \ \ \ \it{rule\_string} = \it{rule\_string} + (\it{x} $\in$ \it{\exmm}(\it{u}) 
\item[] \tab \stab \ $\vee$ \it{x} $\in$ \it{\immm}(\it{u}) 
\item[] \tab \stab \ $\vee$ \it{x} $\in$ \it{\eximm}(\it{u}) 
\item[] \tab \stab \ $\vee$ \it{x} $\in$ \it{\imimm}(\it{u})) 

\STATE \ \  \textbf{else if} \it{op} = assign $\vee$ revoke $\wedge$ \it{symbol} is role 
\item[] \stab \ \ \ and in the form $\bar{x}$ (i.e., the user doesn't hold 
\item[] \stab \ \ \ role \it{x})
\STATE \ \ \ \ \it{rule\_string} = \it{rule\_string} + (\it{x} $\notin$ \it{\exmm}(\it{u}) 
\item[] \tab \stab \ $\wedge$ \it{x} $\notin$ \it{\immm}(\it{u}) $\wedge$ 
\item[] \tab \stab \ \it{x} $\notin$ \it{\eximm}(\it{u}) $\wedge$ \item[] \tab \stab \ \it{x} $\notin$ \it{\imimm}(\it{u})) 
\STATE \ \   \textbf{else}
\STATE \ \ \ \ \it{rule\_string} = \it{rule\_string} + \it{symbol}\item[] \ \ \  /* where a \it{symbol} is a $\wedge$ or $\vee$ logical operator */
\STATE \stab \textbf{end if}
\end{spacing}
\end{algorithmic}
\end{algorithm}

\subsection{URA99 in AURA}
\subsubsection{URA99 Instance}\label{sec:ura99instance}
In this segment, we present an example instance of URA99 model as follows:
\underline{Sets and functions:}
\begin{itemize}
\item \U\ = \{\textbf{u\textsubscript1, u\textsubscript2, u\textsubscript3, u\textsubscript4}\}
\item \R\ = \{\textbf{x\textsubscript{1}, x\textsubscript{2}, x\textsubscript{3}, x\textsubscript{4}, x\textsubscript{5}, x\textsubscript{6}}\}
\item \AR\ = \{\textbf{ar\textsubscript{1}, ar\textsubscript2}\}
\item \UA\ = \{(\textbf{u\textsubscript1}, M\textbf{x\textsubscript1}), (\textbf{u\textsubscript2}, IM\textbf{x\textsubscript1}), (\textbf{u\textsubscript2}, IM\textbf{x\textsubscript2}), \\ 
(\textbf{u\textsubscript1}, IM\textbf{x\textsubscript3})\}
\item \AUA\ = \{(\textbf{u\textsubscript3, ar\textsubscript1}), (\textbf{u\textsubscript4, ar\textsubscript2})\}
\item \RH\ = \{(\textbf{x\textsubscript1, x\textsubscript2}), (\textbf{x\textsubscript2, x\textsubscript3}), (\textbf{x\textsubscript3, x\textsubscript4}), (\textbf{x\textsubscript4, x\textsubscript5}), (\textbf{x\textsubscript5, x\textsubscript6})\}
\item \ARH\ = \{(\textbf{ar\textsubscript{1}, ar\textsubscript2})\}
\item \CR\ = \{\textbf{x\textsubscript{1} $\wedge$ x\textsubscript{2}, $\bar{\textrm{x}}$\textsubscript{1}}\}
\end{itemize}

\vspace{0.17cm}
\noindent
Let \textit{cr\textsubscript{1}} = \textbf{x\textsubscript{1} $\wedge$ x\textsubscript{2}} and,  
 \textit{cr\textsubscript{2}} = \textbf{$\bar{\textrm{x}}$\textsubscript{1}}. \textit{cr\textsubscript{1}} is evaluated as follows:\\
For any user \it{u} $\in$ \U\ undertaken for assignment, \\
((\textit{u}, M\textbf{x\textsubscript1}) $\in$ \UA\ $\vee$ (($\exists$\it{x\textquotesingle} $\geq$ \textbf{x\textsubscript1}). (\textit{u}, M\textit{x\textquotesingle}) $\in$ \UA) $\wedge$ \\
 (\textit{u}, IM\textbf{x\textsubscript1}) $\notin$ \UA)) $\wedge$ ((\textit{u}, M\textbf{x\textsubscript2}) $\in$ \UA\ $\vee$ \\
 (($\exists$\textit{x\textquotesingle} $\geq$ \textbf{x\textsubscript2}). (\textit{u}, M\textit{x\textquotesingle}) $\in$ \UA) $\wedge$ (\textit{u}, IM\textbf{x\textsubscript2}) $\notin$ \UA)

\vspace{0.17cm}
\noindent
\textit{cr\textsubscript{2}} is evaluated as follows:\\
For any user \it{u} $\in$ \U\ undertaken for assignment, \\
(\textit{u}, M\textbf{x\textsubscript1}) $\notin$ \UA\ $\wedge$ (($\exists$\textit{x\textquotesingle} $\geq$ \textbf{x\textsubscript1}). (\textit{u}, M\textit{x\textquotesingle}) $\notin$ \UA) $\wedge$ \\
(\textit{u}, IM\textbf{x\textsubscript1}) $\notin$ \UA\ $\wedge$ (($\exists$\textit{x\textquotesingle} $\geq$ \textbf{x\textsubscript1}). (\textit{u}, IM\textit{x\textquotesingle}) $\notin$ \UA)
 
\vspace{0.17cm}
\noindent
Let \textit{can-assign-M} and \textit{can-assign-IM} in URA99 be as follows:\\
\textit{can-assign-M} = \{(\textbf{ar\textsubscript1}, \textit{cr\textsubscript1,} \{\textbf{x\textsubscript4, x\textsubscript5}\})\}\\
\textit{can-assign-IM}= \{(\textbf{ar\textsubscript1}, \textit{cr\textsubscript2,} \{\textbf{x\textsubscript5, x\textsubscript6}\})\}

\vspace{0.17cm}
\noindent
Unlike URA97, there is a notion of prerequisite condition in URA99 revoke model. We have considered same prerequisite conditions for both grant and revoke models instances for simplicity. Prerequisite conditions for URA99 revoke model are evaluated as follows:

\vspace{0.17cm}
\noindent
\it{cr\textsubscript1} is evaluated as follows:\\
For any user \it{u} $\in$ \U\ that needs to be revoked, \\
((\it{u}, M\textbf{x\textsubscript1}) $\in$ \UA\ $\vee$ (\it{u}, IM\textbf{x\textsubscript1}) $\in$ \UA\ $\vee$ (($\exists$\textit{x\textquotesingle} $\geq$ \textbf{x\textsubscript1}). \\
(\textit{u}, M\textit{x\textquotesingle}) $\in$ \UA) $\vee$  (($\exists$\textit{x\textquotesingle} $\geq$ \textbf{x\textsubscript1}). (\textit{u}, IM\textit{x\textquotesingle}) $\in$ \UA)) $\wedge$ \\
((\it{u}, M\textbf{x\textsubscript2}) $\in$ \UA\ $\vee$ (\it{u}, IM\textbf{x\textsubscript2}) $\in$ \UA\ $\vee$ (($\exists$\textit{x\textquotesingle} $\geq$ \textbf{x\textsubscript2}). \\ 
(\textit{u}, M\textit{x\textquotesingle}) $\in$ \UA) $\vee$  (($\exists$\textit{x\textquotesingle} $\geq$ \textbf{x\textsubscript2}). (\textit{u}, IM\textit{x\textquotesingle}) $\in$ \UA))

\vspace{0.17cm}
\noindent
\it{cr\textsubscript2} is evaluated as follows:\\
For any user \it{u} $\in$ \U\ that needs to be revoked, \\
(\it{u}, M\textbf{x\textsubscript1}) $\notin$ \UA\ $\wedge$ (\it{u}, IM\textbf{x\textsubscript1}) $\notin$ \UA\ $\wedge$  (($\exists$\textit{x\textquotesingle} $\geq$ \textbf{x\textsubscript1}). (\textit{u}, M\textit{x\textquotesingle}) $\notin$ \UA) $\wedge$
 (($\exists$\textit{x\textquotesingle} $\geq$ \textbf{x\textsubscript1}). (\textit{u}, IM\textit{x\textquotesingle}) $\notin$ \UA) 
 
\vspace{0.17cm}
\noindent
\textit{can-revoke-M} and \textit{can-revoke-IM} are as follows:\\
\textit{can-revoke-M} = \{(\textbf{ar\textsubscript1}, \textit{cr\textsubscript1,} \{\textbf{x\textsubscript3, x\textsubscript4, x\textsubscript5}\})\}\\
\textit{can-revoke-IM}= \{(\textbf{ar\textsubscript1}, \textit{cr\textsubscript2,} \{\textbf{x\textsubscript5, x\textsubscript6}\})\}\\

\subsubsection{Equivalent URA99 Instance in AURA}
\label{sec:aura99}
An equivalent AURA instance for aforementioned URA99 example instance is presented in this segment. \\
\underline{Set and functions:} 
\begin{itemize}
\item \U\ = \{\textbf{u\textsubscript1, u\textsubscript2, u\textsubscript3, u\textsubscript4}\}
\item \AU\ = \{\textbf{u\textsubscript1, u\textsubscript2, u\textsubscript3, u\textsubscript4}\}
\item \R\ = \{\textbf{x\textsubscript1, x\textsubscript2, x\textsubscript3, x\textsubscript4, x\textsubscript5, x\textsubscript6}\}
\item \RH\ = \{(\textbf{x\textsubscript1, x\textsubscript2}), (\textbf{x\textsubscript2, x\textsubscript3}), (\textbf{x\textsubscript3, x\textsubscript4}), (\textbf{x\textsubscript4, x\textsubscript5}), (\textbf{x\textsubscript5, x\textsubscript6})\}
\item \it{\adroles}(\textbf{u\textsubscript1}) = \{M\textbf{x\textsubscript1}, IM\textbf{x\textsubscript3}\}, 
\item[] \it{\adroles}(\textbf{u\textsubscript2}) = \{IM\textbf{x\textsubscript1}, IM\textbf{x\textsubscript3}\}
\item \OP\ = \{\textbf{mob-assign, immob-assign, mob-revoke, immob-revoke}\}
\item \UATT = \{\textit{\exmm, \immm, 
\item[]\eximm, \imimm}\}
\end{itemize}
\begin{itemize}
\item \scope(\it{\exmm}) = \R\
\item[] attType(\textit{\exmm}) = set
\item[] \isord(\it{\exmm}) = True
\item[] H\textsubscript{\it{\exmm}} = \RH\
\item \scope(\textit{\immm})= \R\
\item[] attType(\textit{\immm}) = set
\item[] \isord(\it{\immm}) = True 
\item[] H\textsubscript{\it{\immm}} = \RH\
\item \scope(\textit{\eximm}) = \R,
\item[] attType(\textit{\eximm}) = set
\item[] \isord(\it{\eximm}) = True
\item[] H\textsubscript{\it{\eximm}} = \RH\
\item \scope(\textit{\imimm}) = \R
\item[] attType(\textit{\imimm}) = set
\item[] \isord(\it{\imimm}) = True
\item[] H\textsubscript{\it{\imimm}} = \RH\
\item[] \AATT\ = \{\textit{aroles}\}
\item \scope(\textit{aroles}) = \{\textbf{ar\textsubscript{1}, ar\textsubscript{2}}\}
\item[] attType(\it{aroles}) = set, \isord(\it{aroles}) = True,
\item[] H\textsubscript{\it{aroles}} = \{(\textbf{ar\textsubscript{1}, ar\textsubscript{2}})\}
\item \it{aroles}(\textbf{u\textsubscript3}) = \{\textbf{ar\textsubscript1}\}, \it{aroles}(\textbf{u\textsubscript4}) = \{\textbf{ar\textsubscript2}\}
\end{itemize}
\vspace{.15cm}
Authorization rules for assignment and revocation of a user as a mobile member of role can be expressed
\vspace{.17cm}
respectively, as follows:\\
For any user \it{u} $\in$ \U, undertaken for assignment,\\
-- {\isauth}U\textsubscript{\textbf{mob-assign}}(\it{au} : \AR, \it{u} : \U, \\ 
\hspace*{.17cm} \it{r} : \R)  $\equiv$  \\
\hspace*{.17cm} (($\exists$\textit{ar} $\geq$ \textbf{ar\textsubscript1}). \textit{ar} $\in$ \textit{aroles}(\textit{u}) $\wedge$ \it{r} $\in$ \{\textbf{x\textsubscript4, x\textsubscript5}\} $\wedge$ (\textbf{x\textsubscript{1}} $\in$ \\
\hspace*{.17cm} \textit{\exmm}(\textit{u}) $\vee$ (\textbf{x\textsubscript{1}} $\in$ \textit{\immm}(\textit{u}) \\ 
\hspace*{.17cm} $\wedge$ \textbf{x\textsubscript{1}} $\notin$ \textit{\eximm}(\textit{u}))) $\wedge$
 (\textbf{x\textsubscript{2}} $\in$ \\
 \hspace*{.17cm} \textit{\exmm}(\textit{u}) $\vee$ (\textbf{x\textsubscript{2}} $\in$ 
 \textit{\immm}(\textit{u}) \\ 
 \hspace*{.17cm} $\wedge$ \textbf{x\textsubscript{2}} $\notin$ \textit{\eximm}(\textit{u})))) $\vee$  (($\exists$\textit{ar} $\geq$ \textbf{ar\textsubscript1}). \\
 \hspace*{.17cm} \textit{ar} $\in$ \textit{aroles}(\textit{u}) $\wedge$ \textit{r} $\in$ \{\textbf{x\textsubscript5, x\textsubscript6}\} $\wedge$ (\textbf{x\textsubscript{1}} $\notin$\\
 \hspace*{.17cm} \textit{\exmm} (\textit{u}) $\wedge$ \textbf{x\textsubscript{1}} $\notin$ \textit{\immm}(\textit{u}) \\
 \hspace*{.17cm} $\wedge$ \textbf{x\textsubscript{1}} $\notin$ \textit{\eximm}(\textit{u}) $\wedge$ \textbf{x\textsubscript{1}} $\notin$ \\
\vspace{.17cm}
\hspace*{.17cm}\textit{\imimm} (\textit{u})))\\
\noindent
For any user \it{u} $\in$ \U\ that needs to be revoked, \\
-- {\isauth}U\textsubscript{\textbf{mob-revoke}}(\it{au} : \AR, \it{u} : \U, \\ 
\hspace*{.17cm} \it{r} : \R)  $\equiv$ \\
\hspace*{.17cm} (($\exists$\textit{ar} $\geq$ \textbf{ar\textsubscript1}). \textit{ar} $\in$ \textit{aroles}(\textit{u}) $\wedge$ \textit{r} $\in$ \{\textbf{x\textsubscript3, x\textsubscript4, x\textsubscript5}\} $\wedge$ \\
\hspace*{.17cm} ((\textbf{x\textsubscript{1}} $\in$ \textit{\exmm}(\textit{u}) $\vee$ \textbf{x\textsubscript{1}} $\in$ \\
\hspace*{.17cm} \textit{\immm}(\textit{u}) $\vee$ \textbf{x\textsubscript{1}} $\in$ \textit{\eximm}(\textit{u}) \\
\hspace*{.17cm} $\vee$ \textbf{x\textsubscript{1}} $\in$ \textit{\imimm}) $\wedge$ (\textbf{x\textsubscript{2}} $\in$ \\
\hspace*{.17cm} \textit{\exmm}(\textit{u}) $\vee$ \textbf{x\textsubscript{2}} $\in$ \textit{\immm}(\textit{u})\\
\hspace*{.17cm} $\vee$ \textbf{x\textsubscript{2}} $\in$ \textit{\eximm}(\textit{u}) $\vee$ \textbf{x\textsubscript{2}} $\in$ \\
\hspace*{.17cm} \textit{\imimm})) $\vee$ (($\exists$\textit{ar} $\geq$ \textbf{ar\textsubscript1}). \textit{ar} $\in$ \textit{aroles}(\textit{u})\\
\hspace*{.17cm} $\wedge$ \textit{r} $\in$ \{\textbf{x\textsubscript5, x\textsubscript6}\} $\wedge$ 
\textbf{x\textsubscript{1}} $\notin$ \textit{\exmm}(\textit{u}) $\wedge$ 
\textbf{x\textsubscript{1}} $\notin$ \\
\hspace*{.17cm} \textit{\immm}(\textit{u}) $\wedge$ \textbf{x\textsubscript{1}} $\notin$ \textit{\eximm}(\textit{u}) \\ 
\hspace*{.17cm} $\wedge$ \textbf{x\textsubscript{1}} $\notin$ \textit{\imimm}(\textit{u}))

\vspace{.15cm}
Authorization rules for assignment and revocation of a user as an immobile member of role can be expressed
\vspace{.17cm}
respectively, as follows:\\
For any user \it{u} $\in$ \U, undertaken for assignment,\\
-- {\isauth}U\textsubscript{\textbf{immob-assign}}(\it{au} : \AR, \it{u} : \U, \\ 
\hspace*{.17cm} \it{r} : \R)  $\equiv$ \\
\hspace*{.17cm} (($\exists$\textit{ar} $\geq$ \textbf{ar\textsubscript1}). \textit{ar} $\in$ \textit{aroles}(\textit{u}) $\wedge$ \textit{r} $\in$ \{\textbf{x\textsubscript5, x\textsubscript6}\} $\wedge$ 
(\textbf{x\textsubscript{1}} $\in$ \\
\hspace*{.17cm} \textit{\exmm}(\textit{u}) $\vee$ (\textbf{x\textsubscript{1}} $\in$ \textit{\immm}(\textit{u}) \\
\hspace*{.17cm} $\wedge$ \textbf{x\textsubscript{1}} $\notin$ \textit{\eximm}(\textit{u}))) $\wedge$ 
 (\textbf{x\textsubscript{2}} $\in$ \\
 \hspace*{.17cm} \textit{\exmm}(\textit{u}) $\vee$ (\textbf{x\textsubscript{2}} $\in$ \textit{\immm}(\textit{u}) \\
 \hspace*{.17cm} $\wedge$ \textbf{x\textsubscript{2}} $\notin$ \textit{\eximm}(\textit{u})))) $\vee$ (($\exists$\textit{ar} $\geq$ \textbf{ar\textsubscript1}). \\
 \hspace*{.17cm} \it{ar} $\in$ \textit{aroles}(\textit{u}) $\wedge$ \textit{r} $\in$ \{\textbf{x\textsubscript5, x\textsubscript6}\} $\wedge$ (\textbf{x\textsubscript{1}} $\notin$ \\
 \hspace*{.17cm} \textit{\exmm}(\textit{u}) $\wedge$ \textbf{x\textsubscript{1}} $\notin$ \textit{\immm}(\textit{u}) \\
 \hspace*{.17cm} $\wedge$ \textbf{x\textsubscript{1}} $\notin$ \textit{\eximm}(\textit{u}) $\wedge$ \\
 \hspace*{.17cm} \textbf{x\textsubscript{1}} $\notin$ \textit{\imimm}(\textit{u}))) \\
\\
For any user \it{u} $\in$ \U\ that needs to be revoked, \\
-- {\isauth}U\textsubscript{\textbf{immob-revoke}}(\it{au} : \AR, \it{u} : \U, \\ 
\hspace*{.17cm} \it{r} : \R)  $\equiv$ \\
\hspace*{.17cm} (($\exists$\textit{ar} $\geq$ \textbf{ar\textsubscript1}). \textit{ar} $\in$ \textit{aroles}(\textit{u}) $\wedge$ \textit{r} $\in$ \{\textbf{x\textsubscript5, x\textsubscript6}\} $\wedge$\\
\hspace*{.17cm} ((\textbf{x\textsubscript{1}} $\in$ \textit{\exmm}(\textit{u}) $\vee$ \textbf{x\textsubscript{1}} $\in$ \\
\hspace*{.17cm} \textit{\immm}(\textit{u}) $\vee$ \textbf{x\textsubscript{1}} $\in$ \textit{\eximm}(\textit{u}) \\
\hspace*{.17cm} $\vee$ \textbf{x\textsubscript{1}} $\in$ \textit{\imimm}) $\wedge$ (\textbf{x\textsubscript{2}} $\in$ \\
\hspace*{.17cm} \textit{\exmm}(\textit{u}) $\vee$ \textbf{x\textsubscript{2}} $\in$ \textit{\immm}(\textit{u})\\
\hspace*{.17cm} $\vee$ \textbf{x\textsubscript{2}} $\in$ \textit{\eximm}(\textit{u}) $\vee$ \textbf{x\textsubscript{2}} $\in$ \\
\hspace*{.17cm} \textit{\imimm})) $\vee$ (($\exists$\textit{ar} $\geq$ \textbf{ar\textsubscript1}). \textit{ar} $\in$ \textit{aroles}(\textit{u}) \\
\hspace*{.17cm} $\wedge$ \textit{r} $\in$ \{\textbf{x\textsubscript5, x\textsubscript6}\} $\wedge$ \textbf{x\textsubscript{1}} $\notin$ \textit{\exmm}(\textit{u}) $\wedge$ \textbf{x\textsubscript{1}} $\notin$\\
\hspace*{.17cm} \textit{\immm}(\textit{u}) $\wedge$ \textbf{x\textsubscript{1}} $\notin$ \textit{\eximm}(\textit{u}) \\
\hspace*{.17cm} $\wedge$ \textbf{x\textsubscript{1}} $\notin$ \textit{\imimm}(\textit{u})) \\


\subsubsection{Map\textsubscript{URA99}} \label{algosec:ura99}
Algorithm Map\textsubscript{URA99} is an algorithm for mapping a URA99 instance into equivalent AURA instance. Sets and functions from URA99 and AURA are marked with superscripts \t{99} and \t{A}, respectively. Map\textsubscript{URA99} takes URA99 instance as its input. In particular, input for Map\textsubscript{URA99} fundamentally has \U\textsuperscript{\t{99}}, \R\textsuperscript{\t{99}}, \AR\textsuperscript{\t{99}}, \UA\textsuperscript{\t{99}}, \AUA\textsuperscript{\t{99}}, \RH\textsuperscript{\t{99}}, \ARH\textsuperscript{\t{99}}, \it{can-assign-M}\textsuperscript{\t{99}}, \it{can-assign-IM}\textsuperscript{\t{99}}, \it{can-revoke-M}\textsuperscript{\t{99}}, and \it{can-revoke-IM}\textsuperscript{\t{99}}.

Output from Map\textsubscript{URA99} algorithm is an equivalent AURA instance, with primarily consisting of following sets and functions: \U\textsuperscript{\t{A}}, \AU\textsuperscript{\t{A}}, \R\textsuperscript{\t{A}}, \RH\textsuperscript{\t{A}}, \OP\textsuperscript{\t{A}}, \UATT\textsuperscript{\t{A}}, \AATT\textsuperscript{\t{A}}, For each attribute \it{att} $\in$ \UATT\textsuperscript{\t{A}} $\cup$ \AATT\textsuperscript{\t{A}},  \scope(\it{att}), attType(\it{att}), \isord(\it{att}) and H\textsubscript{\it{att}}, For each user \it{u} $\in$ \U, and for each \it{aatt} $\in$ \AATT\textsuperscript{\t{A}}, \it{aatt}(\it{u}), For each user \it{u} $\in$ \U\textsuperscript{\t{A}}, and for each \it{uatt} $\in$ \UATT\textsuperscript{\t{A}}, \it{uatt}(\it{u}), Authorization rule for mobile assign (auth\_mob\_assign), Authorization rule for mobile revoke (auth\_mob\_revoke), Authorization rule for immobile assign (auth\_immob\_assign), and Authorization rule for immobile revoke (auth\_immob\_revoke).

As shown in Algorithm Map\textsubscript{URA99}, there are four main steps required in mapping any instance of URA99 model to AURA instance. In Step 1, sets and functions from URA99 instance are mapped into AURA sets and functions. In Step 2, user attributes and administrative user attribute functions are expressed. There are four regular user attributes, \it{\exmm, \immm, \eximm,} and \it{\imimm}. Each captures, a user's explicit mobile membership, implicit mobile membership, explicit immobile membership and implicit immobile membership on roles, respectively. Admin user attribute \it{aroles} captures admin roles assigned to admin users. Step 3 involves constructing assign-mob-formula and assign-immob-formula in AURA that is equivalent to \it{can-assign-M} and \it{can-assign-IM} in URA99, respectively, in URA99. Both \it{can-assign-M} and \it{can-assign-IM} are set of triples. Each triple bears information on whether an admin role can assign a candidate user to a set of roles as a mobile member in the case of \it{can-assign-M} and, as an immobile member in the case of \it{can-assign-IM}. AURA equivalent for \it{can-assign-M} is given by {\isauth}U\textsubscript{\textbf{mob-assign}}(\it{au} : \AU\textsuperscript{\t{A}}, \it{u} : \U\textsuperscript{\t{A}}, \it{r} : \R\textsuperscript{\t{A}}) and an equivalent translation for \it{can-assign-IM} is given by {\isauth}U\textsubscript{\textbf{immob-assign}}(\it{au} : \AU\textsuperscript{\t{A}}, \it{u} : \U\textsuperscript{\t{A}}, \it{r} : \R\textsuperscript{\t{A}}). Similarly, In Step 4, revoke-mob-formula equivalent to \it{can-revoke-M} and \it{can-revoke-IM} are presented. \it{translate}\textsubscript{99} is a support routine for Map\textsubscript{URA99} that translates prerequisite condition in URA99 into its AURA equivalent. A complete example instance and its corresponding equivalent AURA instances are presented in Section~\ref{sec:ura99instance} and Section~\ref{sec:aura99}, respectively.


\floatname{algorithm}{Algorithm}
\setcounter{algorithm}{2}
\begin{algorithm}[tp]
\caption{Map\textsubscript{URA02}}
\begin{algorithmic}[] 
\begin{spacing}{1.15}
\item[]\hspace{-17pt}\textbf{Input:} URA02 instance 
\item[]\hspace{-17pt}\textbf{Output:} AURA instance
\item[] \textbf{Begin:}
\item[\textbf{Step 1:}] \ \ /* Map basic sets and functions in AURA */
\item[] a. \U\textsuperscript{\t{A}} $\leftarrow$ \U\textsuperscript{\t{02}} ; \AU\textsuperscript{\t{A}} $\leftarrow$ \AU\textsuperscript{\t{02}} 
\item[] b. \OP\textsuperscript{\t{A}} $\leftarrow$ \{\textbf{assign, revoke}\} 
\item[] c. \R\textsuperscript{\t{A}} $\leftarrow$ \R\textsuperscript{\t{02}} 
\item[] d. \RH\textsuperscript{\t{A}} $\leftarrow$ \R\textsuperscript{\t{02}}
\item[] e. For each \it{u} $\in$ \U\textsuperscript{\t{A}}, \it{\adroles}(\it{u}) = $\phi$ ; 
\item[] f. For each (\it{u, r}) $\in$ \UA\textsuperscript{\t{02}}, \it{\adroles}(\it{u}) $\cup$ \it{r}
\item[\textbf{Step 2:}] \stab /* Map attribute functions to AURA */
\item[] a. \UATT\textsuperscript{\t{A}} $\leftarrow$ \{\it{org\_units}\} 
\item[] b. \scope(\it{org\_units}) = \OU\textsuperscript{\t{02}} 
\item[] c. attType(\it{org\_units}) =  set 
\item[] d. \isord(\it{org\_units}) = True ; H\textsubscript{\it{org\_units}} =  OUH\textsuperscript{\t{02}}
\item[] e. For each \it{u} $\in$ \U\textsuperscript{\t{A}}, \it{org\_units}(\it{u}) = $\phi$ ; 
\item[] f. For each (\it{u, orgu}) $\in$ \UUA\textsuperscript{\t{02}}, 
\item[] \tab \it{org\_units}(\it{u}) = \it{org\_units}(\it{u}) $\cup$ \it{orgu}

\item[] g. \AATT\textsuperscript{\t{A}} $\leftarrow$ \{\it{aroles}\} 
\item[] h. \scope(\it{aroles}) = \AR\textsuperscript{\t{02}} 
\item[] i. attType(\it{aroles}) = set
\item[] j. \isord(\it{aroles}) = True ; H\textsubscript{\it{aroles}} $\leftarrow$ \ARH\textsuperscript{\t{02}}
\item[] k. For each \it{u} $\in$ \AU\textsuperscript{\t{A}}, \it{aroles}(\it{u}) = $\phi$ 
\item[] l. For each (\it{u, ar}) in \AUA\textsuperscript{\t{02}}, 
\item[] \tab \it{aroles}(\it{u}) = \it{aroles}(\it{u}) $\cup$ \it{ar}

\item[\textbf{Step 3:}] \stab /* Construct assign rule in AURA */ 
\item[] a. assign\_formula = $\phi$
\item[] b. For each (\it{ar, cr, Z}) $\in$ \it{can\_assign}\textsuperscript{\t{02}}, 
\item[] \stab assign\_formula\textquotesingle\ = assign\_formula $\vee$ 
\item[]\stab (($\exists$\it{ar\textquotesingle\ $\geq$ ar}). \it{ar\textquotesingle} $\in$ \it{aroles}(\it{au}) $\wedge$ 
\item[] \stab \it{r $\in$ Z} $\wedge$
 (\it{translate}(\it{cr}))) 
\item[] c. auth\_assign = 
\item[]\hspace{0.17cm} {\isauth}U\textsubscript{\textbf{assign}}(\it{au} : \AU\textsuperscript{\t{A}}, \it{u} : \U\textsuperscript{\t{A}}, 
\item[]\hspace{0.17cm} \it{r} : \R\textsuperscript{\t{A}}) $\equiv$ assign\_formula\textquotesingle
\item[\textbf{Step 4:}]\stab /* Construct revoke rule in AURA */
\item[] a. revoke\_formula = $\phi$
\item[] b. For each (\it{ar, cr, Z}) $\in$ \it{can\_revoke}\textsuperscript{\t{02}}, 
\item[] \stab revoke\_formula\textquotesingle\ = revoke\_formula $\vee$ \\
\item[] \stab (($\exists$\it{ar\textquotesingle\ $\geq$ ar}). \it{ar\textquotesingle} $\in$ \it{aroles}(\it{au}) $\wedge$ \it{r} $\in$ \it{Z}) 
\item[] c. auth\_revoke = $\forall$\it{au} $\in$ \AU\textsuperscript{\t{A}}, $\forall$\it{u} $\in$ \U\textsuperscript{\t{A}}, $\forall$\it{r} $\in$ \RN\textsuperscript{\t{A}}.
\item[]\hspace{0.17cm} {\isauth}U\textsubscript{\textbf{revoke}}(\it{au} : \AU\textsuperscript{\t{A}}, \it{u} : \U\textsuperscript{\t{A}}, 
\item[]\hspace{0.17cm} \it{r} : \R\textsuperscript{\t{A}}) $\equiv$ revoke\_formula\textquotesingle
\vspace{-0.5cm}
\end{spacing}
\end{algorithmic}
\end{algorithm}
\floatname{algorithm}{Support routine for algorithm}
\setcounter{algorithm}{2}
\begin{algorithm}
\caption{\it{translate}\textsubscript{02}}
\begin{algorithmic}[1]
\begin{spacing}{1.1}
\item[]\hspace{-17pt}\textbf{Input:} A URA02 prerequisite condition (\it{cr}), \\
 \ \ \ \ Case 1, Case 2
\item[]\hspace{-17pt}\textbf{Output:} An equivalent sub-rule for AURA 
\item[] \stab authorization rule.
\STATE \it{rule\_string} = $\phi$ 
\STATE \textbf{Case Of} selection
\STATE \stab \textquotesingle\ Case 1 \textquotesingle\ (\it{cr} is based on roles) :
\STATE \tab \it{translate}\textsubscript{97} 

\STATE \stab \textquotesingle\ Case 2 \textquotesingle\ (\it{cr} is based on org\_units): 
\STATE \stab For each \it{symbol} in \it{cr}
\STATE \stab \ \ \ \textbf{if} \it{symbol} is an organization unit and in the
\item[] \tab \ \  form \it{x} (i.e., the user is a member of 
\item[] \tab \ \  organization unit \it{x})
\STATE \tab \it{rule\_string} = \it{rule\_string} + ($\exists$\textit{x\textquotesingle} $\leq$ \it{x}). \it{x\textquotesingle} 
\item[] \tab \tab $\in$ \textit{org\_units}(\textit{u}) 

\STATE \stab \ \ \ \textbf{else if} \it{symbol} an organization unit and in the 
\item[] \tab \ \ form $\bar{x}$ (i.e., the user is not a member of
\item[] \tab \ \ organization unit \it{x})
\STATE \tab \it{rule\_string} = \it{rule\_string} + ($\exists$\it{x\textquotesingle}\ $\leq$ \it{x}). \it{x\textquotesingle}\ 
\item[] \tab \tab $\notin$ \it{org\_units}(\it{u}) 

\STATE \stab \ \ \ \textbf{else}
\STATE \tab \it{rule\_string} = \it{rule\_string} + \it{symbol} 
\item[] \ /* where a \it{symbol} is a $\wedge$ or $\vee$ logical operator */
\STATE \stab \ \ \  \textbf{end if}

\STATE \textbf{end Case}
\item[] \textbf{End}
\vspace{-0.5cm}
\end{spacing}
\end{algorithmic}
\end{algorithm}

\subsection{URA02 in AURA}
In this section we present an example instance of URA02 model followed by an equivalent AURA instance.
 We also present a translation algorithm, Map\textsubscript{URA02} that converting URA02 instance to AURA instance.
\subsubsection{URA02 Instance}
\label{sec:ura02instance}
In URA02, decision to assign/revoke user-role can be made based on two factors, a user's membership on role or a user's membership in an organization unit. They can be viewed as two different cases. In this example instance we represent roles with \it{r} and organization units with \it{x} for simplicity. \\
\underline{Sets and functions:}
\begin{itemize}
\item \U\ = \{\textbf{u\textsubscript1, u\textsubscript2, u\textsubscript3, u\textsubscript4}\}
\item \R\ = \{\textbf{r\textsubscript1, r\textsubscript2, r\textsubscript3, r\textsubscript4, r\textsubscript5, r\textsubscript6}\}
\item \AR\ = \{\textbf{ar\textsubscript{1}, ar\textsubscript2}\}
\item \UA\ = \{(\textbf{u\textsubscript1, r\textsubscript1}), (\textbf{u\textsubscript1, r\textsubscript2}), (\textbf{u\textsubscript2, r\textsubscript3}), (\textbf{u\textsubscript2, r\textsubscript4})\}
\item \AUA\ = \{(\textbf{u\textsubscript3, ar\textsubscript1}), (\textbf{u\textsubscript4, ar\textsubscript2})\}
\item \RH\ = \{(\textbf{r\textsubscript1, r\textsubscript2}), (\textbf{r\textsubscript2, r\textsubscript3}), (\textbf{r\textsubscript3, r\textsubscript4}), (\textbf{r\textsubscript4, r\textsubscript5}), (\textbf{r\textsubscript5, r\textsubscript6})\}
\item \ARH\ = \{(\textbf{ar\textsubscript1, ar\textsubscript2})\}
\item \OU\ = \{\textbf{x\textsubscript{1}, x\textsubscript{2}, x\textsubscript{3}}\}
\item \OUH\ = \{(\textbf{x\textsubscript3, x\textsubscript{2}}), (\textbf{x\textsubscript{2}, x\textsubscript{1}})\}
\item \UUA\ = \{(\textbf{u\textsubscript1, x\textsubscript1}), (\textbf{u\textsubscript2, x\textsubscript3})\}
\\
\underline{Case 1:}
\item \CR\ = \{\textbf{r\textsubscript{1} $\wedge$ r\textsubscript{2}, r\textsubscript{1} $\vee$ $\bar{\textrm{r}}$\textsubscript{2} $\wedge$ x\textsubscript{3}}\}\\
\noindent
Let \textit{cr\textsubscript{1}} = \textbf{r\textsubscript{1} $\wedge$ r\textsubscript{2}} and, 
 \textit{cr\textsubscript{2}} = \textbf{r\textsubscript{1} $\vee$ $\bar{\textrm{r}}$\textsubscript{2} $\wedge$ r\textsubscript{3}}
\\
\underline{Case 2:}
\item \CR\ = \{\textbf{x\textsubscript{1} $\wedge$ x\textsubscript{2}, x\textsubscript{1} $\vee$ $\bar{\textrm{x}}$\textsubscript{2} $\wedge$ x\textsubscript{3}}\}
\end{itemize}
Let \textit{cr\textsubscript{3}} = \textbf{x\textsubscript{1} $\wedge$ x\textsubscript{2}} and, 
 \textit{cr\textsubscript{4}} = \textbf{x\textsubscript{1} $\vee$ $\bar{\textrm{x}}$\textsubscript{2} $\wedge$ x\textsubscript{3}}
\noindent \\
\underline{Case 1:}\\
\textit{cr\textsubscript{1}} is evaluated as follows:\\
For any user \textit{u} $\in$ \U\ undertaken for assginment,\\
($\exists$\textit{r} $\geq$ \textbf{r\textsubscript{1}}). (\textit{u}, \textit{r}) $\in$ \UA\ $\wedge$ ($\exists$\textit{r} $\geq$ \textbf{r\textsubscript{2}}). (\it{u}, \it{r}) $\in$ \UA

\vspace{0.12cm}
\noindent \textit{cr\textsubscript{2}} is evaluated as follows:\\
For any user \textit{u} $\in$ \U\ undertaken for assginment,\\
($\exists$\it{r} $\geq$ \textbf{r\textsubscript{1}}). (\textit{u}, \textit{r}) $\in$ \UA\ $\vee$ $\neg$(($\forall$\textit{r} $\geq$ \textbf{r\textsubscript{2}}). (\it{u}, \it{r}) $\in$ \UA) $\wedge$ ($\exists$\textit{r} $\geq$ \textbf{r\textsubscript{3}}). (\textit{u}, \textit{r}) $\in$ \UA\

\vspace{0.12cm}
\noindent \underline{Case 2:} \\
\textit{cr\textsubscript{3}} is evaluated as follows:\\
For any user \textit{u} $\in$ \U\ undertaken for assginment,\\
($\exists$\textit{x} $\leq$ \textbf{x\textsubscript{1}}). (\textit{u}, \textit{x}) $\in$ \UUA\ $\wedge$ ($\exists$\textit{x} $\leq$ \textbf{x\textsubscript{2}}). (\textit{u}, \textit{x}) $\in$ \UUA

\vspace{0.12cm}
\noindent\textit{cr\textsubscript{4}} is evaluated as follows:\\
For any user \textit{u} $\in$ \U\ undertaken for assginment,\\
($\exists$\textit{x} $\leq$ \textbf{x\textsubscript{1}}). (\textit{u}, \textit{x}) $\in$ \UUA\ $\vee$ $\neg$(($\forall$\textit{x} $\leq$ \textbf{x\textsubscript{2}}). (\textit{u}, \textit{x}) $\in$ \UUA) $\wedge$ ($\exists$\textit{x} $\leq$ \textbf{x\textsubscript{3}}). (\textit{u}, \textit{x}) $\in$ \UUA

\vspace{0.2cm}
\noindent \textit{can\_assign} and \it{can\_revoke} for respective cases are as follows:\\
\underline{Case 1:}\\
\textit{can\_assign} = \{(\textbf{ar\textsubscript1}, \textit{cr\textsubscript1,} \{\textbf{r\textsubscript4, r\textsubscript5}\}), (\textbf{ar\textsubscript1}, \textit{cr\textsubscript2,} \{\textbf{r\textsubscript6}\})\}\\
\it{can\_revoke} = \{(\textbf{ar\textsubscript1}, \{\textbf{r\textsubscript1, r\textsubscript3, r\textsubscript4}\})\}

\vspace{0.1cm}
\noindent\underline{Case 2:}\\
\textit{can\_assign} = \{(\textbf{ar\textsubscript1}, \textit{cr\textsubscript3,} \{\textbf{r\textsubscript4, r\textsubscript5}\}), (\textbf{ar\textsubscript1}, \textit{cr\textsubscript4,} \{\textbf{r\textsubscript6}\})\}\\
\it{can\_revoke} = \{(\textbf{ar\textsubscript1}, \{\textbf{r\textsubscript1, r\textsubscript3, r\textsubscript4}\})\}

\vspace{0.18cm}
\noindent \subsubsection{Equivalent URA02 Instance in AURA}
\label{sec:aura02}
\begin{itemize}
\item \U\ = \{\textbf{u\textsubscript1, u\textsubscript2, u\textsubscript3, u\textsubscript4}\}
\item \AU\ = \{\textbf{u\textsubscript1, u\textsubscript2, u\textsubscript3, u\textsubscript4}\}
\item \OP\ = \{\textbf{assign, revoke}\}
\item \R\  = \{\textbf{r\textsubscript1, r\textsubscript2, r\textsubscript3, r\textsubscript4, r\textsubscript5, r\textsubscript6}\}
\item \RH\ = \{(\textbf{r\textsubscript1, r\textsubscript2}), (\textbf{r\textsubscript2, r\textsubscript3}), (\textbf{r\textsubscript3, r\textsubscript4}), (\textbf{r\textsubscript4, r\textsubscript5}), (\textbf{r\textsubscript5, r\textsubscript6})\}
\item \it{\adroles}(\textbf{u\textsubscript1}) = \{\textbf{r\textsubscript1, r\textsubscript2}\}, 
\item[]\it{\adroles}(\textbf{u\textsubscript2}) = \{\textbf{r\textsubscript3, r\textsubscript4}\}
\item \UATT\ = \{\textit{org\_units}\}
\item\scope(\textit{org\_units}) =  \{\textbf{x\textsubscript{1}, x\textsubscript{2}, x\textsubscript{3}}\}, 
\item[] attType(\textit{org\_units}) = set
\item[] {\isord}(\it{org\_units}) = True, 
\item[] H\textsubscript{\it{org\_units}} = \{(\textbf{x\textsubscript3, x\textsubscript{2}}), (\textbf{x\textsubscript{2}, x\textsubscript{1}})\}
\item \it{org\_units}(\textbf{u\textsubscript1}) = \{\textbf{x\textsubscript1}\}, \it{org\_units}(\textbf{u\textsubscript2}) = \{\textbf{x\textsubscript3}\}
\item \AATT\ = \{\textit{aroles}\}
\item \scope(\textit{aroles}) = \{\textbf{ar\textsubscript{1}, ar\textsubscript2}\}, 
\item[] attType(\textit{aroles}) = set
{\isord}(\it{aroles}) = True, 
\item[] H\textsubscript{\it{aroles}} =  \{(\textbf{ar\textsubscript{1}, ar\textsubscript2})\}
\item \it{aroles}(\textbf{u\textsubscript3}) = \{\textbf{ar\textsubscript1}\}, \it{aroles}(\textbf{u\textsubscript4}) = \{\textbf{ar\textsubscript2}\}
\end{itemize}
For each \it{op} in \OP, authorization rule for user to role assignment and revocation can be expressed respectively, as follows:

\vspace{0.1cm}
\noindent \underline{Case 1:}\\
For any user \it{u} $\in$ \U, undertaken for assignment,\\
-- {\isauth}U\textsubscript{\textbf{assign}}(\it{au} : \AU, \it{u} : \U, \it{r} : \R) $\equiv$ 
(($\exists$\textit{ar} $\geq$ \textbf{ar\textsubscript1}). \textit{ar} $\in$ \textit{aroles}(\textit{u}) $\wedge$ \it{r} $\in$ \{\textbf{r\textsubscript4, r\textsubscript5}\} $\wedge$ \\
\hspace*{7pt} (($\exists$\textit{r} $\geq$ \textbf{r\textsubscript{1}}). \it{r} $\in$ \textit{\adroles}(\textit{u}) $\wedge$ ($\exists$\textit{r} $\geq$ \textbf{r\textsubscript{2}}). \\
\hspace*{7pt} \it{r} $\in$ \textit{\adroles}(\textit{u}))) $\vee$ 
 (($\exists$\textit{ar} $\geq$ \textbf{ar\textsubscript1}). 
 \textit{ar} $\in$ \textit{aroles}(\textit{u}) 
  \hspace*{7pt} $\wedge$ \it{r} $\in$ \{\textbf{r\textsubscript6}\} $\wedge$ 
(($\exists$\textit{r} $\geq$ \textbf{r\textsubscript{1}}). \it{r} $\in$ \textit{\adroles}(\textit{u}) $\vee$
 \\
 \hspace*{7pt} ($\exists$\textit{r} $\geq$ \textbf{r\textsubscript{2}}). \it{r} $\notin$ \textit{\adroles}(\textit{u}) $\wedge$ ($\exists$\textit{r} $\geq$ \textbf{r\textsubscript{3}}).\\ 
 \hspace*{7pt} \it{r} $\in$ \textit{\adroles}(\textit{u})))\\
\\

 For any user \textit{u} $\in$ \U\ undertaken for revocation,\\
-- {\isauth}U\textsubscript{\textbf{revoke}}(\it{au} : \AU, \it{u} : \U, \\
\hspace*{7pt} \it{r} : \R) $\equiv$ ($\exists$\textit{ar} $\geq$ \textbf{ar\textsubscript1}). \textit{ar} $\in$ \textit{aroles}(\textit{u}) $\wedge$
\\
\hspace*{7pt} \it{r} $\in$ \{\textbf{r\textsubscript1, r\textsubscript3, r\textsubscript4}\}\\
\\
\underline{Case 2:}\\
For any user \it{u} $\in$ \U, undertaken for assignment,\\
-- {\isauth}U\textsubscript{\textbf{assign}}(\it{au} : \AU, \it{u} : \U, \it{r} : \R) $\equiv$ 
(($\exists$\textit{ar} $\geq$ \textbf{ar\textsubscript1}). \textit{ar} $\in$ \textit{aroles}(\textit{u}) $\wedge$ \it{r} $\in$ \{\textbf{r\textsubscript4, r\textsubscript5}\} $\wedge$ \\
\hspace*{7pt} (($\exists$\textit{x} $\leq$ \textbf{x\textsubscript{1}}). \it{x} $\in$ \textit{org\_units}(\textit{u}) $\wedge$ ($\exists$\textit{x} $\leq$ \textbf{x\textsubscript{2}}). \\
\hspace*{7pt} \it{x} $\in$ \textit{org\_units}(\textit{u}))) $\vee$
 (($\exists$\textit{ar} $\geq$ \textbf{ar\textsubscript1}). \textit{ar} $\in$ \textit{aroles}(\textit{u}) $\wedge$\\
  \hspace*{7pt} \it{r} $\in$ \{\textbf{r\textsubscript6}\} $\wedge$ 
(($\exists$\textit{x} $\leq$ \textbf{x\textsubscript{1}}). \it{x} $\in$ \textit{org\_units}(\textit{u}) $\vee$ ($\exists$\textit{x} $\leq$ \textbf{x\textsubscript{2}}). \\ 
\hspace*{7pt} \it{x} $\notin$ \textit{org\_units}(\textit{u}) $\wedge$ ($\exists$\textit{x} $\leq$ \textbf{x\textsubscript{3}}). \it{x} $\in$ \textit{org\_units}(\textit{u})))

\vspace{0.1cm}
\noindent
 For any user \textit{u} $\in$ \U\ undertaken for revocation,\\
\noindent -- {\isauth}U\textsubscript{\textbf{revoke}}(\it{au} : \AU, \it{u} : \U, \\
\hspace*{7pt} \it{r} : \R) $\equiv$ 
($\exists$\textit{ar} $\geq$ \textbf{ar\textsubscript1}). \textit{ar} $\in$ \textit{aroles}(\textit{u}) $\wedge$ \\
\hspace*{7pt} \it{r} $\in$ \{\textbf{r\textsubscript1, r\textsubscript3, r\textsubscript4}\}

\subsubsection{Map\textsubscript{URA02}}

Algorithm Map\textsubscript{URA02} is an algorithm for mapping a URA02 instance into equivalent AURA instance. Sets and functions from URA02 and AURA are marked with superscripts \t{02} and \t{A}, respectively. Map\textsubscript{URA02} takes URA02 instance as its input. In particular, input for Map\textsubscript{URA02} fundamentally has \U\textsuperscript{\t{02}}, \R\textsuperscript{\t{02}}, \AR\textsuperscript{\t{02}}, \UA\textsuperscript{\t{02}}, \AUA\textsuperscript{\t{02}}, \RH\textsuperscript{\t{02}}, \ARH\textsuperscript{\t{02}}, \it{can\_assign}\textsuperscript{\t{02}}, \it{can\_revoke}\textsuperscript{\t{02}}, \OU\textsuperscript{\t{02}}, OUH\textsuperscript{\t{02}}, and \UUA\textsuperscript{\t{02}}.

Output from Map\textsubscript{URA02} algorithm is an equivalent AURA instance, with primarily consisting of following sets and functions: \U\textsuperscript{\t{A}}, \AU\textsuperscript{\t{A}}, \OP\textsuperscript{\t{A}}, \R\textsuperscript{\t{A}}, \RH\textsuperscript{\t{A}}, For each \it{u} $\in$ \U\textsuperscript{\t{A}}, \it{\adroles}(\it{u}), \UATT\textsuperscript{\t{A}}, \AATT\textsuperscript{\t{A}}, For each attribute \it{att} $\in$ \UATT\textsuperscript{\t{A}} $\cup$  \AATT\textsuperscript{\t{A}}, 
\scope(\it{att}), attType(\it{att}), \isord(\it{att}) and H\textsubscript{\it{att}}, For each user \it{u} $\in$ \U\textsuperscript{\t{A}}, \it{aroles}(\it{u}) and \it{org\_units}(\it{u}), Authorization rule for assign (auth\_assign), and Authorization rule to revoke (auth\_revoke)

A shown in Algorithm Map\textsubscript{URA02}, there are four main steps required in mapping any instance of URA02 model to AURA instance. In Step 1, sets and functions from URA02 instance are mapped into AURA sets and functions. In Step 2, user attributes and administrative user attribute functions are expressed. \UATT\ set has one user attribute called \it{org\_units}. This attribute captures a regular user's appointment or association in an organization unit. There are two ways a user assignment decision is made in URA02 which are marked as Case 1 and Case 2 in the model. Case 1 checks for user's existing membership on roles while Case 2 checks for user's membership on organization units. \it{org\_units} captures Case 2. Case 1 is same as URA97. Admin user attribute \it{aroles} captures admin roles assigned to admin users. Step 3 involves constructing assign\_formula in AURA that is equivalent to \it{can\_assign}\textsuperscript{\t{02}} in URA02. \it{can\_assign}\textsuperscript{\t{02}} is a set of triples. Each triple bears information on whether an admin role can assign a candidate user to a set of roles. Equivalent translation in AURA for URA02 is given by {\isauth}U\textsubscript{\textbf{assign}}(\it{au} : \AU\textsuperscript{\t{A}}, \it{u} : \U\textsuperscript{\t{A}}, \it{r} : \R\textsuperscript{\t{A}}). Similarly, In Step 4, revoke\_formula equivalent to \it{can\_revoke}\textsuperscript{\t{02}} is presented. A support routine for Map\textsubscript{URA02}, \it{translate}\textsubscript{02} translates prerequisite condition in URA02 into its equivalent in AURA. A complete example instance and its corresponding equivalent AURA instances were presented in Section~\ref{sec:ura02instance} and Section~\ref{sec:aura02}, respectively.

\subsection{Uni-ARBAC's URA in AURA}
In this section we present an example instance for URA in Uni-ARBAC (URA-Uni) and its equivalent AURA instance. We also present an algorithm that translates any given URA-Uni instance to AURA instance. \\

\subsubsection{Uni-ARBAC's URA Instance}\label{sec:ura-uni}
This segment presents an instance of URA in Uni-ARBAC model.\\
\underline{Sets and functions:}
\begin{itemize}
\item \U\ = \{\textbf{u\textsubscript1, u\textsubscript2, u\textsubscript3, u\textsubscript4}\}
\item \R\ = \{\textbf{r\textsubscript{1}, r\textsubscript{2}, r\textsubscript3}\}
\item \RH\ = \{(\textbf{r\textsubscript1, r\textsubscript2}), (\textbf{r\textsubscript2, r\textsubscript3})\}
\item \UA\ = \{(\textbf{u\textsubscript3, r\textsubscript1}), (\textbf{u\textsubscript4, r\textsubscript3})\}
\end{itemize}
\underline{user-pools sets and relations}
\begin{itemize}
\item \UPH\ = \{(\textbf{up\textsubscript2, up\textsubscript1})\}
\item \UUPA\ = \{\textbf{(u\textsubscript1, up\textsubscript1), (u\textsubscript2, up\textsubscript2), (u\textsubscript3, up\textsubscript1), (u\textsubscript4, up\textsubscript2)}\}
\end{itemize}
\underline{Administrative Units and Partitioned Assignments}
\begin{itemize}
\item \AU\ = \{\textbf{au\textsubscript1, au\textsubscript2}\}
\item \it{roles}(\textbf{au\textsubscript1}) = \{\textbf{r\textsubscript1, r\textsubscript2}\}, \it{roles}(\textbf{au\textsubscript2}) = \{\textbf{r\textsubscript3}\}
\item \it{user\_pools}(\textbf{au\textsubscript1}) = \{\textbf{up\textsubscript1}\}, \it{user\_pools}(\textbf{au\textsubscript2}) = \{\textbf{up\textsubscript2}\}
\end{itemize}
\underline{Derived Function}
\begin{itemize}
\item \it{user\_pools}*(\textbf{au\textsubscript1}) = \{\textbf{up\textsubscript1}\}
\item \it{user\_pools}*(\textbf{au\textsubscript2}) = \{\textbf{up\textsubscript1, up\textsubscript2}\}
\end{itemize}
\underline{Administrative User Assignments}
\begin{itemize}
\item \it{UA\_admin} = \{(\textbf{u\textsubscript1, au\textsubscript1}), (\textbf{u\textsubscript2, au\textsubscript2})\} 
\item \AUH\ = \{(\textbf{au\textsubscript1, au\textsubscript2})\}
\end{itemize}
\underline{User-role assignment condition in uni-ARBAC:}\\
-- \it{can\_manage\_user\_role}(\it{u}\textsubscript1 : \U, \it{u}\textsubscript2: \U, \\
\hspace*{0.17cm} \it{r}: \R) =
($\exists$\it{au\textsubscript{i}, au\textsubscript{j}})[(\it{u}\textsubscript1, \it{au\textsubscript{i}}) $\in$ \it{UA\_admin} $\wedge$ \it{au\textsubscript{i}} \\
\hspace*{0.17cm} $\succeq$\textsubscript{\it{au}} \it{au\textsubscript{j}} $\wedge$ \it{r} $\in$ \it{roles}(\it{au\textsubscript{j}}) $\wedge$ ($\exists$\it{up} $\in$ UP)[(\it{u\textsubscript2, up}) \\
\hspace*{0.17cm} $\in$ \UUPA\ $\wedge$ \it{up} $\in$ \it{user\_pools}*(\it{au\textsubscript{j}})]]\\
\subsubsection{Equivalent AURA instance of URA in Uni-ARBAC}
\label{sec:aura-uni}
This segment represents an equivalent AURA instance for example instance presented in section~\ref{sec:ura-uni}  
\begin{itemize}
\item \U\ = \{\textbf{u\textsubscript1, u\textsubscript2, u\textsubscript3, u\textsubscript4}\}
\item \AU\ = \{\textbf{u\textsubscript1, u\textsubscript2, u\textsubscript3, u\textsubscript4}\}
\item \OP\ = \{\textbf{assign, revoke}\}
\item \R\ = \{\textbf{r\textsubscript{1}, r\textsubscript{2}, r\textsubscript3}\}
\item \RH\ = \{(\textbf{r\textsubscript1, r\textsubscript2}), (\textbf{r\textsubscript2, r\textsubscript3})\} 
\item \it{\adroles}(\textbf{u\textsubscript3}) = \{\textbf{r\textsubscript1}\},
\item[]\it{\adroles}(\textbf{u\textsubscript4}) = \{\textbf{r\textsubscript3}\}
\item \UATT\ = \{\it{\ups, \upau}\}
\item \scope(\it{\ups}) = \{\textbf{up\textsubscript1, up\textsubscript2}\}, 
\item[] attType(\it{\ups}) = set, 
\item[] \isord(\it{\ups}) = True, 
\item[] H\textsubscript{\it{\ups}} = \{(\textbf{up\textsubscript2, up\textsubscript1})\}
\item \scope(\it{\upau}) = \{(\textbf{up\textsubscript1, au\textsubscript1}), 
\item[] (\textbf{up\textsubscript2, au\textsubscript2})\}, attType(\it{\upau}) = set,
\item[] \isord(\it{\upau}) = False, 
\item[] H\textsubscript{\it{\upau}} = $\phi$
\item \it{\ups}(\textbf{u\textsubscript1}) = \{\textbf{up\textsubscript1, up\textsubscript2}\}, \it{\ups}(\textbf{u\textsubscript2}) = \{\textbf{up\textsubscript2}\},\\ \it{\ups}(\textbf{u\textsubscript3}) = \{\textbf{up\textsubscript1, up\textsubscript2}\}, \it{\ups}(\textbf{u\textsubscript4}) = \{\textbf{up\textsubscript2}\}
%
%
\item \textit{\upau}(\textbf{u\textsubscript1}) = \{(\textbf{up\textsubscript1, au\textsubscript1}), (\textbf{up\textsubscript2, au\textsubscript2})\}, \textit{\upau}(\textbf{u\textsubscript2}) = \{(\textbf{up\textsubscript2, au\textsubscript2})\}, \\ 
\textit{\upau}(\textbf{u\textsubscript3}) = \{(\textbf{up\textsubscript1, au\textsubscript1}), (\textbf{up\textsubscript2, au\textsubscript2})\}, \it{\upau}(\textbf{u\textsubscript4}) = \{(\textbf{up\textsubscript2, au\textsubscript2})\}

\vspace{0.15cm}
\item \AATT\ = \{\textit{\au, \aur}\}
\item \scope(\textit{\au}) =  \{\textbf{au\textsubscript1, au\textsubscript2}\}, 
\item[] attType(\it{\au}) = set,
\item[] \isord(\it{\au}) = True, 
\item[] H\textsubscript{\it{\au}} =  \{(\textbf{au\textsubscript1, au\textsubscript2})\}
\item \scope(\textit{\aur}) =  \{(\textbf{au\textsubscript1, r\textsubscript1}), (\textbf{au\textsubscript1, r\textsubscript2}), 
\item[] (\textbf{au\textsubscript2, r\textsubscript3})\}, 
\item[]  attType(\textit{\aur}) = set,
\item[] \isord(\it{\aur}) = False, 
\item[] H\textsubscript{\it{\aur}} = $\phi$
%
\item \it{\au}(\textbf{u\textsubscript1}) = \{\textbf{au\textsubscript1}\}, \it{\au}(\textbf{u\textsubscript2}) = \{\textbf{au\textsubscript2}\}, \it{\au}(\textbf{u\textsubscript3}) = \{\}, \\ 
\it{\au}(\textbf{u\textsubscript4}) = \{\textbf{}\}
\item \it{\aur}(\textbf{u\textsubscript1}) = \{(\textbf{au\textsubscript1, r\textsubscript1}), (\textbf{au\textsubscript1, r\textsubscript2}), 
\item[] (\textbf{au\textsubscript2, r\textsubscript3})\}, \it{\aur}(\textbf{u\textsubscript2}) = \{(\textbf{au\textsubscript2, r\textsubscript3})\},\\
 \textit{\aur}(\textbf{u\textsubscript3}) = \{\}, \it{\aur}(\textbf{u\textsubscript4}) = \{\}
\end{itemize}

\vspace{0.17cm}
\noindent
For each \it{op} in \OP, authorization rule to aassign/revoke a user to/from a role can be expressed as follows:\\
For any user \it{u\textsubscript2} $\in$ \U\ undertaken for assignment,\\
-- {\isauth}U\textsubscript{\textbf{assign}}(\it{u\textsubscript1} : \U, \it{u\textsubscript2} : \U, \\
\hspace*{0.17cm} \it{r} : \R) $\equiv$ \\
\hspace*{0.17cm} $\exists$\it{au\textsubscript1,  au\textsubscript2} $\in$ \scope(\it{\au}). (\it{au\textsubscript1, au\textsubscript2}) \\
\hspace*{0.17cm} $\in$ H\textsubscript{\it{\au}} $\wedge$  (\it{au\textsubscript1} $\in$ \it{\au}(\it{u\textsubscript1}) $\wedge$ 
(\it{au\textsubscript2, r}) $\in$ \\
\hspace*{0.17cm}   \it{\aur}(\it{u\textsubscript1})) $\wedge$ $\exists$\it{up\textsubscript1, up\textsubscript2} $\in$ \scope(\it{\ups}). \\
\hspace*{0.17cm} (\it{up\textsubscript2, up\textsubscript1}) $\in$ H\textsubscript{\it{\ups}} $\wedge$ ((\it{up\textsubscript2, au\textsubscript2}) \\
\hspace*{0.17cm} $\in$ \it{\upau}(\it{u\textsubscript2})) 

\vspace{0.15cm}
\noindent
For any user \it{u\textsubscript2} $\in$ \U\ undertaken for revocation,\\
-- {\isauth}U\textsubscript{\textbf{revoke}}(\it{u\textsubscript1} : \U, \it{u\textsubscript2} : \U, \\
\hspace*{0.17cm} \it{r} : \R) $\equiv$ {\isauth}U\textsubscript{\textbf{assign}}(\it{u\textsubscript1} : \U, \\
\hspace*{0.17cm} \it{u\textsubscript2} : \U, \it{r} : \R)\\
\subsubsection{Map\textsubscript{URA-Uni-ARBAC}}
Map\textsubscript{URA-Uni-ARBAC} represents a translation process of any instance of URA in Uni-ARBAC to AURA instance. For clarity, basic sets from URA in Uni-ARBAC are marked with superscript \t{Uni} and basic sets from AURA are marked with superscript \t{A}.

Map\textsubscript{URA-Uni-ARBAC}  takes URA-Uni instance as input. In particular in involves \U\textsuperscript{\t{Uni}}, \R\textsuperscript{\t{Uni}}, \RH\textsuperscript{\t{Uni}}, \UA\textsuperscript{\t{Uni}}, \UP\textsuperscript{\t{Uni}}, \UPH\textsuperscript{\t{Uni}}, \UUPA\textsuperscript{\t{Uni}}, \AU\textsuperscript{\t{Uni}}, For each \it{au} in \AU\textsuperscript{\t{Uni}}, \it{roles}\textsuperscript{\t{Uni}}(\it{au}), For each \it{au} in \AU\textsuperscript{\t{Uni}}, \it{user\_pools}*(\it{au}), \it{UA\_admin}\textsuperscript{\t{Uni}}, \AUH\textsuperscript{\t{Uni}}, and \it{can\_manage\_user\_role}(\it{u\textsubscript{1}} : \U\textsuperscript{\t{Uni}}, \it{u\textsubscript{2}} : \U\textsuperscript{\t{Uni}}, \it{r} : \R\textsuperscript{\t{Uni}}).

It yields an equivalent instance of AURA as \U\textsuperscript{\t{A}}, \AU\textsuperscript{\t{A}}, \OP\textsuperscript{\t{A}}, \R\textsuperscript{\t{A}}, \RH\textsuperscript{\t{A}}, For each \it{u} $\in$ \U\textsuperscript{\t{A}}, \it{\adroles}(\it{u}), \UATT\textsuperscript{\t{A}}, \AATT\textsuperscript{\t{A}}, For each attribute \it{att} $\in$ \UATT\textsuperscript{\t{A}} $\cup$ \AATT\textsuperscript{\t{A}}, \scope(\it{att}), attType(\it{att}), \isord(\it{att}) and H\textsubscript{\it{att}}, 
For each user \it{u} $\in$ \U\textsuperscript{\t{A}}, and for each \it{att} $\in$ \UATT\textsuperscript{\t{A}} $\cup$ \AATT\textsuperscript{\t{A}}, \it{att}(\it{u}), Authorization rule for assign (auth\_assign), and Authorization rule for revoke (auth\_revoke).

A shown in Algorithm Map\textsubscript{URA-Uni-ARBAC}, there are four main steps required in mapping any instance of URA-Uni model to AURA instance. In Step 1, sets and functions from URA-Uni instance are mapped into AURA sets and functions. In Step 2, user attributes and administrative user attribute functions are expressed. There are two user attribute, \it{userpools} and \it{\upau}. \it{userpools} captures regular user's binding with a group called user-pools. Regular user attribute \it{\upau} provides regular user's association with user-pools, and for each user-pool a user is associated with, user-pool's mapping with admin unit. As a result, this attribute captures a regular user's association with an admin unit. 
We note that we need both the attributes. Although \it{\upau} captures regular user's association with user-pools and corresponding admin units, it cannot capture user association with user-pools which may not have admin unit associated with it. It is the user-pools that are mapped to admin units.   
There are two admin user attributes, \it{\au} and \it{\aur}. \it{\au} captures \it{UA\_admin} relation in Uni-ARBAC, and \it{\aur} captures admin user's mapping with admin unit and for each admin unit an admin user is mapped to, admin unit's associated roles. 

The notion of Uni-ARBAC model is that an admin user to have admin authority (given by \it{UA\_admin} relation) to assign/revoke regular user and role, if both regular user and role are mapped to that admin unit where admin user has admin authority.

Step 3 involves constructing auth\_assign in AURA that is equivalent to \it{can\_manage\_user\_role}(\it{u\textsubscript{1}} : \U\textsuperscript{\t{Uni}}, \it{u\textsubscript{2}} : \U\textsuperscript{\t{Uni}}, \it{r} : \R\textsuperscript{\t{Uni}}) in URA-Uni. Its translation in AURA is given by {\isauth}U\textsubscript{\textbf{assign}}(\it{au} : \AU\textsuperscript{\t{A}}, \it{u} : \U\textsuperscript{\t{A}}, \it{r} : \R\textsuperscript{\t{A}}). Similarly, In Step 4, authorization rule to revoke user-role (auth\_revoke), which is equivalent to \it{can\_manage\_user\_role}(\it{u\textsubscript{1}} : \U\textsuperscript{\t{Uni}}, \it{u\textsubscript{2}} : \U\textsuperscript{\t{Uni}}, \it{r} : \R\textsuperscript{\t{Uni}}) is expressed.

\floatname{algorithm}{Algorithm}
\begin{algorithm}[tp]
\label{algo:urauni}
\caption{Map\textsubscript{URA-Uni-ARBAC}}
\begin{algorithmic} [] 
\begin{spacing}{1.09}
\item[]\hspace{-17pt}\textbf{Input:} Instance of URA in Uni-ARBAC 
\item[]\hspace{-17pt}\textbf{Output:} AURA instance
\item[\textbf{Step 1:}] \ \ /* Map basic sets and functions in AURA */
\item[] a. \U\textsuperscript{\t{A}} $\leftarrow$ \U\textsuperscript{\t{Uni}} ; \AU\textsuperscript{\t{A}} $\leftarrow$ \U\textsuperscript{\t{Uni}} ; 
\item[] b. \OP\textsuperscript{\t{A}} $\leftarrow$ \{\textbf{assign, revoke}\}
\item[] c. \R\textsuperscript{\t{A}} $\leftarrow$ \R\textsuperscript{\t{Uni}}
\item[] d. \RH\textsubscript{A} $\leftarrow$ \RH\textsuperscript{\t{Uni}} ; For each \it{u} $\in$ \U\textsuperscript{\t{A}}, 
\item[] \tab \it{\adroles}(\it{u}) = $\phi$  
\item[] e. For each (\it{u, r}) $\in$ \UA\textsuperscript{\t{Uni}}, \it{\adroles}(\it{u})\textquotesingle\ = \it{\adroles}(\it{u}) $\cup$ \it{r}
\item[\textbf{Step 2:}]   \stab /* Map attribute functions in AURA */
\item[] a. \UATT\textsuperscript{\t{A}} $\leftarrow$ \{\it{\ups, \upau}\}
\item[] b. \scope(\it{\ups}) = \UP\textsuperscript{\t{Uni}}
\item[] c. attType(\it{\ups}) = set  
\item[] d. is\_ordered(\it{\ups}) = True
\item[] e. H\textsubscript{\it{\ups}} = \UPH\textsuperscript{\t{Uni}}
\item[] f. For each \it{u} in \U\textsuperscript{\t{A}}, \it{\ups}(\it{u}) = $\phi$ 
\item[] g. For each (\it{u, up}) $\in$ \UUPA\textsuperscript{\t{Uni}}, 
\item[] \tab \it{\ups}(\it{u})\textquotesingle\ = \it{\ups}(\it{u}) $\cup$ \it{up}
\item[] h. \scope(\it{\upau}) =
\item[] \tab \tab  \U\textsuperscript{\t{Uni}} $\times$ AU\textsuperscript{\t{Uni}}
\item[] i. attType(\it{\upau}) = set
\item[] j. is\_ordered(\it{\upau}) = False 
\item[] k. H\textsubscript{\it{\upau}}
\item[] l. For each \it{u} in \U\textsuperscript{\t{A}},
\item[] \tab  \it{\upau}(\it{u}) = $\phi$ ; 
\item[] m. For each (\it{u, up}) $\in$ \UUPA\textsuperscript{\t{Uni}} 
\item[] \tab \tab and for each \it{au} in \AU\textsuperscript{\t{Uni}},\\
\item[] \stab \textbf{if} \it{up} $\in$ \it{user\_pools}(\it{au}) \textbf{then} 
\item[] \stab \ \ \it{\upau}(\it{u})\textquotesingle\ =
\item[] \tab \it{\upau}(\it{u}) $\cup$ (\it{up, au}) 
\item[] n. \AATT\textsuperscript{\t{A}} $\leftarrow$ \{\it{\au, \aur}\} 
\item[] o. \scope(\it{\au}) = \AU\textsuperscript{\t{Uni}} 
\item[] p. attType(\it{\au}) = set 
\item[] q. is\_ordered(\it{\au}) = True
\item[] r. H\textsubscript{\it{\au}} = AUH\textsuperscript{\t{Uni}}
\item[] s. For each \it{u} in \AU\textsuperscript{\t{A}}, \it{\au}(\it{u}) = $\phi$ 
\item[] t. For each (\it{u, au}) $\in$ \it{UA\_admin}\textsuperscript{\t{Uni}},
\item[] \tab \it{\au}(\it{u})\textquotesingle\ = \it{\au}(\it{u}) $\cup$ \it{au}
\item[] u. \scope(\it{\aur}) = \AU\textsuperscript{\t{Uni}} $\times$ \R\textsuperscript{\t{Uni}} 
\item[] v. attType(\it{\aur}) = set 
\item[] w. is\_ordered(\it{\aur}) = False 
\item[] x. For each \it{u} in \AU\textsuperscript{\t{A}}, \it{\aur}(\it{u}) = $\phi$ 
\item[] y. For each (\it{u, au}) $\in$ \it{UA\_admin}\textsuperscript{\t{Uni}} 
\item[] \ \ \ and for each \it{r} $\in$ \it{roles}\textsuperscript{\t{Uni}}(\it{au}), 
\item[] \ \ \ \ \it{\aur}(\it{u})\textquotesingle\ = \it{\aur}(\it{u}) $\cup$ (\it{au, r})
\vspace{-0.8cm}
\end{spacing}
\end{algorithmic}
\end{algorithm}
\floatname{algorithm}{Continuation of Algorithm}
\setcounter{algorithm}{3}
\begin{algorithm}
\caption{Map\textsubscript{URA-Uni-ARBAC}}
\begin{algorithmic} [1] 
\begin{spacing}{1.1}
\item[\textbf{Step 3:}] \stab /* Construct assign rule in AURA */
\item[] a. can\_manage\_rule = \\
\hspace{0.17cm}($\exists$\it{au\textsubscript1,  au\textsubscript2} $\in$ \scope(\it{\au}). (\it{au\textsubscript1, au\textsubscript2}) \\
\hspace{0.17cm}$\in$ H\textsubscript{\it{\au}} $\wedge$ (\it{au\textsubscript1} $\in$ \it{\au}(\it{u\textsubscript1}) $\wedge$ \\
\hspace{0.17cm}(\it{au\textsubscript2, r}) $\in$  \it{\aur}(\it{u\textsubscript1})) $\wedge$ $\exists$\it{up\textsubscript1, up\textsubscript2} \\
\hspace{0.17cm}$\in$ \scope(\it{\ups}). (\it{up\textsubscript2, up\textsubscript1}) $\in$ H\textsubscript{\it{\ups}} $\wedge$ \\
\hspace{0.17cm}(\it{up\textsubscript2, au\textsubscript2}) $\in$ \it{\upau}(\it{u\textsubscript2}))
\item[] b. auth\_assign = \\ 
\item[] -- {\isauth}U\textsubscript{\textbf{assign}}(\it{u\textsubscript1} : \AU\textsuperscript{\t{A}}, \it{u\textsubscript2} : \U\textsuperscript{\t{A}}, \\
\item[]\hspace{0.17cm} \it{r} : \R\textsuperscript{\t{A}}) $\equiv$ can\_manage\_rule
\item[\textbf{Step 4:}] \stab /* Construct revoke rule for AURA */
\item[] a. auth\_revoke = \\
\item[] -- {\isauth}U\textsubscript{\textbf{revoke}}(\it{u\textsubscript1} : \AU\textsuperscript{\t{A}}, \it{u\textsubscript2} : \U\textsuperscript{\t{A}}, \\
\item[]\hspace{0.17cm} \it{r} : \R\textsuperscript{\t{A}}) $\equiv$ can\_manage\_rule
\vspace{-0.4cm}
\end{spacing}
\end{algorithmic}
\end{algorithm}

\subsection{UARBAC's URA in AURA}
Li and Mao~\cite{uarbac} redefine RBAC model. They propose a notion of class of objects in RBAC. A summary is presented here.
\subsubsection{RBAC Model}
RBAC model has following schema.\\
\underline{RBAC Schema:}\\
RBAC Schemas is given by following tuple.\\
\tab <{\it{C, OBJS, AM}}>
\begin{itemize}
\item \it{C} is a finite set of object classes with predefined classes: \textsf{user} and \textsf{role}.
\item \it{OBJS}(\it{c}) is a function that gives all possible names for objects of the class \it{c} $\in$ \it{C}. Let
\item[] \textbf{\U}\ =  \it{OBJS}(\textsf{user}) and \textbf{\R}\ = \it{OBJS}(\textsf{role})  
\item \it{AM}(\it{c}) is function that maps class \it{c} to a set of access modes that can be applied on objects of class \it{c}. 
\end{itemize}
Access modes for two predefined classes \textsf{user} and \textsf{role} are fixed by the model and are as follows:
\begin{itemize}
\item[] \it{AM}(\textsf{user}) = \{\textsf{empower, admin}\}
\item[] \it{AM}(\textsf{role}) = \{\textsf{grant, empower, admin}\}
\end{itemize}
\underline{RBAC Permissions:}\\
There are two kinds of permissions in this RBAC model:
\begin{enumerate}
\item Object permissions of the form, 
\item[] \ \ {[\it{c, o, a}]}, where \it{c $\in$ C}, \it{o} $\in$ \it{OBJS}(\it{c}), \it{a} $\in$ \it{AM}(\it{c}).
\item Class permissions of the form,
\item[] \ \ {[\it{c, a}]}, where, \it{c $\in$ C}, and \it{a} $\in$ \{\textsf{create}\} $\cup$ \it{AM}(\it{c}).
\end{enumerate}
\underline{RBAC State:}\\
Given an RBAC Schema, an RBAC state is given by,\\ 
\tab <\it{OB, UA, PA, RH}>
\begin{itemize}
\item \it{OB} is a function that maps each class in \it{C} to a finite set of object names of that class that currently exists, i.e., \it{OB}(\it{c}) $\subseteq$ \it{OBJS}(\it{c}). Let 
\item[] \it{OB}(\textsf{user}) = \it{\U}, and \it{OB}(\textsf{role}) = \it{\R}.
\item[] Set of permissions, \it{P}, is given by
\item[] \it{P} = \{[\it{c, o, a}] $\vert$ \it{c $\in$ C} $\wedge$ \it{o} $\in$ \it{OBJS}(\it{c}) $\wedge$ \it{a} $\in$ \it{AM}(\it{c})\} $\cup$ \{[\it{c, a}] $\vert$ \it{c $\in$ C} $\wedge$ \it{a} $\in$ \{\textsf{create}\} $\cup$ \it{AM}(\it{c})\}
\item \it{UA} $\subseteq$ \it{\U} $\times$ \it{\R}, user-role assignment relation.
\item \it{PA} $\subseteq$ \it{P} $\times$ \it{\R}, permission-role assignment relation.
\item \it{\RH} $\subseteq$ \it{\R} $\times$ \it{\R}, partial order in \it{\R}\ denoted by \it{$\succeq$\textsubscript{RH}}.
\end{itemize}
\underline{Administrative permissions in UARBAC:}\\
All the permissions of user \it{u} who performs administrative operations can be calculated as follows:
\begin{itemize}
\item authorized\_perms[\it{u}] = \{\it{p $\in$ P} $\vert$ $\exists$\it{r\textsubscript1, r\textsubscript2} $\in$ R [(\it{u, r\textsubscript1}) $\in$ \it{UA} $\wedge$ (\it{r\textsubscript1 $\succeq$\textsubscript{RH} r\textsubscript2}) $\wedge$ (\it{r\textsubscript2, p}) $\in$ \it{PA}]\}
\end{itemize}
\underline{User-Role Administration}\\
Operations required to assign user \it{u\textsubscript1} to role \it{r\textsubscript1} and to revoke \it{u\textsubscript1} from role \it{r\textsubscript1} are respectively listed below:
\begin{itemize}
\item grantRoleToUser(\it{r\textsubscript1, u\textsubscript1})
\item revokeRoleFromUser(\it{r\textsubscript1, u\textsubscript1})
\end{itemize}
A user at least requires following two permissions to conduct grantRoleToUser(\it{r\textsubscript1, u\textsubscript1}) operation.
\begin{enumerate}
\item {[\textsf{user}, \it{u\textsubscript1}, \textsf{empower}]}
\item {[\textsf{role}, \it{r\textsubscript1}, \textsf{empower}]}
\end{enumerate}
A user at least requires one of the following (three) options to conduct revokeRoleFromUser(\it{r\textsubscript1, u\textsubscript1}) operation.
\begin{enumerate}
\item {[\textsf{user}, \it{u\textsubscript1}, \textsf{empower}] and [\textsf{role}, \it{r\textsubscript1}, \textsf{empower}]}
\item {[\textsf{user}, \it{u\textsubscript1}, \textsf{admin}]} 
\item {[\textsf{role}, \it{r\textsubscript1}, \textsf{admin}]}
\end{enumerate}

\vspace{0.17cm}
\noindent
\subsubsection{Instance of URA in UARBAC}
 \hfill \break
\underline{RBAC Schema}
\begin{itemize}
\item \it{C} = \{\textsf{user, role}\}
\item \it{OBJS}(\textsf{user}) = USERS, \it{OBJS}(\textsf{role}) = ROLES
\item \it{AM}(\textsf{user}) = \{\textsf{empower, admin}\}, \it{AM}(\textsf{role}) = \{\textsf{grant, empower, admin}\}
\end{itemize}
\underline{RBAC State}
\begin{itemize}
\item \it{\U} = \it{OBJ}(\textsf{user}) = \{\textbf{u\textsubscript1, u\textsubscript2, u\textsubscript3, u\textsubscript4}\} 
\item \it{\R} = \it{OBJ}(\textsf{role})= \{\textbf{r\textsubscript{1}, r\textsubscript{2}, r\textsubscript3, r\textsubscript4}\} 
\item \it{P} = \{[\textsf{user}, \textbf{u\textsubscript1}, \textsf{empower}], [\textsf{user}, \textbf{u\textsubscript1}, \textsf{admin}], [\textsf{user}, \textbf{u\textsubscript2}, \textsf{empower}], [\textsf{user}, \textbf{u\textsubscript2}, \textsf{admin}], [\textsf{user}, \textbf{u\textsubscript3}, \textsf{empower}], [\textsf{user}, \textbf{u\textsubscript3}, \textsf{admin}], [\textsf{user}, \textbf{u\textsubscript4}, \textsf{empower}], [\textsf{user}, \textbf{u\textsubscript4}, \textsf{admin}], [\textsf{role}, \textbf{r\textsubscript1}, \textsf{grant}], [\textsf{role}, \textbf{r\textsubscript1}, \textsf{empower}], [\textsf{role}, \textbf{r\textsubscript1}, \textsf{admin}], [\textsf{role}, \textbf{r\textsubscript2}, \textsf{grant}], [\textsf{role}, \textbf{r\textsubscript2}, \textsf{empower}], [\textsf{role}, \textbf{r\textsubscript2}, \textsf{admin}], [\textsf{role}, \textbf{r\textsubscript3}, \textsf{grant}], [\textsf{role}, \textbf{r\textsubscript3}, \textsf{empower}], [\textsf{role}, \textbf{r\textsubscript3}, \textsf{admin}], {[\textsf{role}, \textbf{r\textsubscript4}, \textsf{grant}], [\textsf{role}, \textbf{r\textsubscript4}, \textsf{empower}],  [\textsf{role}, \textbf{r\textsubscript4}, \textsf{admin}], [\textsf{user, empower}], [\textsf{user, admin}], [\textsf{role, empower}], [\textsf{role, grant}], [\textsf{role, admin}]\}}
\item \it{\UA} = \{(\textbf{u\textsubscript1, r\textsubscript1}), (\textbf{u\textsubscript2, r\textsubscript1}), (\textbf{u\textsubscript2, r\textsubscript2}), (\textbf{u\textsubscript2, r\textsubscript3}), (\textbf{u\textsubscript3, r\textsubscript3}), (\textbf{u\textsubscript4, r\textsubscript2})\}
\item \it{\RH} = \{(\textbf{r\textsubscript1, r\textsubscript2}), (\textbf{r\textsubscript2, r\textsubscript3}), (\textbf{r\textsubscript3, r\textsubscript4})\}
\end{itemize}
\noindent
\underline{Administrative permissions of UARBAC's URA:}\\
Following is the list of administrative permissions each user has for user-role assignment:
\begin{itemize}
\item authorized\_perms[\textbf{u\textsubscript1}]  = \{[\textsf{user}, \textbf{u\textsubscript1}, \textsf{empower}], [\textsf{role}, \textbf{r\textsubscript1}, \textsf{grant}], [\textsf{user}, \textbf{u\textsubscript2}, \textsf{empower}], {[\textsf{role}, \textbf{r\textsubscript3}, \textsf{grant}], [\textsf{user}, \textbf{u\textsubscript3}, \textsf{empower}], [\textsf{user}, \textbf{u\textsubscript4}, \textsf{empower}]}, {[\textsf{role}, \textbf{r\textsubscript2}, \textsf{grant}], [\textsf{user}, \textbf{u\textsubscript3}, \textsf{admin}], [\textsf{role}, \textbf{r\textsubscript1}, \textsf{admin}]}, {[\textsf{role}, \textbf{r\textsubscript4}, \textsf{admin}]}\}

\item authorized\_perms[\textbf{u\textsubscript2}]  = \{[\textsf{user}, \textbf{u\textsubscript1}, \textsf{empower}], [\textsf{role}, \textbf{r\textsubscript1}, \textsf{grant}], [\textsf{user}, \textbf{u\textsubscript2}, \textsf{empower}], {[\textsf{role}, \textbf{r\textsubscript2}, \textsf{grant}]}\}
\item authorized\_perms[\textbf{u\textsubscript3}] = \{\}
\item authorized\_perms[\textbf{u\textsubscript4}] = \{[\textsf{role, grant}], [\textsf{user, empower}]\}
\end{itemize}
\underline{User-Role assignment condition in URA-UARBAC:}\\
One can perform following operation to assign a user \it{u\textsubscript1} to a role \it{r\textsubscript1}.
\begin{itemize}
\item grantRoleToUser(\it{r\textsubscript{1}, u\textsubscript{1}}) 
\end{itemize}
To perform aforementioned operation one needs the following two permissions:
\begin{itemize}
\item{ [\textsf{user}, \it{u\textsubscript{2}}, \textsf{empower}]  and  [\textsf{role}, \it{r\textsubscript1}, \textsf{grant}]}
\end{itemize}
\underline{Condition for revoking user-role in URA-UARBAC:}\\
One can perform following operation to revoke a user \it{u\textsubscript1} to a role \it{r\textsubscript1}.
\begin{itemize}
\item revokeRoleFromUser(\it{r\textsubscript{1}, u\textsubscript{1}}) 
\end{itemize}
To perform aforementioned operation one needs the one of the following permissions:
\begin{itemize}
\item {[\textsf{user}, \it{u\textsubscript{1}}, \textsf{empower}]  and  [\textsf{role}, \it{r\textsubscript1}, \textsf{grant}]} \textbf{or},
\item {[\textsf{role}, \it{r\textsubscript1}, \textsf{admin}]} \textbf{or},
\item {[\textsf{user}, \it{u\textsubscript1}, \textsf{admin}]}\\
\end{itemize}
%
\subsubsection{Equivalent AURA instance for URA in UARBAC}
\begin{itemize}
\item \U\ = \{\textbf{u\textsubscript{1}, u\textsubscript{2}, u\textsubscript{3}, u\textsubscript4}\}
\item \AU\ = \{\textbf{u\textsubscript{1}, u\textsubscript{2}, u\textsubscript{3}, u\textsubscript4}\}
\item \OP\ = \{\textbf{assign, revoke}\}
\item \R\ = \{\textbf{r\textsubscript{1}, r\textsubscript{2}, r\textsubscript3}\} 
\item \RH\ = \{(\textbf{r\textsubscript1, r\textsubscript2}), (\textbf{r\textsubscript2, r\textsubscript3}), (\textbf{r\textsubscript3, r\textsubscript4})\}
\item \it{\adroles}(\textbf{u\textsubscript1}) = \{\textbf{r\textsubscript1}\}, 
\item[] \it{\adroles}(\textbf{u\textsubscript2}) = \{\textbf{r\textsubscript1, r\textsubscript2, r\textsubscript3}\}, 
\item[]  \it{\adroles}(\textbf{u\textsubscript3}) = \{\textbf{r\textsubscript3}\}, 
\item[] \it{\adroles}(\textbf{u\textsubscript4}) = \{\textbf{r\textsubscript2}\}
\item \UATT\ = \{\}
\item \AATT\ = \{\textit{\uam, \ram, \classp}\}
\item \scope(\textit{\uam}) = \{(\textbf{u\textsubscript1}, \textsf{empower}), 
\item[] (\textbf{u\textsubscript2}, \textsf{empower}), (\textbf{u\textsubscript3}, \textsf{empower}), 
\item[] (\textbf{u\textsubscript3}, \textsf{admin}), (\textbf{u\textsubscript4}, \textsf{empower})\}, 
\item[] attType(\textit{\uam}) = set, 
\item[] \isord(\it{\uam}) = False, H\textsubscript{\it{\uam}} = $\phi$

\item \scope(\textit{\ram}) = \{(\textbf{r\textsubscript1, grant}), (\textbf{r\textsubscript2}, \textsf{grant}), 
\item[] (\textbf{r\textsubscript3}, \textsf{grant}),
 (\textbf{r\textsubscript1}, \textsf{admin}), (\textbf{r\textsubscript4}, \textsf{admin})\},
 \item[] attType(\textit{\ram}) = set, 
 \item[] \isord(\it{\ram}) = False, H\textsubscript{\it{\ram}} = $\phi$

\item \scope(\textit{\classp}) = \{(\textsf{user}, \textsf{empower}), (\textsf{user}, \textsf{admin}), 
\item[] (\textsf{role}, \textsf{empower}), (\textsf{user}, \textsf{grant}), (\textsf{role}, \textsf{admin})\}, 
\item[] attType(\textit{\classp}) = set, \isord(\it{\uam}) = False, H\textsubscript{\it{classp}} = $\phi$ 
\item \it{\uam}(\textbf{u\textsubscript1}) = \{(\textbf{u\textsubscript1}, \textsf{empower}), (\textbf{u\textsubscript2}, \textsf{empower}), (\textbf{u\textsubscript3}, \textsf{empower}), (\textbf{u\textsubscript4}, \textsf{empower}), (\textbf{u\textsubscript3}, \textsf{admin})\},
\item[] \it{\uam}(\textbf{u\textsubscript2}) = \{(\textbf{u\textsubscript1}, \textsf{empower}), (\textbf{u\textsubscript2}, \textsf{empower})\},
\item[] \it{\uam}(\textbf{u\textsubscript3}) = \{\},
\it{\uam}(\textbf{u\textsubscript4}) = \{\}

\item \it{\ram}(\textbf{u\textsubscript1}) = \{(\textbf{r\textsubscript1}, \textsf{grant}), (\textbf{r\textsubscript2}, \textsf{grant}), (\textbf{r\textsubscript3}, \textsf{grant}), (\textbf{r\textsubscript1}, \textsf{admin}), (\textbf{r\textsubscript4}, \textsf{admin})\},
\item[] \it{\ram}(\textbf{u\textsubscript2}) = \{(\textbf{r\textsubscript1}, \textsf{grant}), (\textbf{r\textsubscript2}, \textsf{grant})\},
\item[] \it{\ram}(\textbf{u\textsubscript3}) = \{\}, \it{\ram}(\textbf{u\textsubscript4}) = \{\}

\item \it{\classp}(u\textsubscript1) = \{\}, \it{\classp}(u\textsubscript2) = \{\}, 
\it{\classp}(u\textsubscript3) = \{\}, 
\item[] \it{\classp}(u\textsubscript4) = \{(\textsf{role, grant}), (\textsf{user, empower})\}
\end{itemize}
For each \it{op} in \OP, authorization rule to assign/revoke user-role can be expressed as follows:\\
For any regular user \it{u\textsubscript2} $\in$ \U\ taken for assignment,\\
-- {\isauth}U\textsubscript{\textbf{assign}}(\it{u\textsubscript1} : \U, \it{u\textsubscript2} : \U, \\
\hspace*{0.17cm}\it{r\textsubscript1} : \R) $\equiv$ \\
\hspace*{0.17cm}((\it{u\textsubscript{2}}, \textsf{empower}) $\in$ \it{\uam}(\it{u\textsubscript{1}}) $\wedge$  (\it{r\textsubscript1}, \textsf{grant}) \\
\hspace*{0.17cm}$\in$ \it{\ram}(\it{u\textsubscript{1}})) $\vee$ ((\it{u\textsubscript{2}}, \textsf{empower}) $\in$ \it{\uam}(\it{u\textsubscript{1}}) $\wedge$ \\
\hspace*{0.17cm}(\textsf{role}, \textsf{grant}) $\in$ \it{\classp}(\it{u\textsubscript{1}})) $\vee$ ((\textsf{user}, \textsf{empower}) \\
\hspace*{0.17cm}$\in$ \it{\classp}(\it{u\textsubscript{1}}) $\wedge$ (\it{r\textsubscript1}, \textsf{grant}) $\in$ \it{\ram}(\it{u\textsubscript{1}})) $\vee$ \\
\hspace*{0.17cm}((\textsf{user}, \textsf{empower}) $\in$ \it{\classp}(\it{u\textsubscript{1}}) $\wedge$ \\
\hspace*{0.17cm}(\textsf{role}, \textsf{grant}) $\in$ \it{\classp}(\it{u\textsubscript{1}}))
 
 \vspace{0.17cm}
 \noindent
 For any regular user \it{u\textsubscript2} $\in$ \U\ taken for revocation,\\
-- {\isauth}U\textsubscript{\textbf{revoke}}(\it{u\textsubscript1} : \U, \it{u\textsubscript2} : \U,\\
\hspace*{0.17cm}\it{r\textsubscript1} : \R) $\equiv$ \\
\hspace*{0.17cm}((\it{u\textsubscript{2}}, \textsf{empower}) $\in$ \it{\uam}(\it{u\textsubscript{1}}) $\wedge$ (\it{r\textsubscript1}, \textsf{grant}) \\
\hspace*{0.17cm}$\in$ \it{\ram}(\it{u\textsubscript{1}})) $\vee$ 
\hspace*{0.17cm}(\it{u\textsubscript{2}}, \textsf{admin}) $\in$ \it{\uam}(\it{u\textsubscript{1}}) $\vee$ \\
\hspace*{0.17cm}(\it{r\textsubscript1}, \textsf{admin}) $\in$ \it{\ram}(\it{u\textsubscript{1}}) $\vee$ (\textsf{user}, \textsf{admin})\\
\hspace*{0.17cm}$\in$ \it{\classp}(\it{u\textsubscript{1}}) $\vee$ (\textsf{role}, \textsf{admin}) $\in$ \it{\classp}(\it{u\textsubscript{1}})
\\
\floatname{algorithm}{Algorithm}
\begin{algorithm}
\caption{Map\textsubscript{URA-UARBAC}}
\label{alg:URA-U}
\begin{algorithmic} [1] 
\begin{spacing}{1.1}
\item[]\hspace{-17pt}\textbf{Input:} Instance of URA in UARBAC 
\item[]\hspace{-17pt}\textbf{Output:} AURA instance 
\item[\textbf{Step 1:}] \ \  /* Map basic sets and functions in AURA */
\item[] a. \U\textsuperscript{\t{A}} $\leftarrow$ \it{\U}\textsuperscript{\t{U}} ; \AU\textsuperscript{\t{A}} $\leftarrow$ \it{\U}\textsuperscript{\t{U}}
\item[] b. \OP\textsuperscript{\t{A}} $\leftarrow$ \{\textbf{assign, revoke}\}
\item[] c. \R\textsuperscript{\t{A}} $\leftarrow$ \it{\R}\textsuperscript{\t{U}} ; \RH\textsuperscript{\t{A}} $\leftarrow$ \it{\RH}\textsuperscript{\t{U}} 
\item[] d. For each \it{u\textsubscript1} $\in$ \U\textsuperscript{\t{A}}, \it{\adroles}\textsuperscript{\t{A}}(\it{u\textsubscript1}) = $\phi$  
\item[] e. For each (\it{u\textsubscript1, r\textsubscript1}) $\in$ \it{\UA}\textsuperscript{\t{U}},
\item[] \ \ \ \ \ \it{\adroles}\textsuperscript{\t{A}}(\it{u\textsubscript1})\textquotesingle\ = \it{\adroles}\textsuperscript{\t{A}}(\it{u\textsubscript1}) $\cup$ \it{r\textsubscript1}
\item[\textbf{Step 2:}]   \stab /* Map attribute functions in AURA */
\item[] a. \UATT\textsuperscript{\t{A}} = $\phi$ ; 
\item[] b. \AATT\textsuperscript{\t{A}} $\leftarrow$ \{\it{\uam, \ram, classp}\}
\item[] c. \scope\textsuperscript{\t{A}}(\it{\uam}) = \it{\U}\textsuperscript{\t{U}} $\times$ \it{AM}\textsuperscript{\t{U}}(\textsf{user})
\item[] d. attType\textsuperscript{\t{A}}(\it{\uam}) = set 
\item[] e. \isord\textsuperscript{\t{A}}(\it{\uam}) = False, H\textsuperscript{\t{A}}\textsubscript{\it{\uam}} = $\phi$
\item[] f. For each \it{u\textsubscript{}} in \AU\textsuperscript{\t{U}}, \it{\uam}(\it{u\textsubscript{}}) = $\phi$ 
\item[] g. For each \it{u} in \it{U}\textsuperscript{\t{U}} and 
\item[] \stab for each [\it{c}, \it{u\textsubscript1}, \it{am}] $\in$ authorized\_perms\textsuperscript{\t{U}}[\it{u}], 
\item[] \tab \it{\uam}(\it{u\textsubscript{}})\textquotesingle\ = \it{\uam}(\it{u\textsubscript{}}) $\cup$ (\it{u\textsubscript{1}, am})
\item[] h. \scope\textsuperscript{\t{A}}(\it{\ram}) = \it{\R}\textsuperscript{\t{U}} $\times$ \it{AM}\textsuperscript{\t{U}}(\textsf{role}) 
\item[] i. attType\textsuperscript{\t{A}}(\it{\ram}) = set 
\item[] j. \isord\textsuperscript{\t{A}}(\it{\ram}) = False, H\textsuperscript{\t{A}}\textsubscript{\it{\ram}} = $\phi$ 
\item[] k. For each \it{u\textsubscript{}} in \U\textsuperscript{\t{A}}, \it{\ram}(\it{u}) = $\phi$ 
\item[] l. For each \it{u} in \it{U}\textsuperscript{\t{U}} 
\item[] \stab for each [\it{c, r\textsubscript1, am}] $\in$ authorized\_perms\textsuperscript{\t{U}}[\it{u}], 
\item[] \tab \it{\ram}(\it{u\textsubscript{}})\textquotesingle\ = \it{\ram}(\it{u\textsubscript{}}) $\cup$ (\it{r\textsubscript1, am})
\item[] m. \scope\textsuperscript{\t{A}}(\it{classp}) = 
\item[] \tab \it{C}\textsuperscript{\t{U}} $\times$ \{\it{AM}(\textsf{role})\textsuperscript{\t{U}} $\cup$ \it{AM}\textsuperscript{\t{U}}(\textsf{user})\}
\item[] n. attType\textsuperscript{\t{A}}(\it{classp}) = set 
\item[] o. \isord\textsuperscript{\t{A}}(\it{classp}) = False, H\textsuperscript{\t{A}}\textsubscript{\it{classp}} = $\phi$ 
\item[] p. For each \it{u\textsubscript{}} in \U\textsuperscript{\t{A}}, \it{classp}(\it{u}) = $\phi$ 
\item[] q. For each \it{u} in \it{U}\textsuperscript{\t{U}} 
\item[] \stab for each [\it{c, a}] $\in$ authorized\_perms\textsuperscript{\t{U}}[\it{u}], 
\item[]\tab \it{classp}(\it{u})\q = \it{classp}(\it{u}) $\cup$ (\it{c, a})
\item[\textbf{Step 3:}] \stab /* Construct assign rule in AURA */
\item[] a. assign\_formula = 
\item[] \ \ ((\it{u\textsubscript{2}}, \textsf{empower}) $\in$ \it{\uam}(\it{u\textsubscript{1}}) $\wedge$  (\it{r\textsubscript1}, \textsf{grant}) 
\item[] \ \ $\in$ \it{\ram}(\it{u\textsubscript{1}})) $\vee$ ((\it{u\textsubscript{2}}, \textsf{empower}) $\in$ \it{\uam}(\it{u\textsubscript{1}}) \item[] \ \ $\wedge$  (\textsf{role}, \textsf{grant}) $\in$ \it{\classp}(\it{u\textsubscript{1}})) $\vee$ ((\textsf{user}, \textsf{empower}) 
\item[] \ \ $\in$ \it{\classp}(\it{u\textsubscript{1}})
$\wedge$  (\it{r\textsubscript1}, \textsf{grant}) $\in$ \it{\ram}(\it{u\textsubscript{1}})) $\vee$
\item[] \ \ ((\textsf{user}, \textsf{empower}) $\in$ \it{\classp}(\it{u\textsubscript{1}}) $\wedge$
\item[] \ \  (\textsf{role}, \textsf{grant}) $\in$ \it{\classp}(\it{u\textsubscript{1}})) 
\item[] b. auth\_assign = 
\item[] \ \ \ \ {\isauth}U\textsubscript{\textbf{assign}}(\it{u\textsubscript1} : \U\textsuperscript{\t{A}}, \it{u\textsubscript2} : \U\textsuperscript{\t{A}}, 
\item[] \stab \ \it{r\textsubscript1} : \R\textsuperscript{A}) $\equiv$  assign\_formula
\vspace{-0.4cm}
\end{spacing}
\end{algorithmic}
\end{algorithm}


\floatname{algorithm}{Continuation of Algorithm}
\setcounter{algorithm}{4}
\begin{algorithm}
\caption{Map\textsubscript{URA-UARBAC}}
\begin{algorithmic} [1] 
\begin{spacing}{1.2}
\item[\textbf{Step 4:}] \stab /* Construct revoke rule for AURA */
\item[] a. revoke\_formula = 
\item[] \ \ ((\it{u\textsubscript{2}}, \textsf{empower}) $\in$ \it{\uam}(\it{u\textsubscript{1}}) $\wedge$ (\it{r\textsubscript1}, \textsf{grant}) 
\item[] \ \ $\in$ \it{\ram}(\it{u\textsubscript{1}})) $\vee$ (\it{u\textsubscript{2}}, \textsf{admin}) $\in$ \it{\uam}(\it{u\textsubscript{1}}) $\vee$
\item[] \ \  (\it{r\textsubscript1}, \textsf{admin}) $\in$ \it{\ram}(\it{u\textsubscript{1}}) $\vee$ (\textsf{user}, \textsf{admin}) 
\item[] \ \ $\in$ \it{\classp}(\it{u\textsubscript{1}}) $\vee$ (\textsf{role}, \textsf{admin}) $\in$ \it{\classp}(\it{u\textsubscript{1}})
\item[] b. auth\_revoke = 
\item[]  \ \ \ \ {\isauth}U\textsubscript{\textbf{revoke}}(\it{u\textsubscript1} : \U\textsuperscript{\t{A}}, \it{u\textsubscript2} : \U\textsuperscript{\t{A}}, 
\item[]  \stab \ \it{r\textsubscript1} : \R\textsuperscript{A}) $\equiv$ revoke\_formula
\end{spacing}
\end{algorithmic}
\end{algorithm}

\subsubsection{Map\textsubscript{URA-UARBAC}}\label{algo:ura-uarbac}
Map\textsubscript{URA-UARBAC} is an algorithm that maps any instance of URA in UARBAC~\cite{uarbac} (URA-U) to its equivalent AURA instance. For clarity, sets and function from UARBAC model are labeled with superscript \t{U}, and
that of AURA with superscript \t{A}. 
Input to Map\textsubscript{URA-UARBAC} consists of 
\it{C}\textsuperscript{\t{U}}, \it{\U}\textsuperscript{\t{U}}, \it{\R}\textsuperscript{\t{U}}, \it{\UA}\textsuperscript{\t{U}},
\it{\RH}\textsuperscript{\t{U}}, \it{AM}\textsuperscript{\t{U}}(\textsf{user}),
\it{AM}\textsuperscript{\t{U}}(\textsf{role}), For each \it{u} $\in$ \it{\U}\textsuperscript{\t{U}},
authorized\_perms\textsuperscript{\t{U}}[\it{u}],

For each \it{u\textsubscript1} $\in$
\it{\U}\textsuperscript{\t{U}} and for each \it{r\textsubscript1} $\in$ \it{\R}\textsuperscript{\t{U}},
grantRoleToUser(\it{u\textsubscript1, r\textsubscript1}) is true if the granter has one of the
following combinations of permissions: 
\begin{itemize}
 \item {[\textsf{user}, \it{u\textsubscript{1}}, \textsf{empower}] and  [\textsf{role}, \it{r\textsubscript1}, \textsf{grant}]}, or
 \item {[\textsf{user}, \it{u\textsubscript{1}}, \textsf{empower}] and [\textsf{role}, \textsf{grant}]}, or
 \item {[\textsf{user}, \textsf{empower}] and  [\textsf{role}, \it{r\textsubscript1}, \textsf{grant}]}, or
 \item {[\textsf{user}, \textsf{empower}] and [\textsf{role}, \textsf{grant}]},  
 \end{itemize}
 For each \it{u\textsubscript1} $\in$ \it{\U}\textsuperscript{\t{U}} and for each \it{r\textsubscript1} $\in$ \it{\R}\textsuperscript{\t{U}}, revokeRoleFromUser(\it{u\textsubscript1, r\textsubscript1}) is true if the granter has either of the following permissions : 
 \begin{itemize}
 \item {[\textsf{user}, \it{u\textsubscript{1}}, \textsf{empower}] and [\textsf{role}, \it{r\textsubscript1}, \textsf{grant}] }or, 
 \item {[\textsf{user}, \it{u\textsubscript1}, \textsf{admin}]} or, 
 \item {[\textsf{role}, \it{r\textsubscript1}, \textsf{admin}]} or,
 \item {[\textsf{user}, \textsf{admin}]} or,
 \item {[\textsf{role}, \textsf{admin}]}
\end{itemize}
Output from Map\textsubscript{URA-UARBAC} is an AURA instance with primarily following
sets and functions: \U\textsuperscript{\t{A}}, \AU\textsuperscript{\t{A}}, \OP\textsuperscript{\t{A}},
\R\textsuperscript{\t{A}}, \RH\textsuperscript{\t{A}}, For each \it{u} $\in$ \U\textsuperscript{\t{A}},
\it{roles}(\it{u}), \UATT\textsuperscript{\t{A}}, \AATT\textsuperscript{\t{A}}, For each attribute
\it{att} $\in$ \UATT\textsuperscript{\t{A}} $\cup$  \AATT\textsuperscript{\t{A}},
\scope\textsuperscript{\t{A}}(\it{att}), attType\textsuperscript{\t{A}}(\it{att}),
\isord\textsuperscript{\t{A}}(\it{att}) and H\textsuperscript{\t{A}}\textsubscript{\it{att}}, For each user
\it{u} $\in$ \U\textsuperscript{\t{A}}, and for each \it{att} $\in$ \UATT\textsuperscript{\t{A}}
$\cup$ \AATT\textsuperscript{\t{A}}, \it{att}(\it{u}), Authorization rule for assign
(auth\_assign), Authorization rule for revoke (auth\_revoke))

There are four primary steps in translating a URA-U instance to AURA instance. Step 1 in Map\textsubscript{URA-UARBAC} involves translating sets and functions from 
URA-U to AURA equivalent sets and functions. In Step 2, user attributes and 
admin user attributes functions are defined. Regular user attributes (\UATT)
is set to null as there is no regular user attributes required. An admin user's authority towards a regular user and a role, defined by \it{access modes}, decides whether she can assign that user to role. AURA defines three admin user attributes: \it{\uam, \ram} and \it{classp}. \it{\uam} attribute captures an admin user's access mode towards a particular regular user. Similarly, \it{\ram} captures an admin user's access mode towards a particular role. An admin user can also have a class level access mode captured by attribute \it{classp}. With class level access mode, an admin user gains authority over an entire class of object. For example [\textsf{grant, role}] admin permission provides an admin user with power to grant any 
role.   
In Step 3, assign\_formula equal to {\isauth}U\textsubscript{\textbf{assign}}(\it{u\textsubscript1} : \U\textsuperscript{\t{A}}, \it{u\textsubscript2} : \U\textsuperscript{\t{A}}, \it{r\textsubscript1} : \R\textsuperscript{A}) for AURA that is equivalent to
grantRoleToUser(\it{u\textsubscript1, r\textsubscript1}) in URA-U is established.
Similarly, in Step 4 revoke\_formula equivalent to revokeRoleFromUser(\it{u\textsubscript1,
r\textsubscript1}) is constructed. 

\section{Mapping Prior PRA Models in ARPA}\label{sec:PRAtranslations}

In this section, we demonstrate that ARPA can intuitively simulate
the features of prior PRA models. In particular, we have developed
concrete algorithms that can convert any instance of PRA97, PRA99, PRA02, the 
PRA model in UARBAC, and the PRA model in Uni-ARBAC into an equivalent instance of ARPA.
The following sections also present example instances of each of the prior PRA 
models and their corresponding instances in AURA/ARPA model followed by a 
formal mapping alorithms.

\subsection{PRA97 in ARPA}
In this section, we present an example instance for PRA97 followed by an equivalent ARPA instance. We also present an algorithm that translates any instance of PRA97 into corresponding equivalent ARPA instance.
\subsubsection{PRA97 Instance}
In this section we present an example instance for PRA97. The sets are as follows:
\begin{itemize}
\item \U\ = \{\textbf{u\textsubscript{1}, u\textsubscript{2}, u\textsubscript{3}, u\textsubscript4}\}
\item \R\ = \{\textbf{x\textsubscript{1}, x\textsubscript{2}, x\textsubscript{3}, x\textsubscript4, x\textsubscript5, x\textsubscript6}\}
\item \AR\ = \{\textbf{ar\textsubscript{1}, ar\textsubscript{2}}\}
\item \P\ = \{\textbf{p\textsubscript{1}, p\textsubscript{2}, p\textsubscript{3}, p\textsubscript4}\}
\item \AUA\ = \{(\textbf{u\textsubscript1, ar\textsubscript1}), (\textbf{u\textsubscript3, ar\textsubscript2})\}
\item \PA\ = \{(\textbf{p\textsubscript1, r\textsubscript1}), (\textbf{p\textsubscript2, r\textsubscript2}), (\textbf{p\textsubscript2, r\textsubscript4}), (\textbf{p\textsubscript3, r\textsubscript3}), (\textbf{p\textsubscript4, r\textsubscript3}), (\textbf{p\textsubscript4, r\textsubscript4})\}
\item \RH\ = \{(\textbf{x\textsubscript{1}, x\textsubscript{2}}), (\textbf{x\textsubscript2, x\textsubscript{3}}), (\textbf{x\textsubscript3, x\textsubscript4}), (\textbf{x\textsubscript4, x\textsubscript5}), (\textbf{x\textsubscript5, x\textsubscript6})\}
\item \ARH\ = \{(\textbf{ar\textsubscript{1}, ar\textsubscript{2}})\}
\item \CR\ = \{\textbf{x\textsubscript{1} $\wedge$ x\textsubscript{2}}, \textbf{$\bar{\textrm{x}}$\textsubscript{1}} $\vee$ \textbf{x\textsubscript3}\}
\end{itemize}
Let \textit{cr\textsubscript{1}} = \textbf{x\textsubscript{1} $\wedge$ x\textsubscript{2}} and, 
\textit{cr\textsubscript{2}} = \textbf{$\bar{\textrm{x}}$\textsubscript1} $\vee$ \textbf{x\textsubscript3}.

\vspace{0.17cm}
\noindent
Prerequisite condition \it{cr\textsubscript{1}} is evaluated as follows:\\
For each \it{p} that is undertaken for assignment, \\
($\exists$\textit{x} $\leq$ \textbf{x\textsubscript{1}})(\textit{p}, \textit{x}) $\in$ \PA\ $\wedge$ ($\exists$\textit{x} $\leq$ \textbf{x\textsubscript{2}})(\textit{p}, \textit{x}) $\in$ \PA

\vspace{0.17cm}
\noindent
\textit{cr\textsubscript{2}} is evaluated as follows:\\
For each \it{p} that is undertaken for assignment, \\
($\exists$\textit{x} $\leq$ \textbf{x\textsubscript{1}})(\textit{p}, \textit{x}) $\notin$ \PA\ $\vee$ ($\exists$\textit{x} $\leq$ \textbf{x\textsubscript{3}})(\textit{p}, \textit{x}) $\in$ \PA

\vspace{0.17cm}
\noindent
Let \textit{can\_assignp} and \textit{can\_revokep} be as follows:\\
\textit{can\_assignp} = \{(\textbf{ar\textsubscript1}, \textit{cr\textsubscript1,} \{\textbf{x\textsubscript4, x\textsubscript5}\}), (\textbf{ar\textsubscript1}, \textit{cr\textsubscript2,} \{\textbf{x\textsubscript6}\})\}\\
\it{can\_revokep} = \{(\textbf{ar\textsubscript1}, \R)\}\\

\subsubsection{Equivalent Example Instance of ARPA for PRA97}
This section presents an equivalent ARPA instance for the aforementioned PRA97 example instance. 
\underline{Set and functions:}
\begin{itemize}
\item \AU\ = \{\textbf{u\textsubscript{1}, u\textsubscript{2}, u\textsubscript{3}, u\textsubscript4}\}
\item \OP\ = \{\textbf{assign, revoke}\}
\item \R\ = \{\textbf{x\textsubscript{1}, x\textsubscript{2}, x\textsubscript{3}\, x\textsubscript4, x\textsubscript5, x\textsubscript6}\}
\item \RH\ = \{(\textbf{x\textsubscript{1}, x\textsubscript{2}}), (\textbf{x\textsubscript2, x\textsubscript{3}}), (\textbf{x\textsubscript3, x\textsubscript4}), (\textbf{x\textsubscript5, x\textsubscript6})\}
\item \P\ = \{\textbf{p\textsubscript1, p\textsubscript2, p\textsubscript3, p\textsubscript4}\}

\item \AATT\ = \{\textit{aroles}\}
\item \scope(\textit{aroles}) = \{\textbf{ar\textsubscript1, ar\textsubscript2}\}
\item[] attType(\textit{aroles}) = set, \isord(\it{aroles}) = True,
\item[] H\textsubscript{\it{aroles}} = \{(\textbf{ar\textsubscript1, ar\textsubscript2})\}
\item \textit{aroles}(\textbf{u\textsubscript1}) = \{\textbf{ar\textsubscript1}\}, \textit{aroles}(\textbf{u\textsubscript2}) = \{\}, 
\item[]\it{aroles}(\textbf{u\textsubscript3}) = \{\textbf{ar\textsubscript2}\}, \textit{aroles}(\textbf{u\textsubscript4}) = \{\}

\item \PATT\ = \{\it{rolesp}\}
\item \scope(\it{rolesp}) = \R, attType(\it{rolesp}) = set, 
\item \isord(\it{rolesp}) = True, H\textsubscript{\it{rolesp}} = \RH
\item \it{rolesp}(\textbf{p\textsubscript1}) = \{\textbf{r\textsubscript1}\}, \it{rolesp}(\textbf{p\textsubscript2}) = \{\textbf{r\textsubscript2, r\textsubscript4}\}, 
\item[]\it{rolesp}(\textbf{p\textsubscript3}) = \{\textbf{r\textsubscript3}\}, \it{rolesp}(\textbf{p\textsubscript4}) = \{\textbf{r\textsubscript3, r\textsubscript4}\}
\end{itemize}

\noindent
Authorization rule for user-role assignment can be expressed as follows:\\
 For any permission \it{p} $\in$ \P\ undertaken for assignment,\\
 -- {\isauth}P\textsubscript{\textbf{assign}}(\it{au} : \AU\textsuperscript{\t{A}}, \it{p} : \P\textsuperscript{\t{A}}, \it{r} : \R\textsuperscript{\t{A}}) \hspace*{0.18cm}$\equiv$ \\
 \hspace*{0.18cm}(($\exists$\textit{ar} $\geq$ \textbf{ar\textsubscript1}). \textit{ar} $\in$ \textit{aroles}(\it{au}) $\wedge$ \it{r} $\in$ \{\textbf{x\textsubscript4, x\textsubscript5}\} $\wedge$ \\
 \hspace*{0.18cm}($\exists$\textit{x} $\leq$ \textbf{x\textsubscript{1}}). \it{x} $\in$ \textit{rolesp}(\textit{p}) $\wedge$ ($\exists$\textit{x} $\leq$ \textbf{x\textsubscript{2}}). \it{x} $\in$ \textit{rolesp}(\textit{p}) $\vee$ \\ 
 \hspace*{0.18cm}($\exists$\textit{ar} $\geq$ \textbf{ar\textsubscript1}). \textit{ar} $\in$ \textit{aroles}(\textit{au})  $\wedge$ \it{r} $\in$ \{\textbf{x\textsubscript6}\} $\wedge$ \\
 \hspace*{0.18cm}($\exists$\textit{x} $\leq$ \textbf{x\textsubscript{1}}). \it{x} $\notin$ \textit{rolesp}(\textit{p}) $\vee$ ($\exists$\textit{x} $\leq$ \textbf{x\textsubscript{3}}). \it{x} $\in$ \textit{rolesp}(\textit{p}) 

\vspace{0.17cm}
\noindent
Authorization rule to revoke a permission from a role can be expressed as follows:\\
For any permission \it{p} $\in$ \P\ undertaken for revocation,\\
 -- {\isauth}P\textsubscript{\textbf{revoke}}(\it{au} : \AU\textsuperscript{\t{A}}, \it{p} : \P\textsuperscript{\t{A}}, \it{r} : \R\textsuperscript{\t{A}}) \\
 \hspace*{0.2cm}$\equiv$ ($\exists$\textit{ar} $\leq$ \textbf{ar\textsubscript1}). \textit{ar} $\in$ \textit{aroles}(\textit{au}) $\wedge$ \it{r} $\in$ \R\\
\subsubsection{Map\textsubscript{PRA97}}
Algorithm~\ref{alg:pra97} presents Map\textsubscript{PRA97}, which is an algorithm for mapping any PRA97 instance into equivalent ARPA instance. Sets and functions from PRA97 and ARPA are marked with superscripts \t{97} and \t{A}, respectively.
Map\textsubscript{PRA97} takes PRA97 instance as its input. In particular, input for Map\textsubscript{PRA97} fundamentally has
\U\textsuperscript{\t{97}}, \R\textsuperscript{\t{97}}, \AR\textsuperscript{\t{97}}, \P\textsuperscript{\t{97}}, \AUA\textsuperscript{\t{97}}, \PA\textsuperscript{\t{97}}, \RH\textsuperscript{\t{97}}, \ARH\textsuperscript{\t{97}}, \it{can\_assignp}\textsuperscript{\t{97}}, and \it{can\_revokep}\textsuperscript{\t{97}}

Output from Map\textsubscript{PRA97} algorithm is an equivalent ARPA instance, with primarily consisting of 
\AU\textsuperscript{\t{A}}, \OP\textsuperscript{\t{A}}, \RN\textsuperscript{\t{A}}, \RNH\textsuperscript{\t{A}}, \P\textsuperscript{\t{A}},
\AATT\textsuperscript{\t{A}}, \PATT\textsuperscript{\t{A}}, 
For each attribute \it{att} $\in$ \AATT\textsuperscript{\t{A}} $\cup$ \PATT\textsuperscript{\t{A}},
\scope\textsuperscript{\t{A}}(\it{att}), attType\textsuperscript{\t{A}}(\it{att}), is\_ordered\textsuperscript{\t{A}}(\it{att}) and H\textsuperscript{\t{A}}\textsubscript{\it{att}},
For each user \it{u} $\in$ \AU\textsuperscript{\t{A}}, and for each \it{att} $\in$ \AATT\textsuperscript{\t{A}}, \it{att}(\it{u}),
For each permission \it{p} $\in$ \P\textsuperscript{\t{A}}, and for each \it{att} $\in$ \PATT\textsuperscript{\t{A}}, \it{att}(\it{p}),
Authorization rule for permission assign (auth\_assign), and Authorization rule for permission revoke (auth\_revoke)
\floatname{algorithm}{Algorithm}
\begin{algorithm}[tp]
\caption{Map\textsubscript{PRA97}}
\label{alg:pra97}
\begin{algorithmic} [] 
\begin{spacing}{1.3}
\item[]\hspace{-17pt}\textbf{Input:} PRA97 instance 
\item[]\hspace{-17pt}\textbf{Output:} ARPA instance 
\item[\textbf{Step 1:}] \ \ /* Map basic sets and functions in ARPA */
\item[] a. \AU\textsuperscript{\t{A}} $\leftarrow$ \U\textsuperscript{\t{97}} ; \OP\textsuperscript{\t{A}} $\leftarrow$ \{\textbf{assign, revoke}\} 
\item[] b. \R\textsuperscript{\t{A}} $\leftarrow$ \R\textsuperscript{\t{97}} ; \RH\textsuperscript{\t{A}} $\leftarrow$ \RH\textsuperscript{\t{97}} 
\item[] c. \P\textsuperscript{\t{A}} $\leftarrow$ \P\textsuperscript{\t{97}} 

\item[\textbf{Step 2:}]   \stab /* Map attribute functions in ARPA */
\item[] a. \AATT\textsuperscript{\t{A}} $\leftarrow$ \{\it{aroles}\} 
\item[] b. \scope \textsuperscript{\t{A}}(\it{aroles}) = \AR\textsuperscript{\t{97}} ; attType\textsuperscript{\t{A}}(\it{aroles}) = set 
\item[] c. is\_ordered\textsuperscript{\t{A}}(\it{aroles}) = True ; H\textsuperscript{\t{A}}\textsubscript{\it{aroles}} $\leftarrow$ \ARH\textsuperscript{\t{97}}  
\item[] d. For each \it{u} $\in$ \AU\textsuperscript{\t{A}}, \it{aroles}(\it{u}) = $\phi$
\item[] e. For each (\it{u, ar}) in \AUA\textsuperscript{\t{97}},
\item[] \tab  \it{aroles}(\it{u})\textquotesingle\ = \it{aroles}(\it{u}) $\cup$ \it{ar}

\item[] f. \PATT\textsuperscript{\t{A}} $\leftarrow$ \{\it{rolesp}\} 
\item[] g. \scope\textsuperscript{\t{A}}(\it{rolesp}) = \R\textsuperscript{\t{A}} 
\item[] h. attType\textsuperscript{\t{A}}(\it{rolesp}) = set
\item[] i. is\_ordered\textsuperscript{\t{A}}(\it{rolesp}) = True ; H\textsuperscript{\t{A}}\textsubscript{\it{rolesp}} $\leftarrow$ \RH\textsuperscript{\t{A}}
\item[] j. For each \it{p} in \P\textsuperscript{\t{A}}, \it{rolesp}(\it{u}) = $\phi$ 
\item[] k. For each (\it{p, r}) in \PA\textsuperscript{\t{97}}, \it{rolesp}(\it{p})\textquotesingle = \it{rolesp}(\it{p}) $\cup$ \it{r}

\item[\textbf{Step 3:}] \stab /* Construct assign rule in ARPA */
\item[] a. assign\_formula = $\phi$ 
\item[] b. For each (\it{ar, cr, Z}) $\in$ \it{can\_assignp}\textsuperscript{\t{97}}, 
\item[] \ \ \ assign\_formula\textquotesingle\ = assign\_formula $\vee$ 
\item[] \stab (($\exists$\textit{ar\textquotesingle\ $\geq$ ar}). \it{ar\textquotesingle} $\in$ \it{aroles}(\it{au}) $\wedge$ \it{r $\in$ Z} $\wedge$ 
\item[] \stab (\it{translatep}\textsubscript{97}(\it{cr})))
\item[] c. auth\_assign = 
\item[] \stab {\isauth}P\textsubscript{\textbf{assign}}(\it{au} : \AU\textsuperscript{\t{A}}, \it{p} : \P\textsuperscript{\t{A}}, 
\item[]\stab \it{r} : \R\textsuperscript{\t{A}}) $\equiv$ assign\_formula\textquotesingle

\item[\textbf{Step 4:}] \stab /* Construct revoke rule for ARPA */
\item[] a. revoke\_formula = $\phi$
\item[] b. For each (\it{ar, cr, Z}) $\in$ \it{can\_revokep}\textsuperscript{\t{97}}
\item[] \ \  revoke\_formula\textquotesingle\ = revoke\_formula $\vee$
\item[] \stab (($\exists$\textit{ar\textquotesingle\ $\geq$ ar}). \it{ar\textquotesingle} $\in$ \it{aroles}(\it{au}) $\wedge$ \it{r} $\in$ \it{Z})
\item[] c. auth\_revoke = 
\item[]\stab {\isauth}P\textsubscript{\textbf{assign}}(\it{au} : \AU\textsuperscript{\t{A}}, \it{p} : \P\textsuperscript{\t{A}}, 
\item[]\stab \it{r} : \R\textsuperscript{\t{A}}) $\equiv$ assign\_formula\textquotesingle
\vspace{-0.3cm}
\end{spacing}
\end{algorithmic}
\end{algorithm}

\floatname{algorithm}{Support routine for algorithm}
\setcounter{algorithm}{5}
\begin{algorithm}
\caption{\it{translatep}\textsubscript{97}}
\begin{algorithmic} [1]
\begin{spacing}{1.2}
\item[]\hspace{-17pt}\textbf{Input:} A PRA97 prerequisite condition, \it{cr}
\item[]\hspace{-17pt}\textbf{Output:} An equivalent sub-rule for ARPA authorization  
\item[] \stab assign rule.
\STATE \it{rule\_string} = $\phi$ 
\STATE For each \it{symbol} in \it{cr},
\STATE \ \ \ \ \textbf{if} \it{symbol} is a role and in the form \it{x} 
\item[] \stab (i.e., the permission has membership on role \it{x}) 
\STATE \stab \stab \it{rule\_string}\textquotesingle\ = \it{rule\_string} + ($\exists$\it{x\textquotesingle}\ $\leq$ \it{x}). \it{x\textquotesingle}\ 
\item[] \tab \tab \stab $\in$ \it{rolesp}(\it{p})
\STATE \ \ \ \ \textbf{else if} \it{symbol} is a role and in the form $\bar{x}$ 
\item[] \stab (i.e., the permission doesn't have membership 
\item[] \stab on role \it{x}) 
\STATE \stab \stab \it{rule\_string}\textquotesingle\ = \it{rule\_string} + ($\exists$\it{x\textquotesingle}\ $\leq$ \it{x}). \it{x\textquotesingle}\ 
\item[] \tab \tab \stab $\notin$ \it{rolesp}(\it{p}) 
\STATE \stab \textbf{else}
\STATE \stab \stab \it{rule\_string}\textquotesingle\ = \it{rule\_string} + \it{symbol} 
\item[] /* where a \it{symbol} is a $\wedge$ or $\vee$ logical operator */
\STATE \stab \textbf{end if}
\vspace{-0.4cm}
\end{spacing}
\end{algorithmic}
\end{algorithm}

As indicated in Map\textsubscript{PRA97}, there are four main steps for mapping. 
In Step 1, sets and functions from PRA97 are mapped into ARPA sets and functions. 
In Step 2, permission attributes and administrative user attribute functions are
expressed. There exists one permission attribute called \it{rolesp}. It captures association 
between a roles and assigned permissions. 
Admin user attribute \it{aroles} captures the association between admin users and admin roles
in PRA97. Step 3 involves constructing assign\_formula in ARPA that is equivalent to
\it{can\_assignp}\textsuperscript{\t{97}}. \it{can\_assignp}\textsuperscript{\t{97}} is a set of triples. Each
triple bears information on whether an admin role can assign a candidate
permission to a set of roles. 

Equivalent translation equivalent to \it{can\_assignp}\textsuperscript{\t{97}} in ARPA is given by {\isauth}P\textsubscript{\textbf{assign}}(\it{au} : \AU\textsuperscript{\t{A}}, \it{p} :
\P\textsuperscript{\t{A}}, \it{r} : \R\textsuperscript{\t{A}}). Similarly, In Step 4, revoke\_formula equivalent to \it{can\_revokep}\textsuperscript{\t{97}} is presented. A support routine \it{translatep}\textsubscript{97} translates prerequisite condition. 

\floatname{algorithm}{Algorithm}
\setcounter{algorithm}{6}
\begin{algorithm}[tp]
\caption{Map\textsubscript{PRA99}}
\label{alg:pra99}
\begin{algorithmic} [] 
\begin{spacing}{1.25}
\item[]\hspace{-17pt}\textbf{Input:} PRA99 instance 
\item[]\hspace{-17pt}\textbf{Output:} ARPA instance

\item[\textbf{Step 1:}] \ \ /* Map basic sets and functions in ARPA */
\item[] a. \AU\textsuperscript{\t{A}} $\leftarrow$ \U\textsuperscript{\t{99}}
\item[] b. \OP\textsuperscript{\t{A}} $\leftarrow$ \{\textbf{mob-assign, mob-revoke, immob-assign, immob-revoke}\} 
\item[] c. \R\textsuperscript{\t{A}} $\leftarrow$ \R\textsuperscript{\t{99}} ; \RH\textsuperscript{\t{A}} $\leftarrow$ \RH\textsuperscript{\t{99}}
\item[] d. \P\textsuperscript{\t{A}} $\leftarrow$ \P\textsuperscript{\t{99}} 

\item[\textbf{Step 2:}] \stab /* Map attribute functions in ARPA */
\item[] a. \AATT\textsuperscript{\t{A}} $\leftarrow$ \{\it{aroles}\} ; \scope\textsuperscript{\t{A}}(\it{aroles}) = \AR\textsuperscript{\t{99}}
\item[] b. attType\textsuperscript{\t{A}}(\it{aroles}) = set ; \isord\textsuperscript{\t{A}}(\it{aroles}) = True 
\item[] c. H\textsuperscript{\t{A}}\textsubscript{\it{aroles}} $\leftarrow$ \ARH\textsuperscript{\t{99}} ; For each \it{u} $\in$ \AU\textsuperscript{\t{A}}, 
\item[] \tab \tab \tab \stab \it{aroles}(\it{u}) = $\phi$ 
\item[] d. For each (\it{u, ar}) in \AUA\textsuperscript{\t{99}}, 
\item[] \tab \it{aroles}(\it{u}) = \it{aroles}(\it{u}) $\cup$ \it{ar}

\item[] e. \PATT\textsuperscript{\t{A}} $\leftarrow$ \{\it{\exmm, \immm,
\item[] \eximm, \imimm}\} 
\item[] f. \scope\textsuperscript{\t{A}}(\it{\exmm}) = \R\textsuperscript{\t{A}}
\item[] g. attType\textsuperscript{\t{A}}(\it{\exmm}) = set  
\item[] h. \isord\textsuperscript{\t{A}}(\it{\exmm}) = True 
\item[] i. H\textsuperscript{\t{A}}\textsubscript{\it{\exmm}} $\leftarrow$ \RH\textsuperscript{\t{A}} ; For each \it{p} in \P\textsuperscript{\t{A}},
\item[] \stab\it{\exmm}(\it{p}) = $\phi$ 
\item[] j. For each (\it{p}, M\it{r}) in \PA\textsuperscript{\t{99}}, 
\item[] \tab \it{\exmm}(\it{p})\textquotesingle = \it{\exmm}(\it{p}) $\cup$ \it{r}
\item[] k. \scope\textsuperscript{\t{A}}(\it{\immm}) = \R\textsuperscript{\t{A}}
\item[] l. attType\textsuperscript{\t{A}}(\it{\immm}) = set 
\item[] m. \isord\textsuperscript{\t{A}}(\it{\immm}) = True
\item[] n. H\textsuperscript{\t{A}}\textsubscript{\it{\immm}} $\leftarrow$ \RH\textsuperscript{\t{A}} 
\item[] o. For each \it{p} in \P\textsuperscript{\t{A}}, \it{\immm}(\it{p}) = $\phi$  
\item[] p. For each (\it{p}, M\it{r}) in \PA\textsuperscript{\t{99}} 
\item[] \ \ \ \  and for each role \it{r}\textquotesingle\ > \it{r},
\item[] \stab \it{\immm}(\it{p})\textquotesingle = \it{\immm}(\it{p}) $\cup$ \it{r}\textquotesingle
\item[] q. \scope\textsuperscript{\t{A}}(\it{\eximm}) = \R\textsuperscript{\t{A}}
\item[] r. attType\textsuperscript{\t{A}}(\it{\eximm}) = set 
\item[] s. \isord\textsuperscript{\t{A}}(\it{\eximm}) = True 
\item[] t. H\textsuperscript{\t{A}}\textsubscript{\it{\eximm}} $\leftarrow$ \RH\textsuperscript{\t{A}} ; For each \it{p} in \P\textsuperscript{\t{A}}, 
\item[] \stab \it{\eximm}(\it{p}) = $\phi$  ; 
\item[] u. For each (\it{p}, IM\it{r}) in \PA\textsuperscript{\t{99}}, 
\item[] \stab \it{\eximm}(\it{p})\textquotesingle = \it{\eximm}(\it{p}) $\cup$ \it{r}
\item[] v. \scope\textsuperscript{\t{A}}(\it{\imimm}) = \R\textsuperscript{\t{A}}

\vspace{-0.4cm}
\end{spacing}
\end{algorithmic}
\end{algorithm}

\floatname{algorithm}{Continuation of Algorithm}
\setcounter{algorithm}{6}
\begin{algorithm}
\caption{Map\textsubscript{PRA99}}

\begin{algorithmic} [1] 
\begin{spacing}{1.14}

\item[] w. attType\textsuperscript{\t{A}}(\it{\imimm}) = set 
\item[] x. \isord\textsuperscript{\t{A}}(\it{\imimm}) = True  
\item[] y. H\textsuperscript{\t{A}}\textsubscript{\it{\imimm}} $\leftarrow$ \RH\textsuperscript{\t{A}} ; For each \it{p} in \P\textsuperscript{\t{A}},
\item[] \tab \it{\imimm}(\it{p}) = $\phi$
\item[] z. For each (\it{p}, IM\it{r}) in \PA\textsuperscript{\t{99}} 
\item[] \ \ \ \ and for each role \it{r}\textquotesingle\ > \it{r},
\item[] \stab \it{\imimm}(\it{p})\textquotesingle = \it{\imimm}(\it{p}) $\cup$ \it{r}\textquotesingle

\item[\textbf{Step 3:}] \tab /* Construct assign rule in ARPA */
\item[] a. assign-mob-formula = $\phi$
\item[] b. For each (\it{ar, cr, Z}) $\in$ \it{can-assignp-M}\textsuperscript{\t{99}}, 
\item[] \stab assign-mob-formula\textquotesingle\ = assign-mob-formula $\vee$ 
\item[]\tab (($\exists$\it{ar\textquotesingle\ $\geq$ ar}). \it{ar\textquotesingle} $\in$ \it{aroles}(\it{au}) $\wedge$ \it{r $\in$ Z} $\wedge$\\
\item[] \tab (\it{translatep}\textsubscript{99}(\it{cr}, mob-assign))) 
\item[] c. auth\_mob\_assign = 
\item[] \stab {\isauth}P\textsubscript{\textbf{mob-assign}}(\it{au} : \AU\textsuperscript{\t{A}}, \it{p} : \P\textsuperscript{\t{A}}, 
\item[] \tab \it{r} : \R\textsuperscript{\t{A}}) $\equiv$ assign-mob-formula\textquotesingle
 
\item[] d. assign-immob-formula = $\phi$
\item[] e. For each (\it{ar, cr, Z}) $\in$ \it{can-assignp-IM}\textsuperscript{\t{99}}, 
\item[]  \ \ \  assign-immob-formula\textquotesingle\ = assign-immob-formula 
\item[] \stab $\vee$ (($\exists$\it{ar\textquotesingle\ $\geq$ ar}). \it{ar\textquotesingle} $\in$ \it{aroles}(\it{au}) $\wedge$ \it{r} $\in$ \it{Z} $\wedge$ 
\item[] \stab (\it{translatep}\textsubscript{99}(\it{cr}, immob-assign)))
\item[] f. auth\_immob\_assign =
\item[]\stab {\isauth}P\textsubscript{\textbf{immob-assign}}(\it{au} : \AU\textsuperscript{\t{A}}, \it{p} : \P\textsuperscript{\t{A}}, 
\item[] \tab \it{r} : \R\textsuperscript{\t{A}}) $\equiv$ assign-immob-formula\textquotesingle

\item[\textbf{Step 4:}] \stab /* Construct revoke rule in ARPA */
\item[] a. revoke-mob-formula = $\phi$
\item[] b. For each (\it{ar, cr, Z}) $\in$ \it{can-revokep-M}\textsuperscript{\t{99}}, 
\item[] \stab revoke-mob-formula\textquotesingle\ = revoke-mob-formula $\vee$
\item[] \tab(($\exists$\it{ar\textquotesingle\ $\geq$ ar}). \it{ar\textquotesingle} $\in$ \it{aroles}(\it{au}) $\wedge$ \it{r} $\in$ \it{Z} $\wedge$ \\
\item[] \tab (\it{translatep}\textsubscript{99}(\it{cr}, mob-revoke)))  
\item[] c. auth\_mob\_revoke = 
\item[] \stab {\isauth}P\textsubscript{\textbf{mob-revoke}}(\it{au} : \AU\textsuperscript{\t{A}}, \it{p} : \P\textsuperscript{\t{A}}, 
\item[] \tab \it{r} : \R\textsuperscript{\t{A}}) $\equiv$  revoke-mob-formula\textquotesingle

\item[] d. revoke-immob-formula = $\phi$
\item[] e. For each (\it{ar, cr, Z}) $\in$ \it{can-revokep-IM}\textsuperscript{\t{99}}, 
\item[] \ \ \ \ revoke-immob-formula\textquotesingle\ = revoke-immob-formula 
\item[] \stab $\vee$ (($\exists$\it{ar\textquotesingle\ $\geq$ ar}). \it{ar\textquotesingle} $\in$ \it{aroles}(\it{au}) $\wedge$ \it{r} $\in$ \it{Z} $\wedge$ \\
\item[] \stab (\it{translatep}\textsubscript{99}(\it{cr}, immob-revoke)))

\item[] f. auth\_immob\_revoke = 
\item[] \stab {\isauth}P\textsubscript{\textbf{immob-revoke}}(\it{au} : \AU\textsuperscript{\t{A}}, \it{p} : \P\textsuperscript{\t{A}}, 
\item[] \tab \it{r} : \R\textsuperscript{\t{A}}) $\equiv$ revoke-immob-formula\textquotesingle
\vspace{-0.4cm}
\end{spacing}
\end{algorithmic}
\end{algorithm}

\floatname{algorithm}{Support routine for algorithm}
\setcounter{algorithm}{6}
\begin{algorithm}
\caption{\it{translatep}\textsubscript{99}}
\begin{algorithmic} [1]
\begin{spacing}{1.1}
\item[]\hspace{-17pt}\textbf{Input:} A PRA99 prerequisite condition (\it{cr}), \\
\it{op} $\in$ \{\textbf{mob-assign, immob-assign, mob-revoke, immob-revoke}\}
\item[]\hspace{-17pt}\textbf{Output:} An equivalent sub-rule for AURA authorization assign rule.
\STATE \it{rule\_string} = $\phi$ 
\STATE For each \it{symbol} in \it{cr}
\STATE \ \ \ \textbf{if} \it{op} = (mob-assign $\vee$ immob-assign) $\wedge$ \it{symbol} 
\item[] \tab \ \ is a role and in the form \it{x} \\
\item[] \stab (i.e., the permission has membership on role \it{x})
\STATE \stab\stab \it{rule\_string} = \it{rule\_string} + (\it{x} $\in$ 
\item[] \tab \it{\exmm}(\it{p}) $\vee$ (\it{x} $\in$ \it{\immm}(\it{p}) 
\item[] \tab  $\wedge$ \it{x} $\notin$ \it{\eximm}(\it{p}))

\item[] \ \ \ \textbf{else if} \it{op} = (mob-revoke $\vee$ immob-revoke) $\wedge$ 
\item[] \tab \ \ \it{symbol} is a role and in the form \it{x} \\
\item[] \stab (i.e., the permission has membership on role \it{x})
\STATE \tab\it{rule\_string} = \it{rule\_string} + (\it{x} $\in$ 
\item[] \tab \it{\exmm}(\it{p}) $\vee$ \it{x} $\in$ \it{\immm}(\it{p}) 
\item[] \tab $\vee$ \it{x} $\in$ \it{\eximm}(\it{p}) 
\item[] \tab $\vee$ \it{x} $\in$ \it{\imimm}(\it{p})) 

\STATE \ \ \  \textbf{else if} \it{op} = (mob-assign $\vee$ immob-assign $\vee$  
\item[] \tab mob-revoke $\vee$ immob-revoke) $\wedge$
\item[] \tab\it{symbol} is role and in the form $\bar{x}$
\item[] \stab(i.e., the permission doesn't have membership
\item[] \stab on role \it{x})
\STATE \stab \it{rule\_string} = \it{rule\_string} + (\it{x} $\notin$ 
\item[] \stab \it{\exmm}(\it{p}) $\wedge$ \it{x} $\notin$ \it{\immm}(\it{p}) $\wedge$ \\
\item[] \stab \it{x} $\notin$ \it{\eximm}(\it{p}) $\wedge$ 
\item[] \stab \it{x} $\notin$ \it{\imimm}(\it{p})) 
\STATE \ \ \ \textbf{else}
\STATE \tab \it{rule\_string} = \it{rule\_string} + \it{symbol} 
\item[] \ \ \ /* where a \it{symbol} is a $\wedge$ or $\vee$ logical operator */
\STATE \ \ \  \textbf{end if}
\vspace{-0.4cm}
\end{spacing}
\end{algorithmic}
\end{algorithm}

\subsection{PRA99 in ARPA}
\subsubsection{PRA99 Instance}
\label{sec:pra99instance}
In this section, an example instance of the PRA99 model is presented.
\underline{Sets and functions:}
\begin{itemize}
\item \U\ = \{\textbf{u\textsubscript1, u\textsubscript2, u\textsubscript3, u\textsubscript4}\}
\item \R\ = \{\textbf{x\textsubscript{1}, x\textsubscript{2}, x\textsubscript{3}, x\textsubscript{4}, x\textsubscript{5}, x\textsubscript{6}}\}
\item \AR\ = \{\textbf{ar\textsubscript{1}, ar\textsubscript2}\}
\item \P\ = \{\textbf{p\textsubscript1, p\textsubscript2, p\textsubscript3, p\textsubscript4}\}
\item \AUA\ = \{(\textbf{u\textsubscript3, ar\textsubscript1}), (\textbf{u\textsubscript4, ar\textsubscript2})\}
\item \PA\ = \{(\textbf{p\textsubscript1}, M\textbf{x\textsubscript1}), (\textbf{p\textsubscript2}, IM\textbf{x\textsubscript3}), (\textbf{p\textsubscript3}, IM\textbf{x\textsubscript2}), (\textbf{p\textsubscript4}, M\textbf{x\textsubscript4})\}
\item \RH\ = \{(\textbf{x\textsubscript1, x\textsubscript2}), (\textbf{x\textsubscript2, x\textsubscript3}), (\textbf{x\textsubscript3, x\textsubscript4}), (\textbf{x\textsubscript4, x\textsubscript5}), (\textbf{x\textsubscript5, x\textsubscript6})\}
\item \ARH\ = \{(\textbf{ar\textsubscript{1}, ar\textsubscript2})\}
\item \CR\ = \{\textbf{x\textsubscript{2}, $\bar{\textrm{x}}$\textsubscript{1}}\}
\end{itemize}
Let \textit{cr\textsubscript{1}} = \textbf{x\textsubscript2} and,  
 \textit{cr\textsubscript{2}} = \textbf{$\bar{\textrm{x}}$\textsubscript{1}}. 

\noindent
 Prerequisite condition \textit{cr\textsubscript{1}} is evaluated as follows:\\
For each \it{p} that is undertaken for assignment, \\
((\textit{p}, M\textbf{x\textsubscript2}) $\in$ \PA\ $\vee$ (($\exists$\textit{x\textquotesingle} $\leq$ \textbf{x\textsubscript2}). (\textit{p}, M\textit{x\textquotesingle}) $\in$ \PA) $\wedge$\\
 (\textit{p}, IM\textbf{x\textsubscript2}) $\notin$ \PA)

\vspace{0.17cm}
\noindent
\textit{cr\textsubscript{2}} is evaluated as follows:\\
For each \it{p} that is undertaken for assignment, \\
(\textit{p}, M\textbf{x\textsubscript1}) $\notin$ \PA\ $\wedge$ (($\exists$\it{x\textquotesingle} $\leq$ \textbf{x\textsubscript1}). (\textit{p}, M\textit{x\textquotesingle}) $\notin$ \PA) $\wedge$ (\textit{p}, IM\textbf{x\textsubscript1}) $\notin$ \PA\ $\wedge$ 
 (($\exists$\textit{x\textquotesingle} $\leq$ \textbf{x\textsubscript1}). (\textit{p}, IM\textit{x\textquotesingle}) $\notin$ \PA)
 
\vspace{0.17cm}
\noindent
Let \textit{can-assignp-M} and \textit{can-assignp-IM} in PRA99 be as follows:\\
\textit{can-assignp-M} = \{(\textbf{ar\textsubscript1}, \textit{cr\textsubscript1,} \{\textbf{x\textsubscript4, x\textsubscript5}\})\}\\
\textit{can-assignp-IM}= \{(\textbf{ar\textsubscript1}, \textit{cr\textsubscript2,} \{\textbf{x\textsubscript3}\})\}

\vspace{0.17cm}
\noindent
For simplicity, same prerequisite conditions are considered for grant and revoke model instances. Prerequisite conditions for PRA99 revoke model are evaluated as follows:

\vspace{0.17cm}
\noindent 
\it{cr\textsubscript1} is evaluated as follows:\\
 ((\it{p}, M\textbf{x\textsubscript2}) $\in$ \PA\ $\vee$ (\it{p}, IM\textbf{x\textsubscript2}) $\in$ \PA\ $\vee$ 
(($\exists$\textit{x\textquotesingle} $\leq$ \textbf{x\textsubscript2}). \\
(\textit{p}, M\textit{x\textquotesingle}) $\in$ \PA) $\vee$  (($\exists$\textit{x\textquotesingle} $\leq$ \textbf{x\textsubscript2}). (\textit{p}, IM\textit{x\textquotesingle}) $\in$ \PA))

\vspace{0.17cm}
\noindent 
\it{cr\textsubscript2} is evaluated as follows:\\
(\it{p}, M\textbf{x\textsubscript1}) $\notin$ \PA\ $\wedge$ (\it{p}, IM\textbf{x\textsubscript1}) $\notin$ \PA\ $\wedge$  (($\exists$\textit{x\textquotesingle} $\leq$ \textbf{x\textsubscript1}). \\
(\textit{p}, M\textit{x\textquotesingle}) $\notin$ \PA) $\wedge$
 (($\exists$\textit{x\textquotesingle} $\leq$ \textbf{x\textsubscript1}). (\textit{p}, IM\textit{x\textquotesingle}) $\notin$ \PA) 
 
\vspace{0.17cm}
\noindent 
Let \textit{can-revokep-M} and \textit{can-revokep-IM} sets be as follows:\\
\textit{can-revokep-M} = \{(\textbf{ar\textsubscript1}, \textit{cr\textsubscript1,} \{\textbf{x\textsubscript3, x\textsubscript4, x\textsubscript5}\})\}\\
\textit{can-revokep-IM}= \{(\textbf{ar\textsubscript1}, \textit{cr\textsubscript2,} \{\textbf{x\textsubscript5, x\textsubscript6}\})\}

\vspace{0.17cm}
\noindent 
\subsubsection{Equivalent PRA99 Instance in ARPA}
\label{sec:arpa99}
This section presents an equivalent ARPA instance for the aforementioned PRA99 example instance.\\
\underline{Sets and functions:}
\begin{itemize}
\item \AU\ = \{\textbf{u\textsubscript1, u\textsubscript2, u\textsubscript3, u\textsubscript4}\}
\item \OP\ = \{\textbf{mob-assign, immob-assign, 
\item[] mob-revoke, immob-revoke}\}
\item \R\ = \{\textbf{x\textsubscript1, x\textsubscript2, x\textsubscript3, x\textsubscript4, x\textsubscript5, x\textsubscript6}\}
\item \RH\ = \{(\textbf{x\textsubscript1, x\textsubscript2}), (\textbf{x\textsubscript2, x\textsubscript3}), (\textbf{x\textsubscript3, x\textsubscript4}), (\textbf{x\textsubscript4, x\textsubscript5}), (\textbf{x\textsubscript5, x\textsubscript6})\}
\item \P\ = \{\textbf{p\textsubscript1, p\textsubscript2, p\textsubscript3, p\textsubscript4}\}

\item \AATT\ = \{\textit{aroles}\}
\item \scope(\textit{aroles}) = \{\textbf{ar\textsubscript{1}, ar\textsubscript{2}}\}, attType(\it{aroles}) = set,
\item[]\isord(\it{aroles}) = True, H\textsubscript{\it{aroles}} = \{(\textbf{ar\textsubscript{1}, ar\textsubscript{2}})\}

\item \it{aroles}(\textbf{u\textsubscript1}) = \{\}, \it{aroles}(\textbf{u\textsubscript2}) = \{\}, 
\item[]\it{aroles}(\textbf{u\textsubscript3}) = \{\textbf{ar\textsubscript1}\}, \it{aroles}(\textbf{u\textsubscript4}) = \{\textbf{ar\textsubscript2}\}

\item \PATT = \{\textit{\exmm, \immm, 
\item[] \eximm, \imimm}\}
\item \scope(\textit{\exmm}) = \R, 
\item[] attType(\textit{\exmm}) = set, 
\item[] \isord(\it{\exmm}) = True, 
\item[] H\textsubscript{\it{\exmm}} = \RH
\item \it{\exmm}(\textbf{p\textsubscript1}) = \{\textbf{x\textsubscript1}\},
\it{\exmm}(\textbf{p\textsubscript2}) = \{\textbf{}\},
\item[] \it{\exmm}(\textbf{p\textsubscript3}) = \{\textbf{}\},
\it{\exmm}(\textbf{p\textsubscript4}) = \{\textbf{x\textsubscript4}\},

\item \scope(\textit{\immm})= \R,
\item[] attType(\textit{\immm}) = set
\item[]\isord(\it{\immm}) = True, 
\item[] H\textsubscript{\it{\immm}} = \RH
\item \it{\immm}(\textbf{p\textsubscript1}) = \{\textbf{}\},
\it{\immm}(\textbf{p\textsubscript2}) = \{\textbf{}\},
\item[]\it{\immm}(\textbf{p\textsubscript3}) = \{\textbf{}\},
\item[]\it{\immm}(\textbf{p\textsubscript4}) = \{\textbf{x\textsubscript1, x\textsubscript2, x\textsubscript3}\}
\item \scope(\textit{\eximm}) = \R, 
\item[] attType(\textit{\eximm}) = set
\item[]\isord(\it{\eximm}) = True,
\item[] H\textsubscript{\it{\eximm}} = \RH

\item \it{\eximm}(\textbf{p\textsubscript1}) = \{\textbf{}\},
\item[] \it{\eximm}(\textbf{p\textsubscript2}) = \{\textbf{x\textsubscript3}\},
\item[] \it{\eximm}(\textbf{p\textsubscript3}) = \{\textbf{x\textsubscript2}\}, 
\item[] \it{\eximm}(\textbf{p\textsubscript4}) = \{\textbf{}\},

\item \scope(\textit{\imimm}) = \R, 
\item[] attType(\textit{\imimm}) = set
\item \isord(\it{\imimm}) = True, 
\item[] H\textsubscript{\it{\imimm}} = \RH
\item \it{\imimm}(\textbf{p\textsubscript1}) = \{\textbf{}\},
\item[] \it{\imimm}(\textbf{p\textsubscript2}) = \{\textbf{x\textsubscript1, x\textsubscript2}\}, 
\item[] \it{\imimm}(\textbf{p\textsubscript3}) = \{\textbf{x\textsubscript1}\}, 
\item[] \it{\imimm}(\textbf{p\textsubscript4}) = \{\textbf{}\}
\end{itemize}
Authorization rule to assign a permission as a mobile member of a role can be expressed as follows:\\
To assigning any permission \it{p} $\in$ \P\ as a mobile member,\\
-- {\isauth}P\textsubscript{\textbf{mob-assign}}(\it{au} : \AU, \it{p} : \P, \\
\hspace*{0.17cm} \it{r} : \R) $\equiv$ \\
\hspace*{0.17cm} (($\exists$\textit{ar} $\geq$ \textbf{ar\textsubscript1}). \textit{ar} $\in$ \textit{aroles}(\textit{u}) $\wedge$ \textit{r} $\in$ \{\textbf{x\textsubscript4, x\textsubscript5}\} $\wedge$\\
 \hspace*{0.17cm} (\textbf{x\textsubscript{2}} $\in$ \textit{\exmm}(\textit{p}) $\vee$ (\textbf{x\textsubscript{2}} $\in$ \textit{\immm}(\textit{p}) $\wedge$ \\
 \hspace*{0.17cm} \textbf{x\textsubscript{2}} $\notin$ \textit{\eximm}(\textit{p})))

\vspace{0.17cm}
\noindent 
Authorization rule to revoke a mobile permission from a role can be expressed as follows:\\
To revoke any mobile permission \it{p} $\in$ \P\ from a role,\\
 -- {\isauth}P\textsubscript{\textbf{mob-revoke}}(\it{au} : \AU, \it{p} : \P, \\
 \hspace*{0.17cm} \it{r} : \R) $\equiv$ \\
\hspace*{0.17cm} (($\exists$\textit{ar} $\geq$ \textbf{ar\textsubscript1}). \textit{ar} $\in$ \textit{aroles}(\textit{u}) $\wedge$ \textit{r} $\in$ \{\textbf{x\textsubscript3, x\textsubscript4, x\textsubscript5}\} $\wedge$ \\
 \hspace*{0.17cm} (\textbf{x\textsubscript{2}} $\in$ \textit{\exmm}(\textit{p}) $\vee$ \textbf{x\textsubscript{2}} $\in$ \textit{\immm}(\textit{p}) $\vee$ \\
 \hspace*{0.17cm} \textbf{x\textsubscript{2}} $\in$ \textit{\eximm}(\textit{p}) $\vee$ \textbf{x\textsubscript{2}} $\in$ \textit{\imimm})) 

\vspace{0.17cm}
\noindent 
Authorization functions to assign any permission \it{p} $\in$ \P\ as an immobile member of role can be expressed as follows: \\
To assign any permission \it{p} $\in$ \P\ as a immobile member,\\
\noindent
-- {\isauth}P\textsubscript{\textbf{immob-assign}}(\it{au} : \AU, \it{p} : \P, \\
\hspace*{0.17cm} \it{r} : \R) $\equiv$ (($\exists$\textit{ar} $\geq$ \textbf{ar\textsubscript1}). \textit{ar} $\in$ \textit{aroles}(\textit{u}) $\wedge$ \\
\hspace*{0.17cm} \textit{r} $\in$ \{\textbf{x\textsubscript3}\} $\wedge$ 
(\textbf{x\textsubscript{1}} $\notin$ \textit{\exmm}(\textit{p}) $\wedge$ \\
\hspace*{0.17cm} \textbf{x\textsubscript{1}} $\notin$ \textit{\immm}(\textit{p}) $\wedge$ \textbf{x\textsubscript{1}} $\notin$ \textit{\eximm}(\textit{p}) $\wedge$ \\
\hspace*{0.17cm} \textbf{x\textsubscript{1}} $\notin$ \textit{\imimm}(\textit{p})))

\vspace{0.17cm}
\noindent 
Authorization rule to revoke any immobile permission from a role can be expressed as follows:\\
To revoke any immobile permission \it{p} $\in$ \P\ from a role,\\
 -- {\isauth}P\textsubscript{\textbf{immob-revoke}}(\it{au} : \AU, \it{p} : \P, \\
 \hspace*{0.17cm} \it{r} : \R) $\equiv$ \\
 \hspace*{0.17cm} (($\exists$\textit{ar} $\geq$ \textbf{ar\textsubscript1}). \textit{ar} $\in$ \textit{aroles}(\textit{p}) $\wedge$ \textit{r} $\in$ \{\textbf{x\textsubscript5, x\textsubscript6}\} $\wedge$ \\
 \hspace*{0.17cm} \textbf{x\textsubscript{1}} $\notin$ \textit{\exmm}(\textit{p}) $\wedge$ \textbf{x\textsubscript{1}} $\notin$ \textit{\immm}(\textit{p}) $\wedge$ \\
 \hspace*{0.17cm} \textbf{x\textsubscript{1}} $\notin$ \textit{\eximm}(\textit{p}) $\wedge$ \textbf{x\textsubscript{1}} $\notin$ \textit{\imimm}(\textit{p})) \\

\subsubsection{Map\textsubscript{PRA99}}
Algorithm~\ref{alg:pra99} is an algorithm for mapping any PRA99 instance into equivalent ARPA instance. Sets and functions from PRA99 and ARPA are marked with textsuperscripts \t{99} and \t{A}, respectively.
Map\textsubscript{PRA99} takes PRA99 instance as its input. In particular, input for Map\textsubscript{PRA99} fundamentally has \U\textsuperscript{\t{99}}, \P\textsuperscript{\t{99}}, \R\textsuperscript{\t{99}}, \AR\textsuperscript{\t{99}}, \PA\textsuperscript{\t{99}}, \AUA\textsuperscript{\t{99}}, \RH\textsuperscript{\t{99}}, \ARH\textsuperscript{\t{99}}, \it{can-assignp-M}\textsuperscript{\t{99}}, \it{can-assignp-IM}\textsuperscript{\t{99}}, \it{can-revokep-M}\textsuperscript{\t{99}}, and \it{can-revokep-IM}\textsuperscript{\t{99}}.
	
Output from Map\textsubscript{PRA99} algorithm is an equivalent ARPA instance, with primarily consisting of 
\AU\textsuperscript{\t{A}}, \OP\textsuperscript{\t{A}}, \R\textsuperscript{\t{A}}, \RH\textsuperscript{\t{A}}, \P\textsuperscript{\t{A}}, \AATT\textsuperscript{\t{A}}, \PATT\textsuperscript{\t{A}},
For each attribute \it{att} $\in$ \AATT\textsuperscript{\t{A}} $\cup$ \PATT\textsuperscript{\t{A}}, 
\scope\textsuperscript{\t{A}}(\it{att}), attType\textsuperscript{\t{A}}(\it{att}), \isord\textsuperscript{\t{A}}(\it{att}) and H\textsuperscript{\t{A}}\textsubscript{\it{att}}, 
For each user \it{u} $\in$ \AU\textsuperscript{\t{A}}, and for each \it{att} $\in$ \AATT\textsuperscript{\t{A}}, \it{att}(\it{u}),
For each permission \it{p} $\in$ \P\textsuperscript{\t{A}}, and for each \it{att} $\in$ \PATT\textsuperscript{\t{A}}, \it{att}(\it{p}),
Authorization rule for mobile assign (auth\_mob\_assign), Authorization rule for mobile revoke (auth\_mob\_revoke), Authorization rule for immobile assign (auth\_immob\_assign), and Authorization rule for immobile revoke (auth\_immob\_revoke)

 As shown in Map\textsubscript{PRA99}, there are four main steps required in mapping any instance of PRA99 model to ARPA instance. In Step 1, sets and functions from PRA99 instance are mapped into ARPA sets and functions. In Step 2, permission attributes and administrative user attribute functions are expressed. There are four permission attributes: \it{\exmm, \immm, \eximm,} and \it{\imimm}. Each captures, a permission's explicit mobile membership, implicit mobile membership, explicit immobile membership and implicit immobile membership on roles, respectively. Admin user attribute \it{aroles} captures admin roles assigned to admin users. Step 3 involves constructing assign-mob-formula and assign-immob-formula in ARPA that is equivalent to \it{can-assignp-M}\textsuperscript{\t{99}} and \it{can-assignp-IM}\textsuperscript{\t{99}}, respectively. Both \it{can-assignp-M}\textsuperscript{\t{99}} and \it{can-assignp-IM}\textsuperscript{\t{99}} are set of triples. Each triple bears information on whether an admin role can assign a candidate permission to a set of roles as a mobile member in the case of \it{can-assignp-M}\textsuperscript{\t{99}} and, as an immobile member in the case of \it{can-assignp-IM}\textsuperscript{\t{99}}. AURA equivalent for \it{can-assignp-M}\textsuperscript{\t{99}} is given by {\isauth}P\textsubscript{\textbf{mob-assign}}(\it{au} : \AU\textsuperscript{\t{A}}, \it{p} : \P\textsuperscript{\t{A}}, \it{r} : \R\textsuperscript{\t{A}}) and an equivalent translation for \it{can-assignp-IM}\textsuperscript{\t{99}} is given by {\isauth}P\textsubscript{\textbf{immob-assign}}(\it{au} : \AU\textsuperscript{\t{A}}, \it{p} : \P\textsuperscript{\t{A}}, \it{r} : \R\textsuperscript{\t{A}}). Similarly, In Step 4, revoke-mob-formula equivalent to \it{can-revokep-M}\textsuperscript{\t{99}} and \it{can-revokep-IM}\textsuperscript{\t{99}} are presented. A support routine \it{translatep}\textsubscript{99} translates prerequisite condition in PRA99 into its ARPA equivalent.

\floatname{algorithm}{Algorithm}
\setcounter{algorithm}{7}
\begin{algorithm}[tp]
\caption{Map\textsuperscript{PRA02}}
\label{alg:pra02}
\begin{algorithmic} [] 
\begin{spacing}{1.1}
\item[]\hspace{-17pt}\textbf{Input:} PRA02 instance 
\item[]\hspace{-17pt}\textbf{Output:} AURA instance
\item[\textbf{Step 1:}] \ \ /* Map basic sets and functions in ARPA */
\item[] a. \AU\textsuperscript{\t{A}} $\leftarrow$ \AU\textsuperscript{\t{02}} ; \OP\textsuperscript{\t{A}} $\leftarrow$ \{\textbf{assign, revoke}\} 
\item[] b. \R\textsuperscript{\t{A}} $\leftarrow$ \R\textsuperscript{\t{02}} 
\item[] c. \RH\textsuperscript{\t{A}} $\leftarrow$ \RH\textsuperscript{\t{02}} ; \P\textsuperscript{\t{A}} $\leftarrow$ \P\textsuperscript{\t{02}}

\item[\textbf{Step 2:}] \stab /* Map attribute functions to ARPA */
\item[] a. \AATT\textsuperscript{\t{A}} $\leftarrow$ \{\it{aroles}\} ; \scope\textsuperscript{\t{A}}(\it{aroles}) = \AR\textsuperscript{\t{02}} 
\item[] b. attType\textsuperscript{\t{A}}(\it{aroles}) = set
\item[] c. \isord\textsuperscript{\t{A}}(\it{aroles}) = True ; H\textsuperscript{\t{A}}\textsuperscript{\it{aroles}} $\leftarrow$ \ARH\textsuperscript{\t{02}}
\item[] d. For each \it{u} $\in$ \AU\textsuperscript{\t{A}}, \it{aroles}(\it{u}) = $\phi$ 
\item[] e. For each (\it{u, ar}) in \AUA\textsuperscript{\t{02}},
\item[] \tab \it{aroles}(\it{u})\textquotesingle\ = \it{aroles}(\it{u}) $\cup$ \it{ar}

\item[] f. \PATT\textsuperscript{\t{A}} $\leftarrow$ \{\it{org\_units, rolesp}\}
\item[] g. \scope\textsuperscript{\t{A}}(\it{org\_units}) = \OU\textsuperscript{\t{02}}
\item[] h. attType\textsuperscript{\t{A}}(\it{org\_units}) =  set 
\item[] i. \isord\textsuperscript{\t{A}}(\it{org\_units}) = True 
\item[] j. H\textsuperscript{\t{A}}\textsuperscript{\it{org\_units}} =  OUH\textsuperscript{\t{02}}
\item[] k. For each \it{p} $\in$ \P\textsuperscript{\t{A}}, \it{org\_units}(\it{p}) = $\phi$ 
\item[] l. For each (\it{p, orgu}) $\in$ \PPA\textsuperscript{\t{02}},
\item[] \tab  \it{org\_units}(\it{p})\textquotesingle\ = \it{org\_units}(\it{p}) $\cup$ \it{orgu}

\item[] m. \scope\textsuperscript{\t{A}}(\it{rolesp}) = \R\textsuperscript{\t{A}} 
\item[] n. attType\textsuperscript{\t{A}}(\it{rolesp}) =  set 
\item[] o. \isord\textsuperscript{\t{A}}(\it{rolesp}) = True 
\item[] p. H\textsuperscript{\t{A}}\textsuperscript{\it{rolesp}} =  \RH\textsuperscript{\t{A}}
\item[] q. For each \it{p} $\in$ \P\textsuperscript{\t{A}}, \it{rolesp}(\it{p}) = $\phi$ 
\item[] r. For each (\it{p, r}) $\in$ \PPA\textsuperscript{\t{02}}, 
\item[] \tab \it{rolesp}(\it{p})\textquotesingle\ = \it{rolesp}(\it{p}) $\cup$ \it{r}

\item[\textbf{Step 3:}] \stab /* Construct assign rule in ARPA */ 
\item[] a. assign\_formula = $\phi$
\item[] b. For each (\it{ar, cr, Z}) $\in$ \it{can\_assignp}\textsuperscript{\t{02}}, 
\item[] \stab assign\_formula\textquotesingle\ = assign\_formula $\vee$ 
\item[] \tab (($\exists$\it{ar\textquotesingle\ $\geq$ ar}). \it{ar\textquotesingle} $\in$ \it{aroles}(\it{au}) $\wedge$ \it{r $\in$ Z} $\wedge$ \\
\item[] \tab (\it{translatep}\textsuperscript{02}(\it{cr}))) 
\item[] c. auth\_assign = {\isauth}\textsuperscript{\textbf{assign}}(\it{au} : \AU\textsuperscript{\t{A}}, 
\item[] \stab \it{p} : \P{\t{A}}, \it{r} : \R{\t{A}}) $\equiv$ assign\_formula\textquotesingle

\item[\textbf{Step 4:}]\stab /* Construct revoke rule in ARPA */
\item[] a. revoke\_formula = $\phi$
\item[] b. For each (\it{ar, cr, Z}) $\in$ \it{can\_revokep}\textsuperscript{\t{02}}, 
\item[] \stab revoke\_formula\textquotesingle\ = revoke\_formula $\vee$ 
\item[] \tab (($\exists$\it{ar\textquotesingle\ $\geq$ ar}). \it{ar\textquotesingle} $\in$ \it{aroles}(\it{au}) $\wedge$ \it{r} $\in$ \it{Z}) 
\item[] c. auth\_revoke = {\isauth}\textsuperscript{\textbf{revoke}}(\it{au} : \AU{\t{A}}, 
\item[] \stab \it{p} : \P{\t{A}}, \it{r} : \R{\t{A}}) $\equiv$ revoke\_formula\textquotesingle
\vspace{-0.4cm}
\end{spacing}
\end{algorithmic}
\end{algorithm}
\floatname{algorithm}{Support routine for algorithm}
\setcounter{algorithm}{7}
\begin{algorithm}[]
\caption{\it{translatep}\textsuperscript{02}}
\begin{algorithmic} [1]
\begin{spacing}{1.1}
\item[]\hspace{-17pt}\textbf{Input:} A PRA02 prerequisite condition (\it{cr}), Case 1, Case 2
\item[]\hspace{-17pt}\textbf{Output:} An equivalent sub-rule for ARPA authorization rule.
\item[] \textbf{Begin:}
\STATE \it{rule\_string} = $\phi$ 
\STATE \textbf{Case Of} selection
\STATE \stab \textquotesingle\ Case 1 \textquotesingle\ (\it{cr} is based on roles) :
\STATE \tab \it{translatep}\textsuperscript{97} 

\STATE \stab \textquotesingle\ Case 2 \textquotesingle\ (\it{cr} is based on org\_units): 
\STATE \stab For each \it{symbol} in \it{cr}
\STATE \stab \ \ \textbf{if} \it{symbol} is an organization unit and in 
\item[] \stab \ \ the form \it{x}
\item[] \stab (i.e., the permission has a membership on 
\item[] \stab organization unit \it{x})
\STATE \tab \it{rule\_string} = \it{rule\_string} + ($\exists$\textit{x\textquotesingle} $\geq$ \it{x}). 
\item[] \tab \it{x\textquotesingle} $\in$ \textit{org\_units}(\textit{p}) 

\STATE \stab \ \ \textbf{else if} \it{symbol} an organization unit 
\item[] \tab and in the form $\bar{x}$ 
\item[]\stab (i.e., the permission doesn't have a membership 
\item[] \stab on organization unit \it{x})
\STATE \tab \it{rule\_string} = \it{rule\_string} + ($\exists$\it{x\textquotesingle}\ $\geq$ \it{x}). 
\item[] \tab \it{x\textquotesingle}\ $\notin$ \it{org\_units}(\it{p}) 

\STATE \stab \ \ \textbf{else}
\STATE \tab \it{rule\_string} = \it{rule\_string} + \it{symbol} \ 
\item[] \ \ \  /* where a \it{symbol} is a $\wedge$ or $\vee$ logical operator */
\STATE \stab\stab \textbf{end if}
\STATE \textbf{end Case}
\vspace{-0.4cm}
\end{spacing}
\end{algorithmic}
\end{algorithm}

\subsection{PRA02 in ARPA}
\subsubsection{PRA02 Instance}\label{sec:pra02instance}
This section basically consists of PRA02 example instance followed by its equivalent ARPA instance. We also present a mapping algorithm, Map\textsubscript{PRA02}. In PRA02, decision about permission-role assignment and revocation is made on the basis of two factors: a permission's membership on role(s) or a permission's membership in organization unit(s). They can be viewed as two different cases. In this example instance we represent roles with \it{r} and organization units with \it{x}, for simplicity and clarity.  \\
\underline{Sets and functions:}
\begin{itemize}
\item \U\ = \{\textbf{u\textsubscript1, u\textsubscript2, u\textsubscript3, u\textsubscript4}\}
\item \R\ = \{\textbf{r\textsubscript1, r\textsubscript2, r\textsubscript3, r\textsubscript4, r\textsubscript5, r\textsubscript6}\}
\item \AR\ = \{\textbf{ar\textsubscript{1}, ar\textsubscript2}\}
\item \P\ = \{\textbf{p\textsubscript1, p\textsubscript2, p\textsubscript3, p\textsubscript4}\}
\item \AUA\ = \{(\textbf{u\textsubscript3, ar\textsubscript1}), (\textbf{u\textsubscript4, ar\textsubscript2})\}
\item \PA\ = \{(\textbf{p\textsubscript1, r\textsubscript1}), (\textbf{p\textsubscript1, r\textsubscript2}), (\textbf{p\textsubscript2, r\textsubscript3}), (\textbf{p\textsubscript2, r\textsubscript4})\}
\item \RH\ = \{(\textbf{r\textsubscript1, r\textsubscript2}), (\textbf{r\textsubscript2, r\textsubscript3}), (\textbf{r\textsubscript3, r\textsubscript4}), (\textbf{r\textsubscript4, r\textsubscript5}), (\textbf{r\textsubscript5, r\textsubscript6})\}
\item \ARH\ = \{(\textbf{ar\textsubscript1, ar\textsubscript2})\}
\item \OU\ = \{\textbf{x\textsubscript{1}, x\textsubscript{2}, x\textsubscript{3}}\}
\item \OUH\ = \{(\textbf{x\textsubscript3, x\textsubscript{2}}), (\textbf{x\textsubscript{2}, x\textsubscript{1}})\}
\item \PPA\ = \{(\textbf{p\textsubscript1, x\textsubscript1}), (\textbf{p\textsubscript2, x\textsubscript2}), (\textbf{p\textsubscript3, x\textsubscript3}), (\textbf{p\textsubscript1, x\textsubscript3})\}
\\
\underline{Case 1:}
\item \CR\ = \{\textbf{r\textsubscript{1} $\wedge$ r\textsubscript{2}, r\textsubscript{1} $\vee$ $\bar{\textrm{r}}$\textsubscript{2} $\wedge$ x\textsubscript{3}}\}\\
Let \textit{cr\textsubscript{1}} = \textbf{r\textsubscript{1} $\wedge$ r\textsubscript{2}} and, 
 \textit{cr\textsubscript{2}} = \textbf{r\textsubscript{1} $\vee$ $\bar{\textrm{r}}$\textsubscript{2} $\wedge$ r\textsubscript{3}}
\\
\underline{Case 2:}
\item \CR\ = \{\textbf{x\textsubscript{1} $\wedge$ x\textsubscript{2}, x\textsubscript{1} $\vee$ $\bar{\textrm{x}}$\textsubscript{2} $\wedge$ x\textsubscript{3}}\}\\
Let \textit{cr\textsubscript{3}} = \textbf{x\textsubscript{1} $\wedge$ x\textsubscript{2}} and, 
 \textit{cr\textsubscript{4}} = \textbf{x\textsubscript{1} $\vee$ $\bar{\textrm{x}}$\textsubscript{2} $\wedge$ x\textsubscript{3}}
\end{itemize}
Prerequisite conditions are evaluated as follows:\\
\underline{Case 1:}

\noindent
\textit{cr\textsubscript{1}} is evaluated as follows:\\
For each \it{p} that is undertaken for assignment, \\
($\exists$\textit{r} $\leq$ \textbf{r\textsubscript{1}}). (\textit{p}, \textit{r}) $\in$ \PA\ $\wedge$ ($\exists$\textit{r} $\leq$ \textbf{r\textsubscript{2}}). (\it{p}, \it{r}) $\in$ \PA

\vspace{0.17cm}
\noindent
\textit{cr\textsubscript{2}} is evaluated as follows:
For each \it{p} that is undertaken for assignment, \\
($\exists$\it{r} $\leq$ \textbf{r\textsubscript{1}}). (\textit{p}, \textit{r}) $\in$ \PA\ $\vee$ $\neg$(($\forall$\textit{r} $\leq$ \textbf{r\textsubscript{2}}). (\it{p}, \it{r}) $\in$ \PA) $\wedge$ ($\exists$\textit{r} $\leq$ \textbf{r\textsubscript{3}}). (\textit{p}, \textit{r}) $\in$ \PA

\vspace{0.17cm}
\noindent
\underline{Case 2:} 

\noindent
\textit{cr\textsubscript{3}} is evaluated as follows:\\
For each \it{p} that is undertaken for assignment, \\
($\exists$\textit{x} $\geq$ \textbf{x\textsubscript{1}}). (\textit{p}, \textit{x}) $\in$ \PPA\ $\wedge$ ($\exists$\textit{x} $\geq$ \textbf{x\textsubscript{2}}). (\textit{p}, \textit{x}) $\in$ \PPA

\vspace{0.17cm}
\noindent
\textit{cr\textsubscript{4}} is evaluated as follows:\\
For each \it{p} that is undertaken for assignment, \\
($\exists$\textit{x} $\geq$ \textbf{x\textsubscript{1}}). (\textit{p}, \textit{x}) $\in$ \PPA\ $\vee$ $\neg$(($\forall$\textit{x} $\geq$ \textbf{x\textsubscript{2}}). (\textit{p}, \textit{x}) $\in$ \PPA) $\wedge$ ($\exists$\textit{x} $\leq$ \textbf{x\textsubscript{3}}). (\textit{p}, \textit{x}) $\in$ \PPA

\vspace{0.17cm}
\noindent
Let \textit{can\_assignp} and \it{can\_revokep} be as follows:\\
\underline{Case 1:}\\
\textit{can\_assignp} = \{(\textbf{ar\textsubscript1}, \textit{cr\textsubscript1,} \{\textbf{r\textsubscript4, r\textsubscript5}\}), (\textbf{ar\textsubscript1}, \textit{cr\textsubscript2,} \{\textbf{r\textsubscript6}\})\}\\
\it{can\_revokep} = \{(\textbf{ar\textsubscript1}, \{\textbf{r\textsubscript1, r\textsubscript3, r\textsubscript4}\})\}

\vspace{0.17cm}
\noindent
\underline{Case 2:}\\
\textit{can\_assignp} = \{(\textbf{ar\textsubscript1}, \textit{cr\textsubscript3,} \{\textbf{r\textsubscript4, r\textsubscript5}\}), (\textbf{ar\textsubscript1}, \textit{cr\textsubscript4,} \{\textbf{r\textsubscript6}\})\}\\
\it{can\_revokep} = \{(\textbf{ar\textsubscript1}, \{\textbf{r\textsubscript1, r\textsubscript3, r\textsubscript4}\})\}

\vspace{0.17cm}
\noindent
\subsubsection{Equivalent PRA02 Instance in ARPA}\label{sec:arpa02}
This segment presents an equivalent ARPA instance for aforementioned example instance.\\
\underline{Sets and functions}
\begin{itemize}
\item \AU\ = \{\textbf{u\textsubscript1, u\textsubscript2, u\textsubscript3, u\textsubscript4}\}
\item \OP\ = \{assign, revoke\}
\item \R\  = \{\textbf{r\textsubscript1, r\textsubscript2, r\textsubscript3, r\textsubscript4, r\textsubscript5, r\textsubscript6}\}
\item \RH\ = \{(\textbf{r\textsubscript1, r\textsubscript2}), (\textbf{r\textsubscript2, r\textsubscript3}), (\textbf{r\textsubscript3, r\textsubscript4}), (\textbf{r\textsubscript4, r\textsubscript5}), (\textbf{r\textsubscript5, r\textsubscript6})\}
\item \P\ = \{\textbf{p\textsubscript1, p\textsubscript2, p\textsubscript3, p\textsubscript4}\}

\item \AATT\ = \{\textit{aroles}\}
\item \scope(\textit{aroles}) = \{\textbf{ar\textsubscript{1}, ar\textsubscript2}\}, attType(\textit{aroles}) = set 
\item[]{\isord}(\it{aroles}) = True, H\textsubscript{\it{aroles}} =  \{(\textbf{ar\textsubscript{1}, ar\textsubscript2})\}
\item \it{aroles}(\textbf{u\textsubscript3}) = \{\textbf{ar\textsubscript1}\}, \it{aroles}(\textbf{u\textsubscript4}) = \{\textbf{ar\textsubscript2}\}

\item \PATT\ = \{\textit{rolesp, org\_units}\}
\item \scope(\textit{rolesp}) =  \R, attType(\textit{rolesp}) = set\\
\item[]{\isord}(\it{rolesp}) = True, H\textsubscript{\it{rolesp}} = \RH,
\item \it{rolesp}(\textbf{p\textsubscript1}) = \{\textbf{r\textsubscript1, r\textsubscript2}\}, \it{rolesp}(\textbf{p\textsubscript2}) = \{\textbf{r\textsubscript3, r\textsubscript4}\}, 
\it{rolesp}(\textbf{p\textsubscript3}) = \{\}, \it{rolesp}(\textbf{p\textsubscript4}) = \{\}

\item \scope(\it{org\_units}) =  \{\textbf{x\textsubscript{1}, x\textsubscript{2}, x\textsubscript{3}}\}, 
\item[] attType(\textit{org\_units}) = set
\item[]{\isord}(\it{org\_units}) = True, 
\item[]H\textsubscript{\it{org\_units}} = \{(\textbf{x\textsubscript3, x\textsubscript{2}}), (\textbf{x\textsubscript{2}, x\textsubscript{1}})\}
\item \it{org\_units}(\textbf{p\textsubscript1}) = \{\textbf{x\textsubscript1, x\textsubscript3}\}, \it{org\_units}(\textbf{p\textsubscript2}) = \{\textbf{x\textsubscript2}\}, \it{org\_units}(\textbf{p\textsubscript3}) = \{\textbf{x\textsubscript3}\}

\end{itemize}
For each \it{op} in \OP, authorization rule for permission to role assignment and revocation can be expressed respectively, as follows:

\vspace{0.17cm}
\noindent
\underline{Case 1:}\\
For any permission \it{p} $\in$ \P\ undertaken for assignment,\\
-- {\isauth}P\textsubscript{\textbf{assign}}(\it{au} : \AU, \it{p} : \P, \it{r} : \R)\\
\hspace*{0.19cm}$\equiv$ \\
\hspace*{0.19cm}(($\exists$\it{ar} $\geq$ \textbf{ar\textsubscript1}). \it{ar} $\in$ \it{aroles}(\it{au}) $\wedge$ \it{r} $\in$ \{\textbf{r\textsubscript4, r\textsubscript5}\} $\wedge$ \\
\hspace*{0.19cm}(($\exists$\it{r} $\leq$ \textbf{r\textsubscript{1}}). \it{r} $\in$ \it{rolesp}(\it{p}) $\wedge$ ($\exists$\it{r} $\leq$ \textbf{r\textsubscript2}). \it{r} $\in$ \it{rolesp}(\it{p}))) \\
\hspace*{0.19cm}$\vee$ (($\exists$\it{ar} $\geq$ \textbf{ar\textsubscript1}). \it{ar} $\in$ \it{aroles}(\it{au}) $\wedge$ \it{r} $\in$ \{\textbf{r\textsubscript6}\} $\wedge$ \\
\hspace*{0.19cm}(($\exists$\it{r} $\leq$ \textbf{r\textsubscript1}). \it{r} $\in$ \it{rolesp}(\it{p}) $\vee$
 ($\exists$\it{r} $\leq$ \textbf{r\textsubscript2}). \it{r} $\notin$ \it{rolesp}(\it{p}) $\wedge$ \\
 \hspace*{0.19cm}($\exists$\it{r} $\leq$ \textbf{r\textsubscript3}). \it{r} $\in$ \it{rolesp}(\it{p})))

\vspace{0.17cm}
\noindent
For any pemrission \it{p} $\in$ \P\ undertaken for revocation, \\
-- {\isauth}P\textsubscript{\textbf{revoke}}(\it{au} : \AU, \it{p} : \P, \it{r} : \R) \\
\hspace*{0.19cm}$\equiv$ ($\exists$\it{ar} $\geq$ \textbf{ar\textsubscript1}). \it{ar} $\in$ \it{aroles}(\it{u}) $\wedge$ \it{r} $\in$ \{\textbf{r\textsubscript1, r\textsubscript3, r\textsubscript4}\}

\vspace{0.17cm}
\noindent
\underline{Case 2:}\\
or any permission \it{p} $\in$ \P\ undertaken for assignment,\\
-- {\isauth}P\textsubscript{\textbf{assign}}(\it{au} : \AU, \it{p} : \P, \it{r} : \R) \\
\hspace*{0.19cm}$\equiv$ \\
\hspace*{0.19cm}(($\exists$\it{ar} $\geq$ \textbf{ar\textsubscript1}). \it{ar} $\in$ \it{aroles}(\it{au}) $\wedge$ \it{r} $\in$ \{\textbf{r\textsubscript4, r\textsubscript5}\} $\wedge$ \\
\hspace*{0.19cm}(($\exists$\it{x} $\geq$ \textbf{x\textsubscript1}). \it{x} $\in$ \it{org\_units}(\it{p}) $\wedge$ ($\exists$\it{x} $\geq$ \textbf{x\textsubscript2}).\\
 \hspace*{0.19cm}\it{x} $\in$ \it{org\_units}(\it{p}))) $\vee$
 (($\exists$\it{ar} $\geq$ \textbf{ar\textsubscript1}). \it{ar} $\in$ \it{aroles}(\it{au}) $\wedge$\\
 \hspace*{0.19cm}\it{r} $\in$ \{\textbf{r\textsubscript6}\} $\wedge$ 
(($\exists$\it{x} $\geq$ \textbf{x\textsubscript1}). \it{x} $\in$ \it{org\_units}(\it{p}) $\vee$
 ($\exists$\it{x} $\geq$ \textbf{x\textsubscript2}). \\
\hspace*{0.19cm}\it{x} $\notin$ \it{org\_units}(\it{p}) $\wedge$ ($\exists$\it{x} $\geq$ \textbf{x\textsubscript3}). \it{x} $\in$ \it{org\_units}(\it{p})))

\vspace{0.17cm}
\noindent 
For any pemrission \it{p} $\in$ \P\ undertaken for revocation, \\
-- {\isauth}P\textsubscript{\textbf{revoke}}(\it{au} : \AU, \it{p} : \P, \it{r} : \R) \\
\hspace*{0.19cm}$\equiv$ ($\exists$\it{ar} $\geq$ \textbf{ar\textsubscript1}). \it{ar} $\in$ \it{aroles}(\it{p}) $\wedge$ \it{r} $\in$ \{\textbf{r\textsubscript1, r\textsubscript3, r\textsubscript4}\}\\

\subsubsection{Map\ss{PRA02}}
Algorithm~\ref{alg:pra02} facilitates mapping for any PRA02 instance to equivalent ARPA instance. Sets and functions from PRA02 and ARPA are marked with superscripts \t{02} and \t{A}, respectively. Map\ss{PRA02} takes PRA02 instance as its input. In particular, input for Map\ss{PRA02} fundamentally has \U\textsuperscript{\t{02}}, \R\textsuperscript{\t{02}}, \AR\textsuperscript{\t{02}}, \P\textsuperscript{\t{02}}, \AUA\textsuperscript{\t{02}}, \PA\textsuperscript{\t{02}}, \RH\textsuperscript{\t{02}}, \ARH\textsuperscript{\t{02}}, \it{can\_assignp}\textsuperscript{\t{02}}, \it{can\_revokep}\textsuperscript{\t{02}}, \OU\textsuperscript{\t{02}}, OUH\textsuperscript{\t{02}}, and \PPA\textsuperscript{\t{02}}
Output from Map\ss{PRA02} is an equivalent ARPA instance, with primarily consisting of \AU\textsuperscript{\t{A}}, \OP\textsuperscript{\t{A}}, \RN\textsuperscript{\t{A}}, \RNH\textsuperscript{\t{A}}, \P\textsuperscript{\t{A}}, \AATT\textsuperscript{\t{A}}, \PATT\textsuperscript{\t{A}},
For each attribute \it{att} $\in$ \AATT\textsuperscript{\t{A}} $\cup$ \PATT\textsuperscript{\t{A}}, 
 Range\textsuperscript{\t{A}}(\it{att}), attType\textsuperscript{\t{A}}(\it{att}), \isord\textsuperscript{\t{A}}(\it{att}) and H\textsuperscript{\t{A}}\ss{\it{att}}, 
 For each user \it{p} $\in$ \P\textsuperscript{\t{A}} and for each \it{att} $\in$ \PATT\textsuperscript{\t{A}}, \it{att}(\it{p}),
 For each user \it{u} $\in$ \AU\textsuperscript{\t{A}} and for each \it{att} $\in$ \AATT\textsuperscript{\t{A}}, \it{att}(\it{u}),
 Authorization rule to assign (auth\_assignp), and Authorization rule to revoke (auth\_revokep)

 As shown in Algorithm Map\ss{PRA02}, there are four main steps required in mapping any instance of PRA02 model to ARPA instance. In Step 1, sets and functions from PRA02 instance are mapped into ARPA sets and functions. In Step 2, permission attributes and administrative user attribute functions are expressed. \PATT\ set has two permission attributes, \it{org\_units} and \it{rolesp}. \it{org\_units} attribute captures a permission's association in an organization unit and \it{rolesp} captures roles to which permission have been assigned to. There are two ways a assignment decision is made in PRA02 which are marked as Case 1 and Case 2 in the model. Case 1 checks for permission's existing membership on roles while Case 2 checks for user's membership on organization units. \it{org\_units} is captured in Case 2. Case 1 is same as PRA97. Admin user attribute \it{aroles} captures admin roles assigned to admin users. Step 3 involves constructing assignp\_formula in ARPA that is equivalent to \it{can\_assignp}\textsuperscript{\t{02}} in PRA02. \it{can\_assignp}\textsuperscript{\t{02}} is a set of triples. Each triple bears information on whether an admin role can assign a candidate permission to a set of roles. Equivalent translation in ARPA for PRA02 is given by {\isauth}U\ss{\textbf{assign}}(\it{au} : \AU\textsuperscript{\t{A}}, \it{u} : \U\textsuperscript{\t{A}}, \it{r} : \R\textsuperscript{\t{A}}). Similarly, In Step 4, revoke\_formula equivalent to \it{can\_revokep}\textsuperscript{\t{02}} is presented. A support routine \it{translate}\ss{02} translates prerequisite condition in PRA02 into its equivalent in ARPA.

\floatname{algorithm}{Algorithm}
\begin{algorithm}[pt]
\caption{Map\textsubscript{PRA-Uni-ARBAC}}
\label{alg:alg2}
\begin{algorithmic} [1] 
\begin{spacing}{1.25}
\item[]\hspace{-17pt}\textbf{Input:} Instance of PRA in Uni-ARBAC
\item[]\hspace{-17pt}\textbf{Output:} AURA instance
\item[\textbf{Step 1:}] \ \ /* Map basic sets and functions in ARPA */
\item[] a. \AU\textsuperscript{\t{A}} $\leftarrow$ \U\textsuperscript{\t{Uni}} ; \OP\textsuperscript{\t{A}} $\leftarrow$ \{\textbf{assign, revoke}\} 
\item[] b. \R\textsuperscript{\t{A}} $\leftarrow$ \R\textsuperscript{\t{Uni}} ; \RH\textsuperscript{A} $\leftarrow$ \RH\textsuperscript{\t{Uni}}
\item[] c.  \P\textsuperscript{\t{A}} $\leftarrow$ \P\textsuperscript{\t{Uni}}
\item[\textbf{Step 2:}]   \stab /* Map attribute functions in ARPA */
\item[] a. \AATT\textsuperscript{\t{A}} $\leftarrow$ \{\it{\au, \aur}\} 
\item[] b. \scope\textsuperscript{\t{A}}(\it{\au}) = \AU\textsuperscript{\t{Uni}} 
\item[] c. attType\textsuperscript{\t{A}}(\it{\au}) = set
\item[] d. is\_ordered\textsuperscript{\t{A}}(\it{\au}) = True, 
\item[] e. H\textsuperscript{\t{A}}\textsubscript{\it{\au}} = AUH\textsuperscript{\t{Uni}}
\item[] f. For each \it{u} in \AU\textsuperscript{\t{A}}, \it{\au}(\it{u}) = $\phi$ 
\item[] g. For each (\it{u, au}) $\in$ \it{TA\_admin}\textsuperscript{\t{Uni}},
\item[] \stab \it{\au}(\it{u})\textquotesingle\ = \it{\au}(\it{u}) $\cup$ \it{au}
\item[] h. attType\textsuperscript{\t{A}}(\it{\aur}) = set 
\item[] i. is\_ordered\textsuperscript{\t{A}}(\it{\aur}) = False, 
\item[] j. H\textsubscript{\it{\aur}} = $\phi$ ; For each \it{u} in \AU\textsuperscript{\t{A}},
\item[] \stab \it{\aur}(\it{u}) = $\phi$ 
\item[] k. For each (\it{u, au}) $\in$ \it{TA\_admin}\textsuperscript{\t{Uni}} and 
\item[] \ \ for each \it{r} $\in$ \it{roles}\textsuperscript{\t{Uni}}(\it{au}), 
\item[] \ \ \ \it{\aur}(\it{u})\textquotesingle\ = \it{\aur}(\it{u}) $\cup$ (\it{au, r})
\item[] l. \PATT\textsuperscript{\t{A}} $\leftarrow$ \{\it{tasks, \tadu}\}
\item[] m. \scope\textsuperscript{\t{A}}(\it{tasks}) = \T\textsuperscript{\t{Uni}} ; attType\textsuperscript{\t{A}}(\it{tasks}) = set  
\item[] n. is\_ordered\textsuperscript{\t{A}}(\it{tasks}) = True ; H\textsuperscript{\t{A}}\textsubscript{\it{tasks}} = \tH\textsuperscript{\t{Uni}}
\item[] o. For each \it{p} in \P\textsuperscript{\t{A}}, \it{tasks}(\it{p}) = $\phi$ ; 
\item[] p. For each (\it{p, t}) $\in$ \PA\textsuperscript{\t{Uni}}, \it{tasks}(\it{p})\textquotesingle\ = \it{tasks}(\it{p}) $\cup$ \it{t}

\item[] q. \scope\textsuperscript{\t{A}}(\it{\tadu}) = \T\textsuperscript{\t{Uni}} $\times$ \AU\textsuperscript{\t{Uni}}
\item[] r. attType\textsuperscript{\t{A}}(\it{\tadu}) = set ; 
\item[] s. is\_ordered\textsuperscript{\t{A}}(\it{\tadu}) = False
\item[] t. H\textsuperscript{\t{A}}\textsubscript{\it{\tadu}} = $\phi$ ; For each \it{p} in \P\textsuperscript{\t{A}}, 
\item[] \stab \it{\tadu}(\it{p}) = $\phi$
\item[] u. For each (\it{p, t}) $\in$ \PA\textsuperscript{\t{Uni}} and 
\item[] \ \ \ for each \it{t} $\in$ \it{tasks}*\textsuperscript{\t{Uni}}(\it{au}),
\item[] \stab \it{\tadu}(\it{p})\q\ = \it{\tadu}(\it{p}) $\cup$ (\it{t, au})
\item[\textbf{Step 3:}] \stab /* Construct assign rule in ARPA */
\item[] a. can\_manage\_rule = \\
\hspace*{1.1ex}$\exists$\it{au\textsubscript1,  au\textsubscript2} $\in$ \scope(\it{\au}). 
(\it{au\textsubscript1, au\textsubscript2}) \\
\hspace*{1.1ex}$\in$ H\textsubscript{\it{\au}} $\wedge$ (\it{au\textsubscript1} $\in$ \it{\au}(\it{u}) $\wedge$ (\it{au\textsubscript2, r}) \\
\hspace*{1.1ex}$\in$ \it{\aur}(\it{u})) $\wedge$
$\exists$\it{t\textsubscript1, t\textsubscript2} $\in$ \scope(\it{tasks}). 
\end{spacing}
\vspace{-0.3pt}
\end{algorithmic}
\end{algorithm}

\floatname{algorithm}{Continuation of Algorithm}
\begin{algorithm}[pt]
\setcounter{algorithm}{8}
\caption{Map\textsubscript{PRA-Uni-ARBAC}}
\begin{algorithmic} [1] 
\begin{spacing}{1.2}
\item []
 \hspace*{1.1ex}{[(\it{t\textsubscript1, t\textsubscript2}) $\in$ \tH\ $\wedge$ $\forall$\it{q} $\in$ $\chi$. \it{t\textsubscript2} $\in$ \it{tasks}(\it{q}) $\wedge$ \\
 \hspace*{1.1ex}$\exists$\it{q}\q\ $\in$ (\P\textsuperscript{\t{A}}\ - $\chi$). \it{t\textsubscript2} $\notin$ \it{tasks}(\it{q}\q) $\wedge$ 
 \hspace*{1.1ex}(\it{t\textsubscript2, au\textsubscript2}) $\in$ \it{tasks\_adminu}(\it{q})]}
\item[] b. auth\_assign = 
{\isauth}P\textsubscript{\textbf{assign}}(\it{u} : \AU\textsuperscript{\t{A}}, \\
\hspace*{1.1ex}$\chi$ : 2\textsuperscript{\P\textsuperscript{\t{A}}}, 
 \it{r} : \R\textsuperscript{\t{A}}) $\equiv$ can\_manage\_rule

\item[\textbf{Step 4:}] \stab /* Construct revoke rule for ARPA */
\item[] a. auth\_revoke = 
 {\isauth}P\textsubscript{\textbf{revoke}}(\it{u} : \AU\textsuperscript{\t{A}}, \\ 
 \hspace*{1.1ex}$\chi$ : 2\textsuperscript{\P\textsuperscript{\t{A}}},
 \it{r} : \R\textsuperscript{\t{A}}) $\equiv$ can\_manage\_rule
 \vspace*{-0.3cm}
%
\end{spacing}
\end{algorithmic}
\end{algorithm}

\subsection{Uni-ARBAC's PRA in ARPA}
\label{algo:uni-arpa}

\subsubsection{Instance of PRA in Uni-ARBAC}
\label{pra-uni-instance}
In this section we take an example instance for PRA in UARBAC (PRA-U) model.\\
\underline{Traditional RBAC Sets \& Relations:}
\begin{itemize}
\item \U\ = \{\textbf{u\textsubscript1, u\textsubscript2, u\textsubscript3, u\textsubscript4}\}
\item \R\ = \{\textbf{r\textsubscript1, r\textsubscript2, r\textsubscript3, r\textsubscript4}\}
\item \P\ = \{\textbf{p\textsubscript1, p\textsubscript2, p\textsubscript3, p\textsubscript4}\}
\item \RH\ = \{(\textbf{r\textsubscript1, r\textsubscript2}), (\textbf{r\textsubscript2, r\textsubscript3})\}
\end{itemize}

\underline{Addeditional RBAC Sets \& Relations:}
\begin{itemize}
\item \T\ = \{\textbf{t\textsubscript1, t\textsubscript2, t\textsubscript{3}, t\textsubscript4}\}
\item \tH\ = \{(\textbf{t\textsubscript1, t\textsubscript2}), (\textbf{t\textsubscript2, t\textsubscript3}), (\textbf{t\textsubscript2, t\textsubscript4})\}
\item \PA\ = \{(\textbf{p\textsubscript1, t\textsubscript1}), (\textbf{p\textsubscript1, t\textsubscript4}), (\textbf{p\textsubscript2, t\textsubscript4}), (\textbf{p\textsubscript4, t\textsubscript3}), (\textbf{p\textsubscript3, t\textsubscript2})\}
\item \TA\ = \{(\textbf{t\textsubscript1, r\textsubscript2}), (\textbf{t\textsubscript2, r\textsubscript1}), (\textbf{t\textsubscript3, r\textsubscript4}), (\textbf{t\textsubscript4, r\textsubscript3})\}
\end{itemize}

\underline{Derived functions}
\begin{itemize}
\item authorized\_perms(\textbf{r\textsubscript1}) = \{\textbf{p\textsubscript3}\}
\item authorized\_perms(\textbf{r\textsubscript2}) = \{\textbf{p\textsubscript1}\}
\item authorized\_perms(\textbf{r\textsubscript3}) = \{\textbf{p\textsubscript1, p\textsubscript2}\}
\item authorized\_perms(\textbf{r\textsubscript4}) = \{\textbf{p\textsubscript4}\}
\end{itemize}
\underline{Administrative Units and Partitioned Assignments}
\begin{itemize}
\item \AU\ = \{\textbf{au\textsubscript1, au\textsubscript2}\}
\item \it{roles}(\textbf{au\textsubscript1}) = \{\textbf{r\textsubscript1, r\textsubscript2}\}, \it{roles}(\textbf{au\textsubscript2}) = \{\textbf{r\textsubscript3}\}
\item \it{tasks}(\textbf{au\textsubscript1}) = \{\textbf{t\textsubscript1, t\textsubscript2}\}, \it{tasks}(\textbf{au\textsubscript2}) = \{\textbf{t\textsubscript3, t\textsubscript4}\}\\
\underline{Derived Function}
\item \it{tasks}*(\textbf{au\textsubscript1}) = \{\textbf{t\textsubscript1, t\textsubscript2, t\textsubscript3, t\textsubscript4}\} 
\item \it{tasks}*(\textbf{au\textsubscript2}) = \{\textbf{t\textsubscript3, t\textsubscript4}\} \\
\underline{Administrative User Assignments}
\item \it{TA\_admin} = \{(\textbf{u\textsubscript1, au\textsubscript1}), (\textbf{u\textsubscript2, au\textsubscript2})\} 
\item \AUH\ = \{(\textbf{au\textsubscript1, au\textsubscript2})\}
\end{itemize}
\underline{Task-role assignment condition in uni-ARBAC:}\\
-- \it{can\_manage\_task\_role}(\it{u} : \U, \it{t}: \T, \it{r}: \R) =\\
\stab ($\exists$\it{au\textsubscript{i}, au\textsubscript{j}})[(\it{u}, \it{au\textsubscript{i}}) $\in$ \it{TA\_admin} $\wedge$ 
\it{au\textsubscript{i}} $\succeq$\textsubscript{\it{au}} \it{au\textsubscript{j}} $\wedge$ \\
\stab \it{r} $\in$ \it{roles}(\it{au\textsubscript{j}}) $\wedge$ \it{t} $\in$ \it{tasks}*(\it{au\textsubscript{j}})] \\

%
\subsubsection{Equivalent ARPA instance of PRA in Uni-ARBAC}
This section presents an equivalent instance for the example instance presented in section~\ref{pra-uni-instance}.\\
\underline{Set and functions:}
\begin{itemize}
\item \AU\ = \{\textbf{u\textsubscript1, u\textsubscript2, u\textsubscript3, u\textsubscript4}\}
\item \OP\ = \{\textbf{assign, revoke}\}
\item \R\ = \{\textbf{r\textsubscript1, r\textsubscript2, r\textsubscript3, r\textsubscript4}\}
\item \RH\ = \{(\textbf{r\textsubscript1, r\textsubscript2}), (\textbf{r\textsubscript2, r\textsubscript3}), (\textbf{r\textsubscript3, r\textsubscript4})\}                
\item \P\ = \{\textbf{p\textsubscript1, p\textsubscript2, p\textsubscript3, p\textsubscript4}\}

\item \AATT\ = \{\textit{\au, \aur}\}
\item \scope(\textit{\au}) =  \{\textbf{au\textsubscript1, au\textsubscript2}\},
\item[] attType(\it{\au}) = set, 
\item[] is\_ordered(\it{\au}) = True, 
\item[] H\textsubscript{\it{\au}} =  \{(\textbf{au\textsubscript1, au\textsubscript2})\}

\item \it{\au}(\textbf{u\textsubscript1}) = \{\textbf{au\textsubscript1}\}, \it{\au}(\textbf{u\textsubscript2}) = \{\textbf{au\textsubscript2}\}, 
\item[] \it{\au}(\textbf{u\textsubscript3}) = \{\}, \it{\au}(\textbf{u\textsubscript4}) = \{\textbf{}\}

\item \scope(\textit{\aur}) =  \{(\textbf{au\textsubscript1, r\textsubscript1}), (\textbf{au\textsubscript1, r\textsubscript2}), 
\item[] (\textbf{au\textsubscript2, r\textsubscript3})\}, 
\item[] attType(\textit{\aur}) = set,
\item[] is\_ordered(\it{\aur}) = False, 
\item[] H\textsubscript{\it{\aur}} = $\phi$ 

\item \it{\aur}(\textbf{u\textsubscript1}) = \{(\textbf{au\textsubscript1, r\textsubscript1}), (\textbf{au\textsubscript1, r\textsubscript2}), 
\item[] (\textbf{au\textsubscript2, r\textsubscript3})\}, \it{\aur}(\textbf{u\textsubscript2}) = \{(\textbf{au\textsubscript2, r\textsubscript3})\},
 \item[] \textit{\aur}(\textbf{u\textsubscript3}) = \{\}, \it{\aur}(\textbf{u\textsubscript4}) = \{\}

\item \PATT\ = \{\it{tasks, task\_adminu}\}
\item \scope(\it{tasks}) = \{\textbf{t\textsubscript1, t\textsubscript2, t\textsubscript3, t\textsubscript4}\}, 
\item[] attType(\it{tasks}) = set, is\_ordered(\it{tasks}) = True,
\item[] H\textsubscript{\it{tasks}} = \tH\ 
\item \it{tasks}(\textbf{p\textsubscript1}) = \{\textbf{t\textsubscript1, t\textsubscript2, t\textsubscript4}\}, \it{tasks}(\textbf{p\textsubscript2}) = \{\textbf{t\textsubscript1, t\textsubscript2, t\textsubscript4}\}, \it{tasks}(\textbf{p\textsubscript3}) = \{\textbf{t\textsubscript1, t\textsubscript2}\}, \it{tasks}(\textbf{p\textsubscript4}) = \{\textbf{t\textsubscript1, t\textsubscript2, t\textsubscript3}\}

\item \scope(\it{task\_adminu}) = \{(\textbf{t\textsubscript1, au\textsubscript1}), (\textbf{t\textsubscript2, au\textsubscript2})\}, 
\item[] attType(\it{task\_adminu}) = set, 
\item[] is\_ordered(\it{task\_adminu}) = False, H\textsubscript{\it{task\_adminu}} = $\phi$

\item \it{task\_adminu}(\textbf{p\textsubscript1}) = \{(\textbf{t\textsubscript1, au\textsubscript1}), (\textbf{t\textsubscript4, au\textsubscript2})\}, 
\item[] \it{\it{task\_adminu}}(\textbf{p\textsubscript2}) = \{(\textbf{t\textsubscript4, au\textsubscript2})\}, 
\item[] \it{task\_adminu}(\textbf{p\textsubscript3}) = \{(\textbf{t\textsubscript2, au\textsubscript1})\}, 
\item[] \it{task\_adminu}(\textbf{p\textsubscript4}) = \{(\textbf{t\textsubscript3, au\textsubscript2})\} 
\end{itemize}
%
Set of permissions that are mapped to each task in \T\ can be expressed as follows:\\
Let each set be represented with $\chi$\textsubscript{\textsubscript{\it{i}}} as shown below.
\begin{enumerate}[label=(\alph*)]
\item $\chi$\textsubscript{\textsubscript{1}} = \{\it{p} $\vert$ \textbf{t\textsubscript1} $\in$ \it{tasks}(\it{p})\}
\item $\chi$\textsubscript{\textsubscript{2}} = \{\it{p} $\vert$ \textbf{t\textsubscript2} $\in$ \it{tasks}(\it{p})\}
\item $\chi$\textsubscript{\textsubscript{3}} = \{\it{p} $\vert$ \textbf{t\textsubscript3} $\in$ \it{tasks}(\it{p})\}
\item $\chi$\textsubscript{\textsubscript{4}} = \{\it{p} $\vert$ \textbf{t\textsubscript4} $\in$ \it{tasks}(\it{p})\}
\end{enumerate}
For each $\chi$\textsubscript{\textsubscript{i}} in \{$\chi$\textsubscript{\textsubscript{1}}, $\chi$\textsubscript{\textsubscript{2}}, $\chi$\textsubscript{\textsubscript{3}}, $\chi$\textsubscript{\textsubscript{4}}\}, authorization rule for whether an admin user in \AU\ is authorized to assign $\chi$\textsubscript{\textsubscript{i}} to a roles \it{r} in \R\ is given below:\\ 
-- {\isauth}P\textsubscript{\textbf{assign}}(\it{u} : \U, $\chi$\textsubscript{\textsubscript{i}} : 2\textsuperscript{\P}, \\
\it{r} : \R) $\equiv$ \\
\hspace*{0.17cm}$\exists$\it{au\textsubscript1,  au\textsubscript2} $\in$ \scope(\it{\au}). (\it{au\textsubscript1, au\textsubscript2}) $\in$ H\textsubscript{\it{\au}} \\
\hspace*{0.17cm} $\wedge$ (\it{au\textsubscript1} $\in$ \it{\au}(\it{u}) $\wedge$ (\it{au\textsubscript2, r}) \\
\hspace*{0.17cm}$\in$  \it{\aur}(\it{u})) $\wedge$ $\exists$\it{t\textsubscript1, t\textsubscript2} $\in$ \scope(\it{tasks}). [(\it{t\textsubscript1, t\textsubscript2}) \\
\hspace*{0.17cm}$\in$ \tH\ $\wedge$ $\forall$\it{q} $\in$ $\chi$. \it{t\textsubscript2} $\in$ \it{tasks}(\it{q}) $\wedge$ $\exists$\it{q}\q\ $\in$ (\P\ - $\chi$). \\
\hspace*{0.17cm}\it{t\textsubscript2} $\notin$ \it{tasks}(\it{q}\q) $\wedge$ (\it{t\textsubscript2, au\textsubscript2}) $\in$ \it{tasks\_adminu}(\it{q})]\\
\\
For each $\chi$\textsubscript{\textsubscript{i}} in \{$\chi$\textsubscript{\textsubscript{1}}, $\chi$\textsubscript{\textsubscript{2}}, $\chi$\textsubscript{\textsubscript{3}}, $\chi$\textsubscript{\textsubscript{4}}\}, authorization function for whether an admin user in \AU\ is authorized to revoke $\chi$ from a roles \it{r} $\in$ \R\ is given below:\\ 
-- {\isauth}P\textsubscript{\textbf{revoke}}(\it{u} : \U, $\chi$\textsubscript{\textsubscript{i}} : 2\textsuperscript{\P}, \\
\hspace*{0.17cm}\it{r} : \R) $\equiv$ {\isauth}P\textsubscript{\textbf{assign}}(\it{u} : \U, \\
\hspace*{0.17cm}$\chi$\textsubscript{\textsubscript{i}} : 2\textsuperscript{\P}, \it{r} : \R)
\\

\subsubsection{Map\textsubscript{PRA-Uni-ARBAC}}

Map\textsubscript{PRA-Uni} depicted as Algorithm~\ref{alg:alg2} translates instance of
PRA in Uni-ARBAC to an equivalent instance of ARPA. Sets and functions from PRA-Uni
and ARPA are marked with superscript \t{Uni} and superscript \t{A}, repectively. 

It takes following sets and functions from PRA-Uni as an input that includes \U\textsuperscript{\t{Uni}}, \R\textsuperscript{\t{Uni}},
\P\textsuperscript{\t{Uni}}, \RH\textsuperscript{\t{Uni}}, \T\textsuperscript{\t{Uni}}, \tH\textsuperscript{\t{Uni}},
\PA\textsuperscript{\t{Uni}}, \AU\textsuperscript{\t{Uni}}, For each \it{au} in \AU\textsuperscript{\t{Uni}},
\it{roles}\textsuperscript{\t{Uni}}(\it{au}) and \it{tasks}*\textsuperscript{\t{Uni}}(\it{au}),
\it{TA\_admin}\textsuperscript{\t{Uni}}, \AUH\textsuperscript{\t{Uni}},
\it{can\_manage\_task\_role}(\it{u} : \U\textsuperscript{\t{Uni}}, \it{t} :
\T\textsuperscript{\t{Uni}}, \it{r} : \R\textsuperscript{\t{Uni}}.
Output is an equivalent ARPA instance which includes \AU\textsuperscript{\t{A}},
\OP\textsuperscript{\t{A}}, \R\textsuperscript{\t{A}}, \RH\textsuperscript{\t{A}}, \P\textsuperscript{\t{A}},
\AATT\textsuperscript{\t{A}}, \PATT\textsuperscript{\t{A}}, For each attribute \it{att} $\in$
\AATT\textsuperscript{\t{A}} $\cup$ \PATT\textsuperscript{\t{A}}, \scope\textsuperscript{\t{A}}(\it{att}),
attType\textsuperscript{\t{A}}(\it{att}), \isord\textsuperscript{\t{A}}(\it{att}) and
H\textsuperscript{\t{A}}\textsubscript{\it{att}}, For each user \it{u} $\in$ \AU\textsuperscript{\t{A}}, and for
each \it{att} $\in$ \AATT\textsuperscript{\t{A}}, \it{att}(\it{u}), For each user \it{p}
$\in$ \P\textsuperscript{\t{A}}, and for each \it{att} $\in$ \PATT\textsuperscript{\t{A}},
\it{att}(\it{t}), Authorization rule for assigning permission
(auth\_assignp), and finally, Authorization rule for revoking permission
(auth\_revokep).

Map\textsubscript{PRA-Uni} consists of four steps to complete translation. 
In Step 1 sets and functions from PRA-Uni instance are mapped to ARPA instance.  
In Uni-ARBAC, both admin users and regular users belong to same set, \U\textsuperscript{Uni}. 
Thus, \U\textsuperscript{Uni} is mapped to \AU\textsuperscript{A}. In Step 2, admin user attributes and pemrission attributes are defined. There are two admin user attributes: \it{\au} and \it{\aur}. \it{\au} captures the \it{TA\_admin}\textsuperscript{\t{Uni}} relation in URA-Uni, and \it{\aur} captures admin user's mapping with admin unit, and the roles mapped to that admin unit. There are two permission attributes: \it{tasks} and \it{task\_adminu}. Attribute \it{tasks} gives a mapping between permission and tasks. That is for each permission \it{p}, \it{tasks}(\it{p}) yields set of tasks it is mapped to. For a given pemission, attribute \it{task\_adminu} gives its mapping with tasks, and admin units that each task is mapped to. Step 3 constructs an assignment rule equivalent to \it{can\_manage\_task\_role}(\it{u} : \U\textsuperscript{\t{Uni}}, \it{t}: \T\textsuperscript{\t{Uni}}, \it{r}: \R\textsuperscript{\t{Uni}}) in Uni-ARBAC. In PRA-Uni, it evaluates if an admin user \it{u} can assign/revoke a task \it{t} to/from a role \it{r} if admin user \it{u} has \it{Task\_admin}\textsuperscript{\t{Uni}} relation with some admin unit \it{au} to which task \it{t} and role \it{r} are mapped. An ARPA equivalent assignment rule auth\_assign is expressed in Step 3 as {\isauth}P\textsubscript{\textbf{assign}}(\it{u} : \U\textsuperscript{\t{A}}, $\chi$ : 2\textsuperscript{\P\textsuperscript{\t{A}}}, \textbf{r} : \R\textsuperscript{\t{A}}) and revoke rule auth\_revoke in Step 4 as {\isauth}P\textsubscript{\textbf{revoke}}(\it{u} : \U\textsuperscript{\t{A}}, $\chi$ : 2\textsuperscript{\P\textsuperscript{\t{A}}}, \textbf{r} : \R\textsuperscript{\t{A}}. Authorization criteria for \textbf{assign} and \textbf{revoke} is identical.

\floatname{algorithm}{Algorithm}
\begin{algorithm}[tp]
\caption{Map\textsubscript{PRA-UARBAC}}
\label{alg:PRA-U}
\begin{algorithmic} [] 
\begin{spacing}{1.08}
\item[]\textbf{Input:} Instance of PRA in UARBAC
\item[] \textbf{Output:} ARPA instance 
\item[\textbf{Step 1:}] \ \ /* Map basic sets and functions in ARPA */
\item[] a. \AU\textsuperscript{\t{A}} $\leftarrow$ \it{\U}\textsuperscript{\t{U}} ; \OP\textsuperscript{\t{A}} $\leftarrow$ \{\textbf{assign, revoke}\}
\item[] b. \R\textsuperscript{\t{A}} $\leftarrow$ \it{\R}\textsuperscript{\t{U}} ; \RH\textsuperscript{\t{A}} $\leftarrow$ \it{\RH}\textsuperscript{\t{U}}
\item[] c. \P\textsuperscript{\t{A}} $\leftarrow$ \it{\P}\textsuperscript{\t{U}}

\item[\textbf{Step 2:}]   \stab /* Map attribute functions in ARPA */
\item[] a. \AATT\textsuperscript{\t{A}} $\leftarrow$ \{\it{\oam, \ram, classp}\} 
\item[] b. \scope(\it{\oam}) = \it{\OI}\textsuperscript{\t{U}} $\times$ \it{AM}\textsuperscript{\t{U}}(\textsf{file}) 
\item[] c. attType(\it{\oam}) = set 
\item[] d. \isord(\it{\oam}) = False, H\textsubscript{\it{\oam}} = $\phi$
\item[] e. For each \it{u\textsubscript{}} in \AU\textsuperscript{\t{U}}, \it{\oam}(\it{u\textsubscript{}}) = $\phi$ 
\item[] f. For each \it{u} in \it{U}\textsuperscript{\t{U}} and for 
\item[] \stab each [\it{c}, \it{o\textsubscript1}, \it{a}] $\in$ authorized\_perms\textsuperscript{\t{U}}[\it{u}], 
\item[] \tab \stab \it{\oam}(\it{u\textsubscript{}})\q\ = \it{\oam}(\it{u\textsubscript{}}) $\cup$ (\it{o, a})

\item[] g. \scope(\it{\ram}) = \it{\R}\textsuperscript{\t{U}} $\times$ \it{AM}\textsuperscript{\t{U}}(\textsf{role}) ; attType(\it{\ram}) = set 
\item[] h. \isord(\it{\ram}) = False, H\textsubscript{\it{\ram}} = $\phi$
\item[] i. For each \it{u\textsubscript{}} in \AU\textsuperscript{\t{A}}, \it{\ram}(\it{u}) = $\phi$ 
\item[] j. For each \it{u} in \it{U}\textsuperscript{\t{U}} for each 
\item[] \stab {[\it{c, r\textsubscript1, a}]} $\in$ authorized\_perms\textsuperscript{\t{U}}[\it{u}], 

\item[] \tab \stab \it{\ram}(\it{u\textsubscript{}})\textquotesingle\ = \it{\ram}(\it{u\textsubscript{}}) $\cup$ (\it{r, a})

\item[] k. \scope(\it{classp}) = \it{C}\textsuperscript{\t{U}} $\times$ \{\it{AM}\textsuperscript{\t{U}}(\textsf{file}) $\cup$ \it{AM}\textsuperscript{\t{U}}(\textsf{role})\} 
\item[] l. attType(\it{classp}) = set 
\item[] m. \isord(\it{classp}) = False, H\textsubscript{\it{classp}} = $\phi$ 
\item[] n. For each \it{u\textsubscript{}} in \AU\textsuperscript{\t{A}}, \it{\oam}(\it{u}) = $\phi$ 
\item[] o. For each \it{u} in \it{U}\textsuperscript{\t{U}} for 
\item[] \stab each [\it{c, a}] $\in$ authorized\_perms\textsuperscript{\t{U}}[\it{u}], 
\item[] \tab \stab \it{classp}(\it{u\textsubscript{}})\textquotesingle\ = \it{\oam}(\it{u\textsubscript{}}) $\cup$ (\it{c, a})
\item[] p. \PATT\textsuperscript{\t{A}} = \{\it{object\_id}\}
\item[] q. \scope(\it{object\_id}) = \it{\R}\textsuperscript{\t{U}} $\cup$ \it{\P}\textsuperscript{\t{U}} \item[] r. attType(\it{object\_id}) = atomic
\item[] s. \isord(\it{object\_id}) = False, H\textsubscript{\it{object\_id}} = $\phi$

\item[] t. For each \it{u} in \it{U}\textsuperscript{\t{U}} and for
\item[] \stab each \it{p} $\in$ authorized\_perms\textsuperscript{\t{U}}[\it{u}], where \it{p} is of the 
\item[] \tab form [\it{c}, \it{o\textsubscript{i}}, \it{a}] or [\it{c}, \it{a}], 
\item[] \tab \stab \it{object\_id}(\it{p}) = \it{o\textsubscript{i}} or $\phi$

\item[\textbf{Step 3:}] \stab /* Construct assign rule in ARPA */
\item[] a. assign\_formula = \\
\item[]\ \ ((\it{object\_id}(\it{p}), \textsf{admin}) $\in$ \it{\oam}(\it{au}) $\wedge$  
\item[] \ \ (\it{r}, \textsf{empower}) $\in$ \it{\ram}(\it{au})) $\vee$ ((\textsf{file, admin}) 
\item[] \ \ $\in$ \it{classp}(\it{au}) $\wedge$ (\it{r}, \textsf{empower}) $\in$ \it{\ram}(\it{au})) 
\item[] \ \ $\vee$ ((\it{object\_id}(\it{p}), \textsf{admin}) $\in$ \it{\oam}(\it{au}) $\wedge$ 
\item[] \ \ (\textsf{role, empower}) $\in$ \it{classp}(\it{au})) $\vee$ (\textsf{file, admin}) 
\item[] \ \ $\in$ \it{classp}(\it{au}) $\wedge$ (\textsf{role, empower}) $\in$ \it{classp}(\it{au})) 
\vspace{-0.4cm}
\end{spacing}
\end{algorithmic}
\end{algorithm}
\floatname{algorithm}{Continuation from Algorithm}
\setcounter{algorithm}{9}
\begin{algorithm}
\caption{Map\textsubscript{PRA-UARBAC}}

\begin{algorithmic} [1] 
\begin{spacing}{1.1}

\item[] b. auth\_assign = {\isauth}P\textsubscript{\textbf{assign}}(\it{au} : \AU\textsuperscript{\t{A}}, 
\item[] \stab \it{p} : \P\textsuperscript{\t{A}}, \it{r} : \R\textsuperscript{\t{A}}) $\equiv$  assign\_formula

\item[\textbf{Step 4:}] \stab /* Construct revoke rule for ARPA */
\item[] a. revoke\_formula = 
\item[] \ \ (\it{object\_id}(\it{p}), \textsf{admin}) $\in$ \it{\oam}(\it{au}) $\vee$  
\item[] \ \ (\it{r}, \textsf{admin})  $\in$ \it{\ram}(\it{au})) $\vee$ (\textsf{file}, \textsf{admin}) 
\item[] \ \ $\in$ \it{classp}(\it{au}) $\vee$  (\textsf{role}, \textsf{admin}) $\in$ \it{classp}(\it{au}))
\item[] b. auth\_revoke = {\isauth}P\textsubscript{\textbf{revoke}}(\it{au} : \AU\textsuperscript{\t{A}}, 
\item[] \stab \it{p} : \P\textsuperscript{\t{A}}, \it{r} : \R\textsuperscript{\t{A}}) $\equiv$ revoke\_formula
\vspace{-0.4cm}
\end{spacing}
\end{algorithmic}
\end{algorithm}

\subsection{UARBAC's PRA in ARPA}
\subsubsection{RBAC Model}
UARBAC model is designed with a notion of class objects. Thus, includes class level administrative permissions as well. Following is RBAC schema is presented as follows:\\
\underline{RBAC Schema:}\\
RBAC Schemas is given by following tuple.\\
\tab <{\it{C, OBJS, AM}}>
\begin{itemize}
\item \it{C} is a finite set of object classes with predefined classes: \textsf{user} and \textsf{role}.
\item \it{OBJS}(\it{c}) is a function that gives all possible names for objects of the class \it{c} $\in$ \it{C}.
\item[] Let \textbf{\U}\ =  \it{OBJS}(\textsf{user}) and \textbf{\R}\ = \it{OBJS}(\textsf{role})  
\item \it{AM}(\it{c}) is function that maps class \it{c} to a set of access modes that can be applied on objects of class \it{c}. 
\end{itemize}
Access modes for two predefined classes \textsf{user} and \textsf{role} are fixed. By observation we find it relevant to consider files as resource objects. We take file as example resource object to which we will define access. 
\vspace{0.17cm}
\noindent
\begin{itemize}
\item[] \it{AM}(\textsf{user}) = \{\textsf{empower, admin}\}
\item[] \it{AM}(\textsf{role}) = \{\textsf{grant, empower, admin}\}
\item[] \it{AM}(\textsf{file}) = \{\textsf{read, write, append, execute, 
\item[] admin}\}
\end{itemize}
\underline{RBAC Permissions:}\\
There are two kinds of permissions in this RBAC model:
\begin{itemize}
\item Object permissions of the form, 
\item[] \stab {[\it{c, o, a}]}, where \it{c $\in$ C}, \it{o} $\in$ \it{OBJS}(\it{c}), \it{a} $\in$ \it{AM}(\it{c}).
\item Class permissions of the form,
\item[] \ \ \ {[\it{c, a}]}, where, \it{c $\in$ C}, and \it{a} $\in$ \{\textsf{create}\} $\cup$ \it{AM}(\it{c}).
\end{itemize}

\underline{RBAC State:}\\
Given an RBAC Schema, an RBAC state is given by,\\ 
\stab <\it{OB, UA, PA, RH}>
\begin{itemize}
\item \it{OB} is a function that maps each class in \it{C} to a finite set of object names of that class that currently exists, i.e., \it{OB}(\it{c}) $\subseteq$ \it{OBJS}(\it{c}).
\item[] \stab Let \it{OB}(\textsf{user}) = \it{\U}, \it{OB}(\textsf{role}) = \it{\R} and, let \it{OB}(\textsf{file}) = \it{FILES}
\item[] Set of permissions, \it{P}, is given by
\item[] \it{P} = \{[\it{c, o, a}] $\vert$ \it{c $\in$ C} $\wedge$ \it{o} $\in$ \it{OBJS}(\it{c}) $\wedge$ \it{a} $\in$ \it{AM}(\it{c})\} $\cup$ \{[\it{c, a}] $\vert$ \it{c $\in$ C} $\wedge$ \it{a} $\in$ \{\textsf{create}\} $\cup$ \it{AM}(\it{c})\}
\item \it{UA} $\subseteq$ \it{\U} $\times$ \it{\R}, user-role assignment relation.
\item \it{PA} $\subseteq$ \it{P} $\times$ \it{\R}, permission-role assignment relation.
\item \it{\RH} $\subseteq$ \it{\R} $\times$ \it{\R}, partial order in \it{\R}\ denoted by \it{$\succeq$\textsubscript{RH}}.
\end{itemize}
\underline{Administrative permissions in UARBAC:}\\
All the permissions of user \it{u} who performs administrative operations can be calculated as follows:
\begin{itemize}
\item authorized\_perms[\it{u}] = \{\it{p $\in$ P} $\vert$ $\exists$\it{r\textsubscript1, r\textsubscript2} $\in$ R [(\it{u, r\textsubscript1}) $\in$ \it{UA} $\wedge$ (\it{r\textsubscript1 $\succeq$\textsubscript{RH} r\textsubscript2}) $\wedge$ (\it{r\textsubscript2, p}) $\in$ \it{PA}]\}
\end{itemize}
\underline{Permission-Role Administration}\\
Operations required to assign object permission [\it{c, o\textsubscript1, a\textsubscript1}] to role \it{r\textsubscript1} and to revoke object permission [\it{c, o\textsubscript1, a\textsubscript1}] from role \it{r\textsubscript1} are respectively listed below:
\begin{itemize}
\item grantObjPermToRole([\it{c, o\textsubscript1, a\textsubscript1}], \it{r\textsubscript1})
\item revokeObjPermFromRole([\it{c, o\textsubscript1, a\textsubscript1}], \it{r\textsubscript1})
\end{itemize}
An admin user requires one of the following two permissions to conduct grantObjPermToRole([\it{c, o\textsubscript1, a\textsubscript1}], \it{r\textsubscript1}) operation.
\begin{enumerate}
\item {[\it{c, o\textsubscript1}, \textsf{admin}]} and {[\textsf{role}, \it{r\textsubscript1}, \textsf{empower}]} or,
\item {[\it{c, o\textsubscript1}, \textsf{admin}]} and {[\textsf{role}, \textsf{empower}]} or, 
\item {[\it{c}, \textsf{admin}]} and {[\textsf{role}, \it{r\textsubscript1}, \textsf{empower}]} or, 
\item {[\it{c}, \textsf{admin}]} and {[\textsf{role}, \textsf{empower}]}
\end{enumerate}
An admin user requires one of the following permission(s) (single or a pair) to conduct revokeObjPermFromRole([\it{c, o\textsubscript1, a\textsubscript1}], \it{r\textsubscript1}) operation.
\begin{enumerate}
\item {[\it{c, o\textsubscript1}, \textsf{admin}]} and {[\textsf{role}, \it{r\textsubscript1}, \textsf{empower}]} or,
\item {[\it{c, o\textsubscript1}, \textsf{admin}]} or,
\item {[\textsf{role}, \it{r\textsubscript1}, \textsf{admin}]} or,
\item {[\it{c}, \textsf{admin}]} or,
\item {[\textsf{role}, \textsf{admin}]}\\
\end{enumerate}

\subsubsection{Instance of PRA in UARBAC}
 \hfill \break
\underline{RBAC Schema}\\
Let us consider objects, to which users need access via roles, to be of class \textsf{file}.
\begin{itemize} 
\item \it{C} = \{\textsf{user, role, file}\}
\item \it{OBJS}(\textsf{user}) = USERS,
\item \it{OBJS}(\textsf{role}) = ROLES
\item \it{OBJS}(\textsf{file}) = FILES
\end{itemize} 
\noindent
Access modes for \textsf{role} and \textsf{file} class are as follows:
\begin{itemize} 
\item \it{AM}(\textsf{role}) = \{\textsf{grant, empower, admin}\}
\item \it{AM}(\textsf{file}) = \{\textsf{read, write, append, execute, 
\item[] admin}\}
\end{itemize}
\underline{RBAC State}
\begin{itemize} 
 \item \it{\U} = \it{OBJ}(\textsf{user})= \{\textbf{u\textsubscript{1}, u\textsubscript{2}, u\textsubscript3, u\textsubscript4}\}
\item \it{\R} = \it{OBJ}(\textsf{role})= \{\textbf{r\textsubscript{1}, r\textsubscript{2}, r\textsubscript3, r\textsubscript4}\} 
 \item \it{O} = \it{OBJ}(\textsf{file})= \{\textbf{o\textsubscript{1}, o\textsubscript{2}, o\textsubscript3}\}

\item \it{P} = \{[\textsf{role}, \textbf{r\textsubscript1}, \textsf{grant}], [\textsf{role}, \textbf{r\textsubscript1}, \textsf{empower}],  [\textsf{role}, \textbf{r\textsubscript1}, \textsf{admin}], [\textsf{role}, \textbf{r\textsubscript2}, \textsf{grant}], [\textsf{role}, \textbf{r\textsubscript2}, \textsf{empower}], [\textsf{role}, \textbf{r\textsubscript2}, \textsf{admin}], [\textsf{role}, \textbf{r\textsubscript3}, \textsf{grant}], [\textsf{role}, \textbf{r\textsubscript3}, \textsf{empower}], [\textsf{role}, \textbf{r\textsubscript3}, \textsf{admin}], {[\textsf{role}, \textbf{r\textsubscript4}, \textsf{grant}], [\textsf{role}, \textbf{r\textsubscript4}, \textsf{empower}],  [\textsf{role}, \textbf{r\textsubscript4}, \textsf{admin}], 
[\textsf{file}, \textbf{o\textsubscript1}, \textsf{read}], [\textsf{file}, \textbf{o\textsubscript1}, \textsf{write}], [\textsf{file}, \textbf{o\textsubscript1}, \textsf{append}], [\textsf{file}, \textbf{o\textsubscript1}, \textsf{execute}], [\textsf{file}, \textbf{o\textsubscript1}, \textsf{admin}]

[\textsf{file}, \textbf{o\textsubscript2}, \textsf{read}], [\textsf{file}, \textbf{o\textsubscript2}, \textsf{write}], [\textsf{file}, \textbf{o\textsubscript2}, \textsf{append}], [\textsf{file}, \textbf{o\textsubscript2}, \textsf{execute}], [\textsf{file}, \textbf{o\textsubscript2}, \textsf{admin}]

[\textsf{file}, \textbf{o\textsubscript3}, \textsf{read}], [\textsf{file}, \textbf{o\textsubscript3}, \textsf{write}], [\textsf{file}, \textbf{o\textsubscript3}, \textsf{append}], [\textsf{file}, \textbf{o\textsubscript3}, \textsf{execute}], [\textsf{file}, \textbf{o\textsubscript3}, \textsf{admin}]

 [\textsf{file, read}], [\textsf{file, write}], [\textsf{file, append}], [\textsf{file, execute}], [\textsf{file, admin}]\}}
 \stab \ \
\item \it{\PA} = \{([\textsf{file}, \textbf{o\textsubscript1}, \textsf{read}], \textbf{r\textsubscript1}), ([\textsf{file}, \textbf{o\textsubscript2}, \textsf{execute}], \textbf{r\textsubscript1}), ([\textsf{file}, \textbf{o\textsubscript2}, \textsf{execute}], \textbf{r\textsubscript2}), ([\textsf{file}, \textbf{o\textsubscript3}, \textsf{admin}], \textbf{r\textsubscript3}), ([\textsf{file}, \textbf{o\textsubscript1}, \textsf{read}], \textbf{r\textsubscript3}), ([\textsf{file}, \textbf{o\textsubscript3}, \textsf{write}], \textbf{r\textsubscript2})\}
\item \it{\RH} = \{(\textbf{r\textsubscript1, r\textsubscript2}), (\textbf{r\textsubscript2, r\textsubscript3}), (\textbf{r\textsubscript3, r\textsubscript4})\}
\end{itemize}
\underline{Authorized permissions in UARBAC}\\
Following is the list of authorized permissions admin each user has, which includes administrative permissions for permission-role assignment:
\begin{itemize}
\item authorized\_perms[\textbf{u\textsubscript1}]  = \{[\textsf{file}, \textbf{o\textsubscript1}, \textsf{read}], 
[\textsf{role}, \textbf{r\textsubscript1}, \textsf{grant}], [\textsf{file}, \textbf{o\textsubscript1}, \textsf{write}], {[\textsf{role}, \textbf{r\textsubscript3}, \textsf{grant}], [\textsf{file}, \textbf{o\textsubscript2}, \textsf{admin}], [\textsf{file}, \textbf{o\textsubscript3}, \textsf{append}]},{[\textsf{role}, \textbf{r\textsubscript2}, \textsf{grant}], [\textsf{file}, \textbf{o\textsubscript3}, \textsf{admin}], [\textsf{role}, \textbf{r\textsubscript1}, \textsf{admin}]}, {[\textsf{role}, \textbf{r\textsubscript4}, \textsf{admin}]}\}

\item authorized\_perms[\textbf{u\textsubscript2}]  = \{[\textsf{file}, \textbf{o\textsubscript1}, \textsf{append}], [\textsf{role}, \textbf{r\textsubscript1}, \textsf{grant}], [\textsf{file}, \textbf{o\textsubscript2}, \textsf{admin}], {[\textsf{role}, \textbf{r\textsubscript2}, \textsf{grant}]}\}
\item authorized\_perms[\textbf{u\textsubscript3}] = \{[\textsf{file}, \textsf{admin}]\}
\item authorized\_perms[\textbf{u\textsubscript4}] = \{\}
\end{itemize}
\underline{Permission-Role assignment condition in UARBAC's PRA:}\\
One can perform following operation to assign a permission [\textsf{file}, \it{o\textsubscript1}, \it{a}] to a role \it{r\textsubscript1}.
\begin{itemize}
\item grantObjPermToRole([\textsf{file}, \it{o\textsubscript1}, \it{a}], \it{r\textsubscript1}) 
\end{itemize}
To perform aforementioned operation one needs the following two permissions:
\begin{itemize}
\item{ [\textsf{file}, \it{o\textsubscript1}, \textsf{admin}]  and  [\textsf{role}, \it{r\textsubscript1}, \textsf{empower}]}
\end{itemize}
\underline{Condition for revoking permission-role in UARBAC's PRA:}\\
One can perform following operation to revoke a user [\textsf{file}, \it{o\textsubscript1}, \it{a}] to a role \it{r\textsubscript1}.
\begin{itemize}
\item revokeObjPermFromRole([\textsf{file}, \it{o\textsubscript1}, \it{a}], \it{r\textsubscript1}) 
\end{itemize}
To perform aforementioned operation one needs one of the following permissions:
\begin{itemize}
\item {[\textsf{file}, \it{o\textsubscript1}, \textsf{admin}]} \textbf{or},
\item {[\textsf{role}, \it{r\textsubscript1}, \textsf{admin}]}\\
\end{itemize}
\subsubsection{Equivalent ARPA instance of PRA in UARBAC}
This section presents an equivalent ARPA instance for the example instance depicted in previous section.\\
\underline{Sets and functions}
\begin{itemize}
\item \AU\ = \{\textbf{u\textsubscript{1}, u\textsubscript{2}, u\textsubscript{3}, u\textsubscript4}\}
\item \OP\ = \{assignp, revokep\}
\item \R\ = \{\textbf{r\textsubscript{1}, r\textsubscript{2}, r\textsubscript3, r\textsubscript4}\} 
\item \RH\ = \{(\textbf{r\textsubscript1, r\textsubscript2}), (\textbf{r\textsubscript2, r\textsubscript3}), (\textbf{r\textsubscript3, r\textsubscript4})\}

\item \P\ = \{[\textsf{role}, \textbf{r\textsubscript1}, \textsf{grant}], [\textsf{role}, \textbf{r\textsubscript1}, \textsf{empower}],  [\textsf{role}, \textbf{r\textsubscript1}, \textsf{admin}], [\textsf{role}, \textbf{r\textsubscript2}, \textsf{grant}], [\textsf{role}, \textbf{r\textsubscript2}, \textsf{empower}], [\textsf{role}, \textbf{r\textsubscript2}, \textsf{admin}], [\textsf{role}, \textbf{r\textsubscript3}, \textsf{grant}], [\textsf{role}, \textbf{r\textsubscript3}, \textsf{empower}], [\textsf{role}, \textbf{r\textsubscript3}, \textsf{admin}], {[\textsf{role}, \textbf{r\textsubscript4}, \textsf{grant}], [\textsf{role}, \textbf{r\textsubscript4}, \textsf{empower}],  [\textsf{role}, \textbf{r\textsubscript4}, \textsf{admin}], 
[\textsf{file}, \textbf{o\textsubscript1}, \textsf{read}], [\textsf{file}, \textbf{o\textsubscript1}, \textsf{write}], [\textsf{file}, \textbf{o\textsubscript1}, \textsf{append}], [\textsf{file}, \textbf{o\textsubscript1}, \textsf{execute}], [\textsf{file}, \textbf{o\textsubscript1}, \textsf{admin}]

[\textsf{file}, \textbf{o\textsubscript2}, \textsf{read}], [\textsf{file}, \textbf{o\textsubscript2}, \textsf{write}], [\textsf{file}, \textbf{o\textsubscript2}, \textsf{append}], [\textsf{file}, \textbf{o\textsubscript2}, \textsf{execute}], [\textsf{file}, \textbf{o\textsubscript2}, \textsf{admin}]

[\textsf{file}, \textbf{o\textsubscript3}, \textsf{read}], [\textsf{file}, \textbf{o\textsubscript3}, \textsf{write}], [\textsf{file}, \textbf{o\textsubscript3}, \textsf{append}], [\textsf{file}, \textbf{o\textsubscript3}, \textsf{execute}], [\textsf{file}, \textbf{o\textsubscript3}, \textsf{admin}]

 [\textsf{file, read}], [\textsf{file, write}], [\textsf{file, append}], [\textsf{file, execute}], [\textsf{file, admin}]\}}

\item \AATT\ = \{\textit{\oam, \ram, classp}\} 
\item \scope(\textit{\oam}) = \{(\textbf{o\textsubscript1}, \textsf{read}), (\textbf{o\textsubscript1}, \textsf{write}), (\textbf{o\textsubscript1}, \textsf{execute}), (\textbf{o\textsubscript1}, \textsf{append}), (\textbf{o\textsubscript1}, \textsf{admin}), (\textbf{o\textsubscript2}, \textsf{read}), 
(\textbf{o\textsubscript2}, \textsf{write}), (\textbf{o\textsubscript2}, \textsf{append}), (\textbf{o\textsubscript2}, \textsf{execute}), (\textbf{o\textsubscript2}, \textsf{admin}), (\textbf{o\textsubscript3}, \textsf{read}), (\textbf{o\textsubscript3}, \textsf{write}), (\textbf{o\textsubscript3}, \textsf{append}), (\textbf{o\textsubscript3}, \textsf{execute}), (\textbf{o\textsubscript3}, \textsf{admin})\}, 
\item[] attType(\textit{\oam}) = set, 
\item[] \isord(\it{\oam}) = False, H\textsubscript{\it{\oam}} = $\phi$ 

\item \scope(\textit{\ram}) = \{(\textbf{r\textsubscript1}, \textsf{grant}),
(\textbf{r\textsubscript1},\textsf{empower}), 
\item[](\textbf{r\textsubscript1}, \textsf{admin}), (\textbf{r\textsubscript2}, \textsf{grant}), (\textbf{r\textsubscript2}, \textsf{empower}), (\textbf{r\textsubscript2}, \textsf{admin}), (\textbf{r\textsubscript3}, \textsf{grant}), (\textbf{r\textsubscript3}, \textsf{empower}), (\textbf{r\textsubscript3}, \textsf{admin}), (\textbf{r\textsubscript4}, \textsf{grant}), (\textbf{r\textsubscript4}, \textsf{empower}), (\textbf{r\textsubscript4}, \textsf{admin})\}, 
\item[] attType(\textit{\ram}) = set, 
\item[] \isord(\it{\ram}) = False, H\textsubscript{\it{\ram}} = $\phi$

\item \scope(\textit{classp}) =  \{(\textsf{file, read}), (\textsf{file}, \textsf{write}), (\textsf{file}, \textsf{append}),
 (\textsf{file}, \textsf{execute}), (\textsf{file}, \textsf{admin}), (\textsf{role, grant}), (\textsf{role}, \textsf{empower}), (\textsf{role}, \textsf{admin})\}, 
\item[] attType(\textit{classp}) = set, \isord(\it{classp}) = False, 
\item[] H\textsubscript{\it{classp}} = $\phi$

\item \it{\oam}(\textbf{u\textsubscript1}) = \{(\textbf{o\textsubscript1}, \textsf{read}), (\textbf{o\textsubscript1}, \textsf{write}), (\textbf{o\textsubscript2}, \textsf{admin}), (\textbf{o\textsubscript3}, \textsf{append}), (\textbf{o\textsubscript3}, \textsf{admin})\}
\item \it{\oam}(\textbf{u\textsubscript2}) = \{(\textbf{o\textsubscript1}, \textsf{append}), (\textbf{o\textsubscript2}, \textsf{admin})\}
\item \it{\oam}(\textbf{u\textsubscript3}) = \{\}, \it{\oam}(\textbf{u\textsubscript4}) = \{\}

\item \it{\ram}(\textbf{u\textsubscript1}) = \{(\textbf{r\textsubscript1}, \textsf{grant}), (\textbf{r\textsubscript2}, \textsf{grant}), (\textbf{r\textsubscript3}, \textsf{grant}), (\textbf{r\textsubscript1}, \textsf{admin}), (\textbf{r\textsubscript4}, \textsf{admin})\}
\item \it{\ram}(\textbf{u\textsubscript2}) = \{(\textbf{r\textsubscript1}, \textsf{grant}), (\textbf{r\textsubscript2}, \textsf{grant})\}
\item \it{\ram}(\textbf{u\textsubscript3}) = \{\}, \it{\ram}(\textbf{u\textsubscript4}) = \{\}

\item \it{classp}(\textbf{u\textsubscript1}) = \{\}
\item \it{classp}(\textbf{u\textsubscript2}) = \{\}
\item \it{classp}(\textbf{u\textsubscript3}) = \{(\textsf{file, admin})\}
\item \it{classp}(\textbf{u\textsubscript4}) = \{\}

\item \PATT\ = \{\it{object\_id}\}
\item \scope(\textit{object\_id}) = \{\textbf{r\textsubscript1, r\textsubscript2, r\textsubscript3, r\textsubscript4, o\textsubscript1, o\textsubscript2, o\textsubscript3}\}, attType(\it{object\_id}) = atomic, 
\item[] \isord(\it{object\_id}) = False, H\textsubscript{\it{\oam}} = $\phi$
\end{itemize}

\noindent
\it{object\_id} for each permission \it{p} in \P\ is given by, 
\it{object\_id}(\it{p}) = \it{o\textsubscript{i}}.

%
For each \it{op} in \OP, authorization function for assignment and revocation of permission of the form \\
\it{p} = [\textsf{file}, \it{o, a}] to role \it{r} can be expressed as follows:\\

\noindent
For any permission \it{p} $\in$ \P\ undertaken for assignment,\\
-- {\isauth}P\textsubscript{\textbf{assign}}(\it{u} : \AU, \it{p} : \P, \\
\hspace*{0.19cm}\it{r} : \R) $\equiv$\\ 
\hspace*{0.19cm}((\it{object\_id}(\it{p}), \textsf{admin}) $\in$ \it{\oam}(\it{u}) $\wedge$  (\it{r}, \textsf{empower}) \\
\hspace*{0.19cm}$\in$ \it{\ram}(\it{u})) $\vee$ ((\textsf{file, admin}) $\in$ \it{classp}(\it{u}) $\wedge$ \\
\hspace*{0.19cm}(\it{r}, \textsf{empower}) $\in$ \it{\ram}(\it{u})) $\vee$ ((\it{object\_id}(\it{p}), \textsf{admin}) \\
\hspace*{0.19cm}$\in$ \it{\oam}(\it{u}) $\wedge$ (\textsf{role, empower}) $\in$ \it{classp}(\it{u})) \\
\hspace*{0.19cm}$\vee$ (\textsf{file, admin}) $\in$ \it{classp}(\it{u}) \\
\hspace*{0.19cm}$\wedge$ (\textsf{role, empower}) $\in$ \it{classp}(\it{u}))

 \vspace{0.17cm}
 \noindent
 For any permission \it{p} $\in$ \P\ undertaken for revocation,\\
-- {\isauth}P\textsubscript{\textbf{revoke}}(\it{u} : \AU, \it{p} : \P, \it{r} : \R) \\
\hspace*{0.19cm}$\equiv$\\ 
\hspace*{0.19cm}(\it{object\_id}(\it{p}), \textsf{admin}) $\in$ \it{\oam}(\it{u}) $\vee$  (\it{r}, \textsf{admin}) \\
\hspace*{0.19cm}$\in$ \it{\ram}(\it{u}) $\vee$ (\textsf{file}, \textsf{admin}) $\in$ \it{classp}(\it{u}) $\vee$  \\
\hspace*{0.19cm}(\textsf{role}, \textsf{admin}) $\in$ \it{classp}(\it{u})\\

\subsubsection{Map\textsubscript{PRA-UARBAC}}
Algorithm~\ref{alg:PRA-U} maps 
any instance of PRA in UARBAC~\cite{uarbac} (PRA-U) to its equivalent ARPA instance. For clarity, 
sets and function from UARBAC model are labeled with superscript \t{U}, and
that of ARPA with superscript \t{A}. 

Map\textsubscript{PRA-UARBAC} takes following sets and functions as input from PRA-U model. 
\it{C}\textsuperscript{\t{U}}, \it{\U}\textsuperscript{\t{U}}, \it{\R}\textsuperscript{\t{U}}, \it{\P}\textsuperscript{\t{U}}, \it{\PA}\textsuperscript{\t{U}}, \it{\RH}\textsuperscript{\t{U}}, \it{AM}\textsuperscript{\t{U}}(\textsf{role}), \it{AM}\textsuperscript{\t{U}}(\textsf{file}),
For each \it{u} $\in$ \it{\U}\textsuperscript{\t{U}}, authorized\_perms[\it{u}], 
For each [\textsf{file}, \it{o\textsubscript1, a}] $\in$ \it{\P}\textsuperscript{\t{U}} and for each \it{r\textsubscript1} $\in$ \it{\R}\textsuperscript{\t{U}}, grantObjPermToRole([\textsf{file}, \it{o\textsubscript1, a}], \it{r\textsubscript1}) is true if the granter has one of the following combination of permissions:
\begin{enumerate}
\item {[\textsf{file}, \it{o\textsubscript1}, \textsf{admin}] and [\textsf{role}, \it{r\textsubscript1}, \textsf{empower}]}, or 
\item {[\textsf{file}, \it{o\textsubscript1}, \textsf{admin}] and [\textsf{role}, \textsf{empower}]}, or 
\item {[\textsf{file}, \textsf{admin}] and [\textsf{role}, \it{r\textsubscript1}, \textsf{empower}]}, or 
\item {[\textsf{file}, \textsf{admin}] and [\textsf{role}, \textsf{empower}]}
\end{enumerate}

\noindent
For each [\textsf{file}, \it{o\textsubscript1, a}] $\in$ \it{\P}\textsuperscript{\t{U}} and for each \it{r\textsubscript1} $\in$ \it{\R}\textsuperscript{\t{U}}, revokeObjPermFromRole([\textsf{file}, \it{o\textsubscript1, a}], \it{r\textsubscript1}) is true if the granter has either of the following permissions:
\begin{enumerate}
\item {[\textsf{file}, \it{o\textsubscript1}, \textsf{admin}]} or, 
\item {[\textsf{role}, \it{r\textsubscript1}, \textsf{admin}]} or,
\item {[\textsf{file}, \textsf{admin}]} or, 
\item {[\textsf{role}, \textsf{admin}]}
\end{enumerate}

Output from Map\textsubscript{PRA-UARBAC} algorithm is an equivalent ARPA instance, with primarily consisting of \AU\textsuperscript{\t{A}}, \OP\textsuperscript{\t{A}}, \R\textsuperscript{\t{A}}, \RH\textsuperscript{\t{A}}, \P\textsuperscript{\t{A}}, \AATT\textsuperscript{\t{A}}, \PATT\textsuperscript{\t{A}},
For each attribute \it{att} $\in$  \AATT\textsuperscript{\t{A}} $\cup$ \PATT\textsuperscript{\t{A}},
\scope(\it{att}), attType(\it{att}), \isord(\it{att}) and H\textsubscript{\it{att}}, 
For each user \it{u} $\in$ \AU\textsuperscript{\t{A}}, and for each \it{att} $\in$ \AATT\textsuperscript{\t{A}} $\cup$ \PATT\textsuperscript{\t{A}}, \it{att}(\it{u}),
Authorization rule for assign (auth\_assign), and Authorization rule for revoke (auth\_revoke)

Step 1 in Map\textsubscript{PRA-UARBAC} involves translating sets and functions from 
PRA-U to ARPA equivalent sets and functions. In Step 2, permission attributes and 
admin user attributes functions are defined. There exists one permission attributes \it{object\_id}, which captures id of an object for given permission. Note that a permission defines class type, object id and access mode. ARPA defines three admin user attributes: \it{\oam, \ram} and \it{classp}. \it{\oam} attribute captures an admin user's access mode towards an object. Similarly, \it{\ram} captures an admin user's access mode towards a role. An admin user can also have a class level access mode captured by attribute \it{classp}. With class level access mode, an admin user gains authority over an entire class of objects. For example [\textsf{grant, role}] admin permission provides an admin user with power to grant
any role.

In Step 3, assign\_formula for ARPA that is equivalent to
grantObjPermToRole([\textsf{file}, \it{o\textsubscript1, a}], \it{r\textsubscript1}) in PRA-U is established.
Equivalent assign\_formula is expressed as {\isauth}P\textsubscript{\textbf{assign}}(\it{au\textsubscript1} : \AU\textsuperscript{\t{A}}, \it{p} : \P\textsuperscript{\t{A}}, \it{r\textsubscript1} : \R\textsuperscript{\t{A}}) using attributes of permissions and admin user.
Step 4 establishes revoke\_formula equivalent to revokeRoleFromUser(\it{u\textsubscript1,
r\textsubscript1}). It is expressed as {\isauth}P\textsubscript{\textbf{revoke}}(\it{au\textsubscript1} : \AU\textsuperscript{\t{A}}, \it{p} : \P\textsuperscript{\t{A}}, \it{r\textsubscript1} : \R\textsuperscript{\t{A}}) using attributes of permissions and admin user.

\section{Conclusion}\label{sec:conclude}
In this paper, we presented our design for attribute based administration
of RBAC (AARBAC). We developed AURA model for user-role assignment and ARPA
model for permission-role assignment. We then supported these models 
with their extensive example instances and mapping algorithms. Both the models utilized attributes
of RBAC components in making assignment or revocation decision. Role-role
assignment (RRA) is an essential part of RBAC administration. We view
attribute based RRA as our immediate future work. 

One of the motivations
behind our design approach for AURA and ARPA models was to make them
sufficient enough to represent prior ARBAC models. For that matter, we have
presented Map\textsubscript{URA97}, Map\textsubscript{URA99}, Map\textsubscript{URA02}, MAP\textsubscript{URA-Uni-ARBAC} and Map\textsubscript{URA-UARBAC} algorithms that demonstrated m
mapping of prior URA model instances into their respective
equivalent AURA instances. Similarly, we presented Map\textsubscript{PRA97}, Map\textsubscript{PRA99}, Map\textsubscript{PRA02}, MAP\textsubscript{PRA-Uni-ARBAC} and Map\textsubscript{PRA-UARBAC}
algorithms that demonstrated mapping of prior PRA model instances into their equivalent
ARPA instances. 
We note that our models are not limited to expressing prior ARBAC models and
carries the potential to express more.

\section*{Acknowledgment}
This work is partially supported by NSF grants CNS-1423481 and CNS-1553696.

\ifCLASSOPTIONcaptionsoff
  \newpage
\fi

\bibliographystyle{IEEEtran}
\bibliography{bib/References}

\end{document}